\documentclass[
  reprint,
  superscriptaddress,
  amsmath,amssymb,
  aps,
]{revtex4-1}

\usepackage[dvipdfmx]{graphicx}
\usepackage{mediabb} 
\usepackage{lineno} 
\usepackage{diagbox}
\usepackage{hyperref}

\usepackage{dcolumn} 
\usepackage{bm}
\usepackage{footmisc}
\usepackage{url}
\usepackage{color}
\usepackage{multirow}
\usepackage{afterpage}
\usepackage{rotating}
\usepackage[export]{adjustbox}

\begin{document}

\title{Diffuse Supernova Neutrino Background Search at Super-Kamiokande}

\newcommand{\AFFicise}{\affiliation{Institute For Interdisciplinary Research in Science and Education, ICISE, Quy Nhon, 55121, Vietnam }}
\newcommand{\AFFtohoku}{\affiliation{Department of Physics, Faculty of Science, Tohoku University, Sendai, Miyagi, 980-8578, Japan }}
\newcommand{\AFFicrr}{\affiliation{Kamioka Observatory, Institute for Cosmic Ray Research, University of Tokyo, Kamioka, Gifu 506-1205, Japan}}
\newcommand{\AFFkashiwa}{\affiliation{Research Center for Cosmic Neutrinos, Institute for Cosmic Ray Research, University of Tokyo, Kashiwa, Chiba 277-8582, Japan}}
\newcommand{\AFFicrronly}{\affiliation{Institute for Cosmic Ray Research, University of Tokyo, Kashiwa, Chiba 277-8582, Japan}}
\newcommand{\AFFipmu}{\affiliation{Kavli Institute for the Physics and
Mathematics of the Universe (WPI), The University of Tokyo Institutes for Advanced Study,
University of Tokyo, Kashiwa, Chiba 277-8583, Japan }}
\newcommand{\AFFmad}{\affiliation{Department of Theoretical Physics, University Autonoma Madrid, 28049 Madrid, Spain}}
\newcommand{\AFFubc}{\affiliation{Department of Physics and Astronomy, University of British Columbia, Vancouver, BC, V6T1Z4, Canada}}
\newcommand{\AFFbu}{\affiliation{Department of Physics, Boston University, Boston, MA 02215, USA}}
\newcommand{\AFFuci}{\affiliation{Department of Physics and Astronomy, University of California, Irvine, Irvine, CA 92697-4575, USA }}
\newcommand{\AFFcsu}{\affiliation{Department of Physics, California State University, Dominguez Hills, Carson, CA 90747, USA}}
\newcommand{\AFFcnm}{\affiliation{Institute for Universe and Elementary Particles, Chonnam National University, Gwangju 61186, Korea}}
\newcommand{\AFFduke}{\affiliation{Department of Physics, Duke University, Durham NC 27708, USA}}
\newcommand{\AFFfukuoka}{\affiliation{Junior College, Fukuoka Institute of Technology, Fukuoka, Fukuoka 811-0295, Japan}}
\newcommand{\AFFgifu}{\affiliation{Department of Physics, Gifu University, Gifu, Gifu 501-1193, Japan}}
\newcommand{\AFFgist}{\affiliation{GIST College, Gwangju Institute of Science and Technology, Gwangju 500-712, Korea}}
\newcommand{\AFFuh}{\affiliation{Department of Physics and Astronomy, University of Hawaii, Honolulu, HI 96822, USA}}
\newcommand{\AFFicl}{\affiliation{Department of Physics, Imperial College London , London, SW7 2AZ, United Kingdom }}
\newcommand{\AFFkek}{\affiliation{High Energy Accelerator Research Organization (KEK), Tsukuba, Ibaraki 305-0801, Japan }}
\newcommand{\AFFkobe}{\affiliation{Department of Physics, Kobe University, Kobe, Hyogo 657-8501, Japan}}
\newcommand{\AFFkyoto}{\affiliation{Department of Physics, Kyoto University, Kyoto, Kyoto 606-8502, Japan}}
\newcommand{\AFFliv}{\affiliation{Department of Physics, University of Liverpool, Liverpool, L69 7ZE, United Kingdom}}
\newcommand{\AFFmiyagi}{\affiliation{Department of Physics, Miyagi University of Education, Sendai, Miyagi 980-0845, Japan}}
\newcommand{\AFFnagoya}{\affiliation{Institute for Space-Earth Environmental Research, Nagoya University, Nagoya, Aichi 464-8602, Japan}}
\newcommand{\AFFkmi}{\affiliation{Kobayashi-Maskawa Institute for the Origin of Particles and the Universe, Nagoya University, Nagoya, Aichi 464-8602, Japan}}
\newcommand{\AFFpol}{\affiliation{National Centre For Nuclear Research, 02-093 Warsaw, Poland}}
\newcommand{\AFFsuny}{\affiliation{Department of Physics and Astronomy, State University of New York at Stony Brook, NY 11794-3800, USA}}
\newcommand{\AFFokayama}{\affiliation{Department of Physics, Okayama University, Okayama, Okayama 700-8530, Japan }}
\newcommand{\AFFosaka}{\affiliation{Department of Physics, Osaka University, Toyonaka, Osaka 560-0043, Japan}}
\newcommand{\AFFox}{\affiliation{Department of Physics, Oxford University, Oxford, OX1 3PU, United Kingdom}}
\newcommand{\AFFqmul}{\affiliation{School of Physics and Astronomy, Queen Mary University of London, London, E1 4NS, United Kingdom}}
\newcommand{\AFFregina}{\affiliation{Department of Physics, University of Regina, 3737 Wascana Parkway, Regina, SK, S4SOA2, Canada}}
\newcommand{\AFFseoul}{\affiliation{Department of Physics, Seoul National University, Seoul 151-742, Korea}}
\newcommand{\AFFsheff}{\affiliation{Department of Physics and Astronomy, University of Sheffield, S3 7RH, Sheffield, United Kingdom}}
\newcommand{\AFFshizuokasc}{\affiliation{Department of Informatics in
Social Welfare, Shizuoka University of Welfare, Yaizu, Shizuoka, 425-8611, Japan}}
\newcommand{\AFFstfc}{\affiliation{STFC, Rutherford Appleton Laboratory, Harwell Oxford, and Daresbury Laboratory, Warrington, OX11 0QX, United Kingdom}}
\newcommand{\AFFskk}{\affiliation{Department of Physics, Sungkyunkwan University, Suwon 440-746, Korea}}
\newcommand{\AFFtokyo}{\affiliation{The University of Tokyo, Bunkyo, Tokyo 113-0033, Japan }}
\newcommand{\AFFtodai}{\affiliation{Department of Physics, University of Tokyo, Bunkyo, Tokyo 113-0033, Japan }}
\newcommand{\AFFtit}{\affiliation{Department of Physics,Tokyo Institute of Technology, Meguro, Tokyo 152-8551, Japan }}
\newcommand{\AFFtus}{\affiliation{Department of Physics, Faculty of Science and Technology, Tokyo University of Science, Noda, Chiba 278-8510, Japan }}
\newcommand{\AFFtoronto}{\affiliation{Department of Physics, University of Toronto, ON, M5S 1A7, Canada }}
\newcommand{\AFFtriumf}{\affiliation{TRIUMF, 4004 Wesbrook Mall, Vancouver, BC, V6T2A3, Canada }}
\newcommand{\AFFtokai}{\affiliation{Department of Physics, Tokai University, Hiratsuka, Kanagawa 259-1292, Japan}}
\newcommand{\AFFtsinghua}{\affiliation{Department of Engineering Physics, Tsinghua University, Beijing, 100084, China}}
\newcommand{\AFFynu}{\affiliation{Department of Physics, Yokohama National University, Yokohama, Kanagawa, 240-8501, Japan}}
\newcommand{\AFFllr}{\affiliation{Ecole Polytechnique, IN2P3-CNRS, Laboratoire Leprince-Ringuet, F-91120 Palaiseau, France }}
\newcommand{\AFFbari}{\affiliation{ Dipartimento Interuniversitario di Fisica, INFN Sezione di Bari and Universit\`a e Politecnico di Bari, I-70125, Bari, Italy}}
\newcommand{\AFFnapoli}{\affiliation{Dipartimento di Fisica, INFN Sezione di Napoli and Universit\`a di Napoli, I-80126, Napoli, Italy}}
\newcommand{\AFFroma}{\affiliation{INFN Sezione di Roma and Universit\`a di Roma ``La Sapienza'', I-00185, Roma, Italy}}
\newcommand{\AFFpadova}{\affiliation{Dipartimento di Fisica, INFN Sezione di Padova and Universit\`a di Padova, I-35131, Padova, Italy}}
\newcommand{\AFFkeio}{\affiliation{Department of Physics, Keio University, Yokohama, Kanagawa, 223-8522, Japan}}
\newcommand{\AFFwinnipeg}{\affiliation{Department of Physics, University of Winnipeg, MB R3J 3L8, Canada }}
\newcommand{\AFFkcl}{\affiliation{Department of Physics, King's College London, London, WC2R 2LS, UK }}
\newcommand{\AFFwarwick}{\affiliation{Department of Physics, University of Warwick, Coventry, CV4 7AL, UK }}
\newcommand{\AFFral}{\affiliation{Rutherford Appleton Laboratory, Harwell, Oxford, OX11 0QX, UK }}
\newcommand{\AFFwu}{\affiliation{Faculty of Physics, University of Warsaw, Warsaw, 02-093, Poland }}
\newcommand{\AFFbcit}{\affiliation{Department of Physics, British Columbia Institute of Technology, Burnaby, BC, V5G 3H2, Canada }}

\AFFicrr
\AFFkashiwa
\AFFicrronly
\AFFmad
\AFFbu
\AFFbcit
\AFFuci
\AFFcsu
\AFFcnm
\AFFduke
\AFFllr
\AFFfukuoka
\AFFgifu
\AFFgist
\AFFuh
\AFFicl
\AFFbari
\AFFnapoli
\AFFpadova
\AFFroma
\AFFkeio
\AFFkcl
\AFFkek
\AFFkobe
\AFFkyoto
\AFFliv
\AFFmiyagi
\AFFnagoya
\AFFkmi
\AFFpol
\AFFsuny
\AFFokayama
\AFFox
\AFFral
\AFFseoul
\AFFsheff
\AFFshizuokasc
\AFFstfc
\AFFskk
\AFFtokai
\AFFtokyo
\AFFtodai
\AFFipmu
\AFFtit
\AFFtus
\AFFtoronto
\AFFtriumf
\AFFtsinghua
\AFFwu
\AFFwarwick
\AFFwinnipeg
\AFFynu

\author{K.~Abe}
\AFFicrr
\AFFipmu
\author{C.~Bronner}
\AFFicrr
\author{Y.~Hayato}
\AFFicrr
\AFFipmu
\author{K.~Hiraide}
\author{M.~Ikeda}
\author{S.~Imaizumi}
\AFFicrr
\author{J.~Kameda}
\AFFicrr
\AFFipmu
\author{Y.~Kanemura}
\author{Y.~Kataoka}
\author{S.~Miki}
\AFFicrr
% ...


\collaboration{Super-Kamiokande Collaboration}
\noaffiliation

\date{\today}

\begin{abstract}
    A new search for the diffuse supernova neutrino background (DSNB) flux has been conducted at Super-Kamiokande (SK), with a $22.5\times2970$-kton$\cdot$day exposure from its fourth operational phase IV. The new analysis improves on the existing background reduction techniques and systematic uncertainties and takes advantage of an improved neutron tagging algorithm to lower the energy threshold compared to the previous phases of SK. This  allows for setting the world's most stringent upper limit on the extraterrestrial $\bar{\nu}_e$ flux, for neutrino energies below 31.3~MeV. The SK-IV results are combined with the ones from the first three phases of SK to perform a joint analysis using $22.5\times5823$ kton$\cdot$days of data. This analysis has the world's best sensitivity to the DSNB $\bar{\nu}_e$ flux, comparable to the predictions from various models. For neutrino energies larger than 17.3~MeV, the new combined $90\%$ C.L. upper limits on the DSNB $\bar{\nu}_e$ flux lie around $2.7$~cm$^{-2}$$\cdot$$\text{sec}^{-1}$, strongly disfavoring the most optimistic predictions. Finally, potentialities of the gadolinium phase of SK and the future Hyper-Kamiokande experiment are discussed.  
\end{abstract}

\maketitle

\section{Introduction}
\label{sec:intro}
\subsection{Diffuse supernova neutrino background}
\label{subsec:dsnbsignal}
Core-collapse supernovae (CCSNe) are among the most cataclysmic phenomena in the Universe and are essential elements of dynamics of the cosmos. Their underlying mechanism is however still poorly understood, as characterizing it would require intricate knowledge of the core of the collapsing star. Information about this core could be accessed by detecting neutrinos emitted by supernova bursts, whose luminosity and energy spectra closely track the different steps of the CCSN mechanism in a neutrino-heating scenario (see Refs.~\cite{bib:fukugitatext,bib:kotake2006,bib:nakazato2013}, for example). Existing neutrino experiments, however, are mostly sensitive to supernova bursts occurring in our galaxy and its immediate surroundings, that are extremely rare (a few times per century in our galaxy~\cite{bib:snrate1,bib:snrate2}). Learning about the aggregate properties of supernovae in the Universe hence requires observing the accumulation of neutrinos from all distant supernovae. The integrated flux of these neutrinos forms the diffuse supernova neutrino background, or DSNB (DSNB neutrinos are also referred to as supernova relic neutrinos (SRNs) in various publications). 

\begin{figure*}[htbp]
  \begin{center}
   \includegraphics[clip,width=15.0cm]{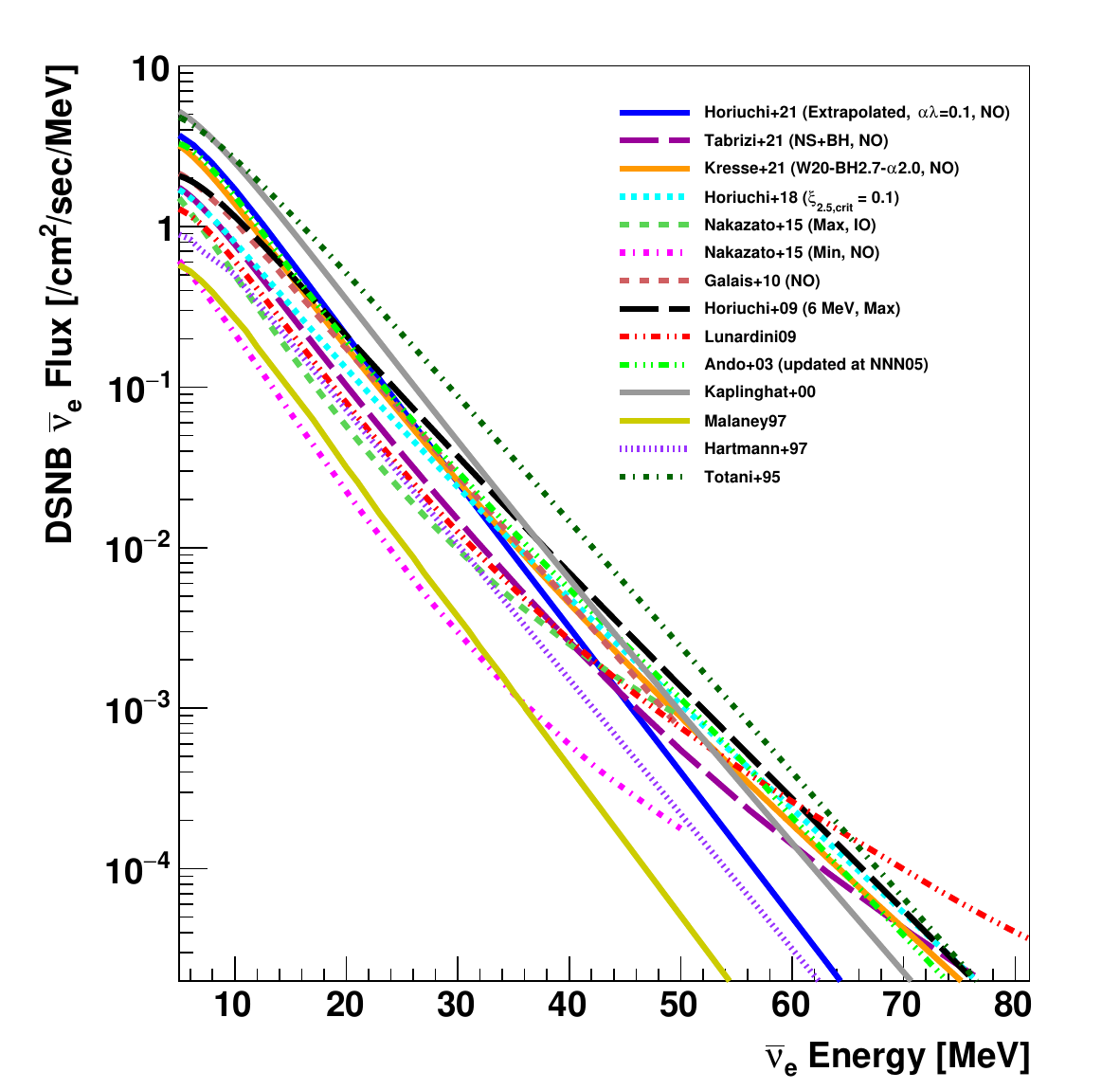}
  \end{center}
  \vspace{-5pt}
  \caption{DSNB $\bar{\nu}_e$ flux predictions from various theoretical models 
           (Horiuchi$+$21~\cite{bib:horiuchi21}, Tabrizi$+$21~\cite{bib:tabrizi21}, 
           Kresse$+$21~\cite{bib:kresse21}, Horiuchi$+$18~\cite{bib:horiuchi18}, 
           Nakazato$+$15~\cite{bib:nakazato15}, Galais$+$10~\cite{bib:galais10}, 
           Horiuchi$+$09~\cite{bib:horiuchi09}, Lunardini09~\cite{bib:lunardini09}, 
           Ando$+$09~\cite{bib:ando03}, Kaplinghat$+$00~\cite{bib:kaplinghat00}, 
           Malaney97~\cite{bib:malaney97}, Hartmann$+$97~\cite{bib:hartmann97}, and 
           Totani$+$95~\cite{bib:totani95}).
           Refer to each publication for the detailed descriptions of model.
           In the legend, ``NO" and ``IO" represent neutrino normal and inverted mass orderings assumed in the calculation, respectively. For the Horiuchi$+$09 model with a 6 MeV temperature, only the maximal flux prediction is shown.
           The prediction for the Galais$+$10 model here is extrapolated up to 50~MeV as the original publication was served up to 40~MeV. 
           The prediction by Nakazato$+$15 is only available up to 50~MeV.
           The values of the flux used in this analysis for the Ando$+$03 model are the ones released at the NNN05 conference~\cite{bib:nnn05}.  The corresponding flux is larger by a factor of 2.56 than in the original publication~\cite{bib:ando03}.
}
  \label{fig:srnmodels}
  \end{figure*}

The DSNB is composed of neutrinos of all flavors whose energies have been redshifted when propagating to Earth. Its spectrum therefore contains unique information not only on the supernova neutrino emission process but also on the star formation and Universe expansion history. Spectra predicted from various models are shown in Fig.~\ref{fig:srnmodels}. Their overall normalization is mostly determined by the supernova rate, related to the cosmic star formation rate. While this redshift-dependent rate also impacts the DSNB spectral shape, the latter is mainly affected by the effective energies of supernova neutrinos.
Other factors shaping the DSNB spectrum include the neutrino mass ordering, the initial mass function of progenitors, the equation-of-state of neutron stars, the fraction of black-hole-forming supernovae, the revival time of the shock wave, etc. Possible combinations of these factors were systematically studied in the Nakazato$+$15 model~\cite{bib:nakazato15}, the maximal and minimal fluxes of which are shown in the figure. The minimal prediction from this model and the Malaney97 gas infall model~\cite{bib:malaney97} give the lower bounds on the DSNB flux. Conversely the Totani$+$95 model~\cite{bib:totani95} can be considered an upper bound on the expected DSNB flux, on par with the most optimistic predictions of the Kaplinghat$+$00 model~\cite{bib:kaplinghat00}, also shown in the figure. Between these bounds, the DSNB flux can vary over one order of magnitude and its spectral shape bears the imprint of a wide range of physical effects. Black-hole-forming supernovae notably make the DSNB spectrum harder, as can be seen with the Lunardini09 model~\cite{bib:lunardini09}, that assumes that $17\%$ of core-collapse supernovae lead to black-hole formation. Similarly, the Horiuchi$+$18 model~\cite{bib:horiuchi18} incorporates a fraction of black-hole-forming supernovae determined using the stars' compactness. Aside from accounting for black-hole formation, the Kresse$+$21 model~\cite{bib:kresse21}, on the other hand, uses state-of-the-art supernova simulations to model contributions from helium stars to the DSNB. Another recent study (Horiuchi$+$21~\cite{bib:horiuchi21}) discusses the impact of interactions in binary systems, such as mergers, and mass transfer, on the DSNB flux. Detecting the DSNB would hence provide valuable insights into a wide array of physical processes.
\vspace{1cm}\\

\subsection{Experimental searches}
\label{subsec:expsearches}

\begin{figure}[htbp]
    \centering
    \includegraphics[width=8.0cm]{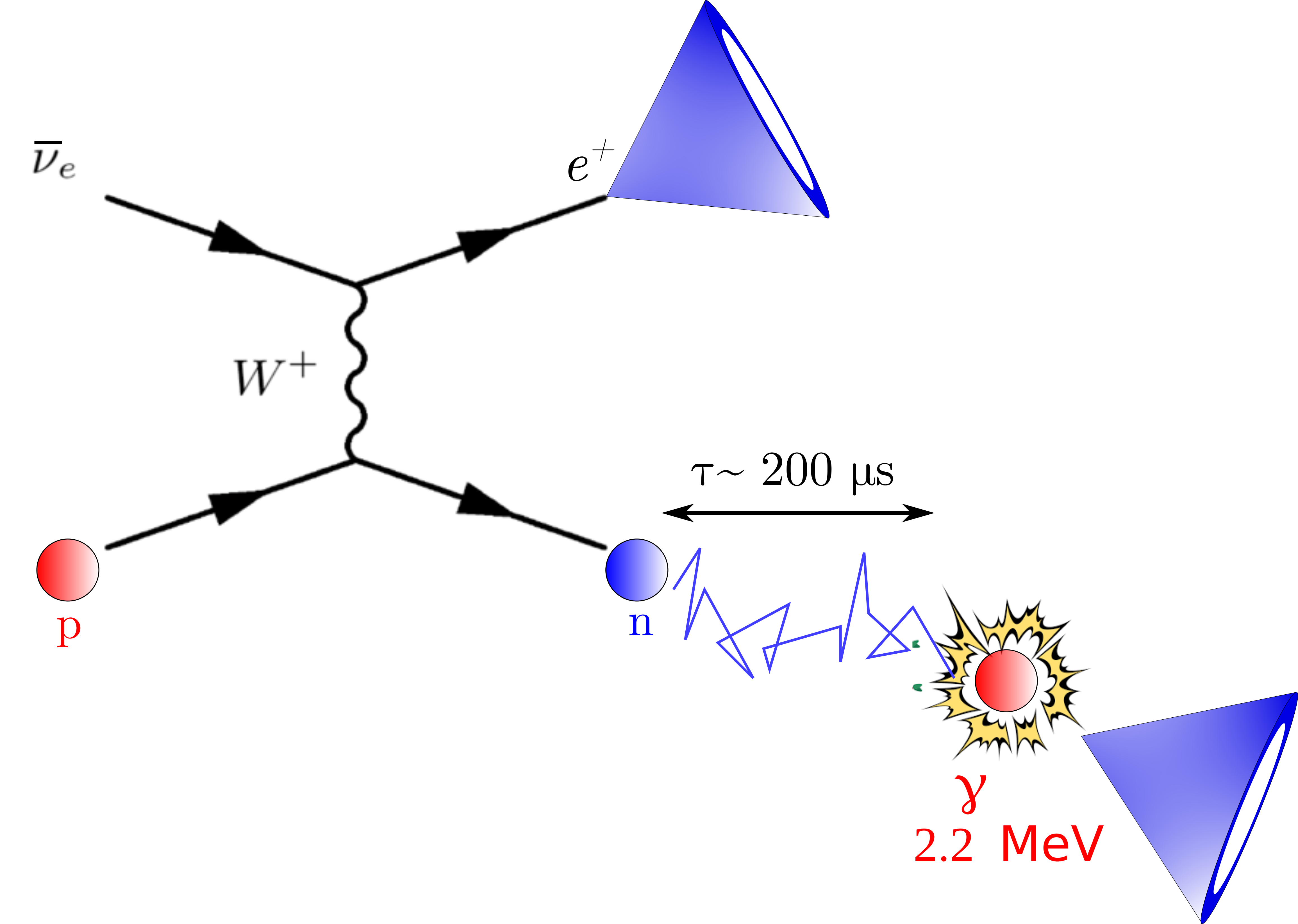}
    \caption{Schematic illustration of an IBD process and the subsequent 
             neutron capture on another proton. 
             The characteristic neutron capture time in water is 
             $\tau = 204.8\pm 0.4~\mu$s~\cite{ncaptime}.}
    \label{fig:ibdscheme}
  \end{figure}

While a significant fraction of DSNB neutrinos are expected to have energies lower than 10~MeV ---as shown in Fig.~\ref{fig:srnmodels}---, the $\mathcal{O}(1)$~MeV region is hard to probe by current experiments due to overwhelming backgrounds from reactor antineutrinos, solar neutrinos, and radioactivity. Experimentally the DSNB signal has therefore been searched for at energies of $\mathcal{O}(10)$~MeV.   
At these energies, the dominant detection channel in most experiments is the inverse beta decay (IBD) of electron antineutrinos ($\bar{\nu}_{e}+p \rightarrow e^{+}+n$), and contributions from subleading channels such as electron neutrino elastic scattering or charged current interactions on oxygen can be neglected. This process produces an easily identifiable positron and a neutron (Fig.~\ref{fig:ibdscheme}). Here the neutron does not emit Cherenkov or scintillation light directly but its capture on a proton leads to emission of a $2.2$~MeV photon. In pure water, the characteristic timescale for neutron moderation and capture is of $204.8 \pm 0.4~\mu$s~\cite{ncaptime}. Pairing the ``prompt'' positron signal with the ``delayed'' photon emission from neutron capture is hence a key component of many analyses. 

Up to now, no evidence for the DSNB signal has been confirmed and upper limits have been set by various underground experiments, such as Super-Kamiokande (SK)~\cite{bib:sksrn1,bib:sksrn123,bib:sksrn4}, KamLAND~\cite{bib:kamlandsrn,bib:kamlandsrnnew}, SNO~\cite{bib:snosrn,bib:snosrnnew} and Borexino~\cite{bib:borexinosrn}. Note that, unlike other searches, the SNO analysis is sensitive to electron neutrinos and, due to the irreducible solar neutrino background, its effective energy threshold lies around the {\it hep} solar neutrino flux endpoint, around 19~MeV. Among all past analyses, the SK and KamLAND experiments placed the most stringent upper limits on the DSNB $\bar{\nu}_e$ flux for neutrino energies above about 9~MeV while Borexino set the tightest constraints at lower energies. At SK the first DSNB search was carried out in 2003 using a $22.5\times1496$-kton$\cdot$day data set~\cite{bib:sksrn1}. Using spectral shape fitting for signal and atmospheric neutrino backgrounds, it placed an upper limit on the DSNB flux for a wide variety of models in the 19.3$-$83.3~MeV neutrino energy range. This analysis already allowed to disfavor the most optimistic DSNB predictions, in particular the Totani$+$95 model~\cite{bib:totani95}, and constrain the parameter space of the Kaplinghat$+$00 model~\cite{bib:kaplinghat00}. In 2012, an improved analysis was performed at SK, using a $22.5\times2853$-kton$\cdot$day exposure, a lower neutrino energy threshold of $17.3$~MeV, and new event selection cuts allowing a 50\% increase in signal efficiency~\cite{bib:sksrn123}.
Below 17.3~MeV, large backgrounds from cosmic muon spallation and solar neutrino interactions make it extremely difficult to search for DSNB antineutrinos based on positron identification alone. 
A search for extragalactic antineutrinos performed at KamLAND in 2021~\cite{bib:kamlandsrnnew} probed neutrino energies ranging from $8.3$ to $30.8$~MeV by investigating coincident positron and neutron capture signals. While neutron identification in a liquid scintillator detector such as KamLAND is significantly easier than in pure water, KamLAND's small fiducial volume only allowed for an exposure of $6.72$~kton$\cdot$year. In 2015, a new SK analysis including neutron identification led to stronger constraints down to $13.3$~MeV neutrino energies~\cite{bib:sksrn4}. Since the SK triggers did not allow to record the neutron capture signal until 2008, this analysis could only be performed on a small part of the SK data set, with a total livetime of $22.5\times960$~kton$\cdot$days. Due to this low exposure and the low neutron tagging efficiency in water, this search however yielded weaker limits than the SK-I,II,III analysis~\cite{bib:sksrn123} above 17.3~MeV.

In this study, we draw on the previous SK analyses to present two DSNB searches for antineutrino energies ranging from $9.3$ to $81.3$~MeV, with significantly improved background modeling and reduction techniques. In the $9.3$ to $31.3$~MeV range, we derive differential upper limits on the $\bar{\nu}_e$ flux independently from the DSNB model, following the strategy outlined in Ref.~\cite{bib:sksrn4} using a $22.5\times2970$-kton$\cdot$day data set. In the $17.3$ to $81.3$~MeV range we constrain a wide variety of DSNB models using spectral fits analogous to the ones described in Ref.~\cite{bib:sksrn123}. We then combine the results of this analysis with the ones obtained in Ref.~\cite{bib:sksrn123} for the former SK phases, thus analyzing $22.5\times5823$-kton$\cdot$days of data. This unprecedented exposure will allow to probe the DSNB with an unmatched sensitivity.

The rest of this article proceeds as follows. First, the SK detector and the specific features of its data acquisition system are described in Section~\ref{sec:sk}. Then, details about modeling of the DSNB signal and the different backgrounds are given in Sections~\ref{sec:signalmodeling} and \ref{sec:background}, respectively. We then describe the data reduction process in Section~\ref{sec:reduc}. The procedures associated with the DSNB model-independent and spectral analyses are described in Sections~\ref{sec:mianalysis} and \ref{sec:spectral}, respectively. Finally we discuss the current constraints on the DSNB flux and future opportunities at SK with gadolinium and Hyper-Kamiokande in Section~\ref{sec:discussion} before concluding in Section~\ref{sec:conclusion}.

\section{Super-Kamiokande}
\label{sec:sk}
Super-Kamiokande is a 50-kton water Cherenkov detector located in the Kamioka mine, Japan. It is structured by a cylindrical stainless steel tank with a diameter of $39.3$~m and height of $41.4$~m and consists of two parts: an outer detector (OD) that serves as a muon veto and an inner detector (ID) where neutrino detection takes place. In order to reduce backgrounds due to radioactivity near the detector wall, most analyses consider only events reconstructed at least 2~m away from the ID wall, thus defining a 22.5-kton fiducial volume (FV). It is located 1000~m underground, that allows reduction of the cosmic ray muon flux by a factor of $10^5$. In order to ensure a high-quality data taking, conditions inside SK are tightly controlled; water is constantly recirculated and purified, and the ID wall is covered with 11,129 20-inch photomultipliers (PMTs) with a $3$-ns time resolution, corresponding to a $40\%$ photocathode coverage. These features allow SK to detect particles with energies ranging from a few MeV to a few TeVs. The OD includes 1,885 8-inch PMTs, facing outwards, to detect the Cherenkov light from muons. Further detailed descriptions of the SK detector and its calibration can be found in Refs.~\cite{bib:superk,bib:skidpmt,bib:skcalib,bib:sklinac,bib:skdt}. 

SK is currently undergoing its sixth data taking phase since it started functioning in 1996. The first phase lasted $1497$~days and ended for a scheduled maintenance. Due to an accident following the maintenance, which resulted in a loss of $60$\% of the ID PMTs, SK operated with a reduced photocathode coverage for $794$~days (phase II). The coverage was brought back to its nominal value for phase III, that lasted $562$~days. SK's previous spectral analysis of the DSNB~\cite{bib:sksrn123} used data from all these three phases. Since 2008, the front-end electronics has been replaced \cite{bib:skqbee} and a new trigger system that allows for neutron tagging has been set up \cite{bib:skntagfirst}. SK operated with this new electronics during phase IV, for 2970~days, until being stopped for refurbishment work in May 2018. After the maintenance, SK operated for a short time with pure water for calibration and monitoring purposes (phase V) since January 2019, before being loaded with gadolinium (phase VI) in July 2020. This study will primarily focus on phase IV, and will present a combined analysis of the SK-I to IV data. Additionally, we will briefly discuss the opportunities offered by the gadolinium lodaded SK and by the future Hyper-Kamiokande experiments.

Data processing in SK makes use of multiple triggers, corresponding to different thresholds on the number of PMT hits found in a $200$-ns window.  In SK-I to III, all PMT hits in a 1.3-$\mu$s window around the main activity peak were stored, using hardware triggers. Since SK-IV, the new data acquisition system acquires every PMT signal and a software trigger system defines events. To identify positrons from IBD processes, we use the super-high-energy (``SHE'') trigger, which requires 58 PMT hits (70 hits before September 2011) in $200$~ns. The data in a $[-5,+35]$-$\mu$s window around the main activity peak is collected with this trigger. This window is too short to contain the delayed neutron capture signal after IBDs. In order to identify this signal, a 500-$\mu$s ($350~\mu$s before November 2010) after trigger (``AFT'') window follows each SHE trigger that is not associated with a pre-determined OD trigger.
The trigger conditions for the different periods in SK-IV are summarized in Table~\ref{tab:runsummary}.
In the rest of this paper, we will describe the analysis of hit patterns in these large SHE+AFT windows to identify the coincident production of a positron and a neutron. 

The prompt positron event is reconstructed using the dedicated solar neutrino~\cite{bib:sksolar1,bib:sksolar2,bib:sksolar3,bib:sksolar4} and muon decay electron vertex and direction fitter, which is then used to reconstruct the event energy~\cite{bib:bonsai}. For this SK phase, we follow the convention introduced in Ref.~\cite{bib:sksolar4} and subtract the 0.511~MeV electron mass from this energy to obtain the electron equivalent kinetic energy $E_{\rm rec}$. This quantity can also be interpreted as the total reconstructed positron energy, and will be used to present the results of this study. 
  \begin{table}[htbp]
  \begin{center}
  \caption{Summary on the SHE and AFT trigger conditions and livetime for the different 
          three periods in SK-IV.}
  \label{tab:runsummary}
  \vspace{+3truept}
  \begin{tabular}{c c r} \hline \hline
    SHE-threshold & AFT-window & Livetime \\ \hline
    70~hits & 350~$\mu$s & 25.0~days \\ 
    70~hits & 500~$\mu$s & 869.8~days \\ 
    58~hits & 500~$\mu$s & 2075.3~days \\ \hline \hline
  \end{tabular}
  \end{center}
  \end{table}
\section{Signal modeling}
\label{sec:signalmodeling}
\subsection{DSNB spectra and kinematics}
\label{subsec:dsnbspectra}
In this analysis we consider both the DSNB models whose fluxes are shown in Fig.~\ref{fig:srnmodels} and models where we vary specific physical parameters. For these parameterized models we compute the DSNB flux using the following formula:

\begin{align}
        \nonumber
        \Phi(E_\nu) = &\frac{c}{H_0}\int \sum_{s}  R_{\rm SN}(z, s)\, \sum_{\nu_i,\bar\nu_i} F_{i}\left(E_\nu (1 + z), s\right) \\
        &\times \frac{dz}{\sqrt{\Omega_M(1+z)^3 + \Omega_\Lambda}},
    \label{eq:dsnb}
\end{align}

\noindent 
where $E_\nu$ is the neutrino energy, $c$ is the speed of light in vacuum, $H_0$ is the Hubble constant, $z$ is the redshift, $R_{\rm SN}(z)$ is the redshift-dependent supernova rate, and $F_i$ is the supernova neutrino emission spectrum for a given flavor $i$. The index $s$ represents the different possible classes of supernovae ---e.g. supernovae collapsing into neutron stars or black-holes--- associated with specific neutrino emission spectra. The last factor accounts for the Universe expansion, with $\Omega_M$ and $\Omega_\Lambda$ being the matter and dark energy contributions to the energy density of the Universe, respectively. We evaluate the supernova rate using a phenomenological model described in Ref.~\cite{bib:horiuchi09}, which extracts the redshift dependence of the cosmic star formation history by fitting observations and uses the Salpeter initial mass function~\cite{Salpeter:1955it} to obtain the fraction of supernova progenitors.  We then consider blackbody neutrino emission spectra, where the effective temperature $T_\nu$ takes into account neutrino oscillation effects. 
For a given neutrino flavor $i$ the spectrum for a model with effective temperature $T_\nu$ is thus given by:

\begin{align}
    F_i(E_\nu) &= f_{\nu, i}E_\nu^{\rm tot}\,\frac{120}{7\pi^4}\,\frac{E_{\nu,i}^2}{T_{\nu,i}^4}\left(e^{E_{\nu, i}/T_{\nu, i}} + 1\right)^{-1}, 
    \label{eq:blackbody}
\end{align}

\noindent 
where $E_\nu^{\rm tot}$ is the total energy of the neutrinos emitted by the supernovae and $f_{\nu,i}$ represents the fraction of neutrinos or antineutrinos with flavor $i$ and can be roughly approximated by $1/6$ for a given species. The blackbody model therefore only depends on the neutrino temperatures and luminosities. In what follows, we will refer to blackbody models with a neutrino effective temperature $T$ as Horiuchi$+$09 $T$ MeV models. Note that, due to neutrino scattering in the densest regions of the collapsing star, supernova neutrino emission is better modeled using a more sharply peaked, ``pinched'',  Fermi-Dirac spectrum~\cite{Keil:2002in}; however, the current exposure at SK does not allow to probe this pinching effect. In this analysis we neglect the effects of degeneracies between the pinching parameter and e.g. the neutrino temperature on the final constraints on neutrino emission spectra. 

This analysis exclusively focuses on the detection of electron antineutrinos via IBD processes. In the rest of this paper, we model these processes using the Strumia-Vissani IBD calculation~\cite{bib:ibdxsec}, that approximates the IBD cross section more precisely than the Beacom-Vogel calculation used for previous SK analyses~\cite{bib:sksrn123}.

\subsection{Detector simulation}
\label{subsec:signalsimu}

SK-IV has been the longest data taking phase of the experiment, lasting about $10$ years. During this term, the PMT gain has steadily increased by about
15\% over the whole SK-IV period. Additionally, since the energy scale change due to
the gain change for this analysis was about 3 \%, that effect was
corrected in order to maintain the stability of the detector. A key ingredient of our DSNB study is therefore a simulation of antineutrino IBD processes that accounts for the evolution of the detector's properties throughout the entire SK-IV period. 

In this study we simulate IBD processes uniformly distributed in the entire inner detector, in order to account for events close to the ID wall being misreconstructed inside the fiducial volume. For each vertex we generate a set of three momentum vectors corresponding to an antineutrino, a positron, and a neutron. In a view of the wide diversity of DSNB models, we generate uniformly distributed positron energies in 1$-$90~MeV and renormalize events later to model specific spectra. This procedure will also allow to use this simulation to model backgrounds associated to other IBD processes or $\beta$ decays, as will be mentioned later. We then simulate detection signals using a dedicated simulation, based on GEANT3~\cite{bib:geant3}, that reproduces the properties of the water in SK, as well as models the PMT properties and electronic response. Detector properties such as water transparency and PMT noise are being measured daily, allowing to accurately model the time-dependent noise contamination of the positron signal. Neutron capture, however, produces a $2.2$~MeV $\gamma$ ray whose light yield is below the SK trigger threshold and particularly difficult to distinguish from PMT dark noise and low energy radioactive decays. In order to develop robust neutron identification techniques we therefore collect noise samples from data using a random wide trigger, and inject them into simulation results, after the positron activity peak.
%

\section{Background Sources}
\label{sec:background}
\subsection{Atmospheric neutrinos}
\label{subsec:atmos}
%
Atmospheric neutrinos are an important background in the present DSNB searches. 
Below about 15~MeV, neutral-current quasielastic (NCQE) interactions, which induce nuclear $\gamma$ ray emission~\cite{bib:ankowski,bib:t2kncqe1to3,bib:skncqe,bib:t2kncqe1to9}, form the main atmospheric neutrino background. 
On the other hand, charged-current quasielastic (CCQE) interactions and pion production dominate at higher energies.
A typical visible signature for these interactions is the Cherenkov light from electrons produced by muon and pion decays. Hence, the associated reconstructed energies will follow a Michel spectrum for reconstructed energies of about 15 to 50~MeV. 
Atmospheric neutrino events are simulated using the HKKM 2011 flux 
\cite{bib:hkkm2006,bib:hkkm2011,bib:hkkm2011web} as an input to the neutrino event 
generator NEUT 5.3.6~\cite{bib:neut} to model interactions. 
Details about the model parameters used in this analysis can be found in Ref.~\cite{bib:sksolarantinu}. 
%
%
%
%
The nuclear deexcitation $\gamma$'s are simulated based on the spectroscopic factors for the $p_{\rm 1/2}$, $p_{\rm 3/2}$, and $s_{\rm 1/2}$ states calculated in Ref.~\cite{bib:ankowski}. 
More detailed descriptions of the excited states can be found in Refs.~\cite{bib:t2kncqe1to3,bib:t2kncqe1to9} for the T2K NCQE measurement. 
One difference between our procedure and Ref.~\cite{bib:t2kncqe1to9} is treatment of the {\it others} state ---a state affected by short-range correlations or with a very high excitation energy---. This state is treated as the highest excitation state ($s_{\rm 1/2}$) in Ref.~\cite{bib:t2kncqe1to9} while it is included in the ground state ($p_{\rm 1/2}$) in the present analysis as it is from the latest atmospheric neutrino analysis in SK~\cite{bib:skatmnu2019}. The systematic uncertainty regarding this treatment is considered in the scaling factor used in Section~\ref{subsec:bkgestimate}. 

The detector simulation for atmospheric neutrino events is performed using the same GEANT3-based simulation tool as for the signal, but without corrections for the PMT gain shift on the primary event. 
The associated systematic error is estimated in Section~\ref{subsec:bkgestimate} and found to be subdominant. Finally, random trigger data are overlaid on top of neutron capture signals in order to simulate background for neutron tagging, following the procedure described in Section~\ref{subsec:signalsimu}.
Note that for this last step the time evolution of the detector properties is taken into account. 
%

\subsection{Cosmic ray muon spallation}
\label{subsec:spallation}
%
In spite of its 2700-m water-equivalent overburden, SK is still exposed to cosmic ray muons with a rate of $\sim$2~Hz. Interactions of these muons in water produce electromagnetic and hadronic showers. Spallation of oxygen nuclei induced by the muons or by secondary particles can then lead to the production of radioactive isotopes, whose decays can be misidentified as IBD events in the DSNB search window.
%
Below about 20~MeV, the associated background is $10^6$ times higher than the DSNB flux predictions, making spallation reduction an essential aspect of this analysis. 
Since most spallation isotope decays only produce $\beta$ and $\gamma$ rays, spallation backgrounds can be significantly reduced using neutron tagging. Nonetheless, accidental pairing between prompt $\beta$ or $\gamma$ events and PMT hits due to dark noise or intrinsic radioactivity will allow a sizable number of spallation events to pass as IBDs. Furthermore, a few isotopes undergo a $\beta$+$n$ decay that mimics the IBD signal, and thus cannot be removed using neutron tagging. Efficient spallation reduction therefore requires both dedicated spallation reduction techniques tailored to the properties of the different isotopes, and neutron tagging cuts.
%
%

\begin{figure}[htbp]
  \begin{center}
   \includegraphics[clip,width=9.5cm]{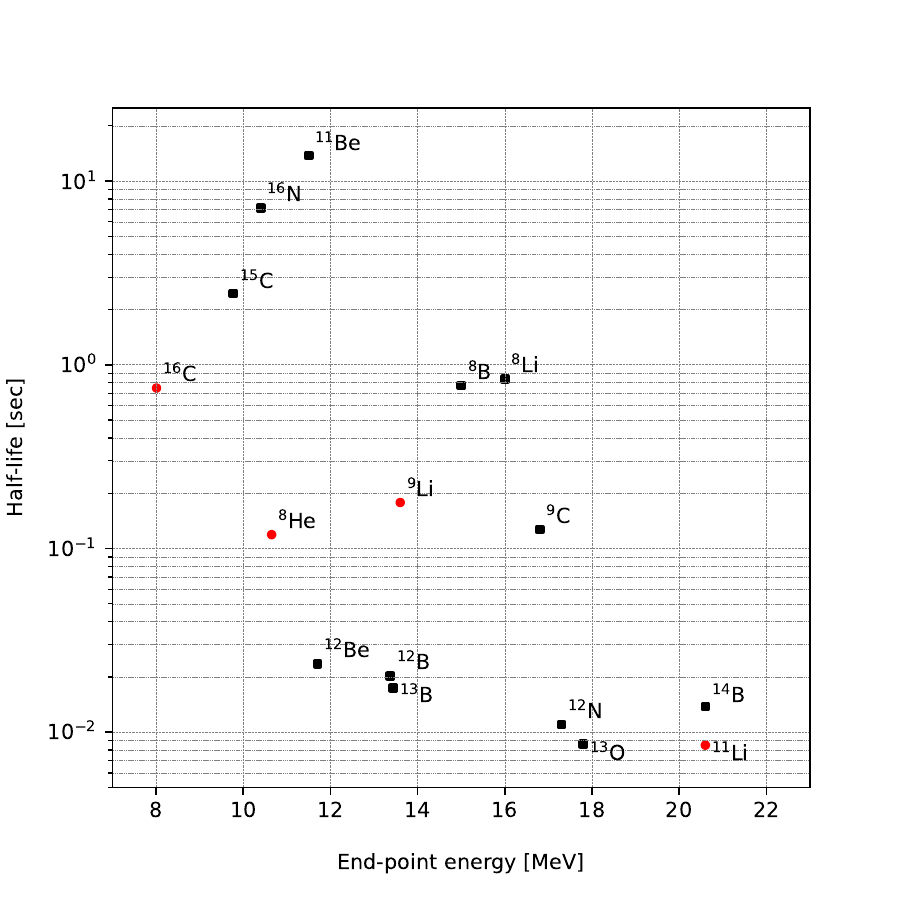}
  \end{center}
  \vspace{-15truept}
  \caption{End-point energies and half-lives of the $\beta$ decay isotopes produced 
           by cosmic ray muon spallation in water, whose decay products are found in the DSNB search window. 
           The isotopes represented by a red circle (a black square) decay with
           (without) neutrons.}
  \label{fig:spall_products}
  \end{figure}

Relevant spallation isotopes and their visibility in water have been studied with the FLUKA simulation \cite{bib:fluka2011} in Refs.~\cite{bib:liandbeacom1,bib:liandbeacom2,bib:liandbeacom3}. 
The isotope lifetimes and the end-point energies of their decays are summarized in Fig.~\ref{fig:spall_products}.
%
%
The highest end-point is 20.6~MeV, from ${\rm ^{14}B}$ and
${\rm ^{11}Li}$. The spallation cuts used in this analysis will therefore be applied up to $23.5$~MeV  reconstructed kinetic electron energy in order to account for energy resolution effects.
%
Note also that the isotope half-lives range from $\mathcal{O}(0.01)$~sec to a few tens of seconds. Given the high rate of cosmic ray muons, pairing a given isotope decay with its parent process is a particularly difficult endeavour.

\begin{figure}[htbp]
    \centering
    \includegraphics[width=\linewidth]{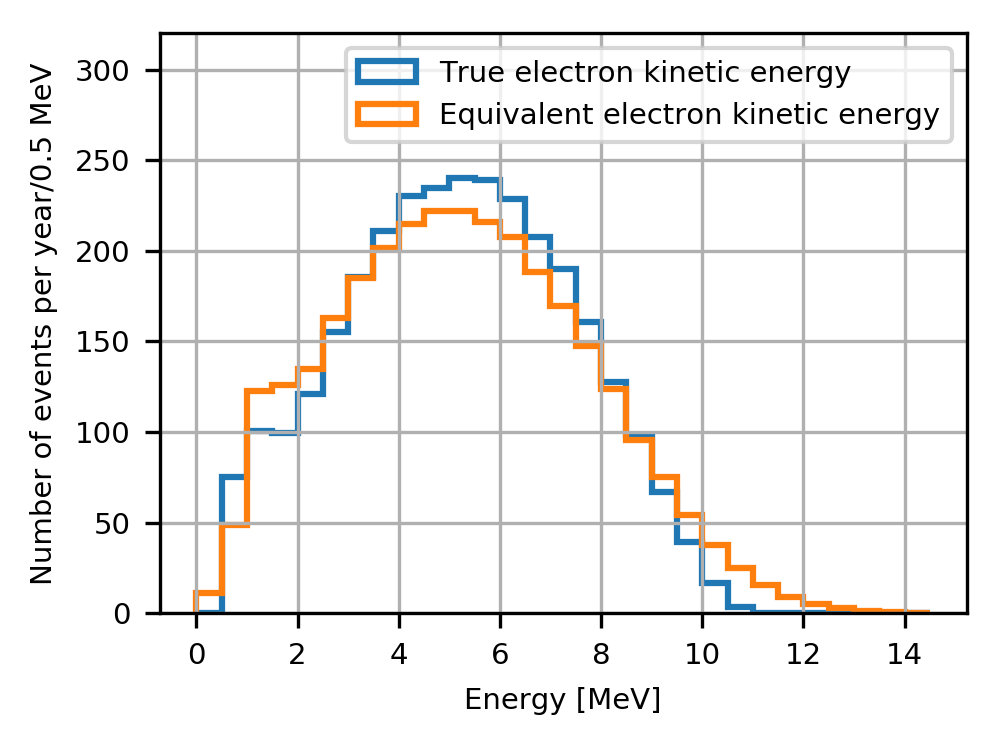}
    \caption{True electron kinetic energy (blue) and reconstructed kinetic energy (orange) spectra from the $^9$Li $\beta$+$n$ decay.}
    \label{fig:li9spec}
\end{figure}

Isotopes that undergo $\beta$$+$$n$ decays cannot be rejected using neutron tagging and therefore need to be modeled with particular care. This is notably the case for $^8$He, $^{11}$Li, $^{16}$C, and $^9$Li. 
Here $^{11}$Li is particularly short-lived and therefore easy to eliminate using 
mild spallation cuts. Its production yield is also expected to be particularly low (around $10^{-9}~\mu^{-1}\cdot$g$^{-1}\cdot$cm$^2$~\cite{bib:liandbeacom1}). 
In addition, $^8$He and $^{16}$C will remain subdominant due to their low production yields (around 0.23 and $0.02 \times 10^{-7}~\mu^{-1}\cdot$g$^{-1}\cdot$cm$^2$~\cite{bib:liandbeacom1}) and endpoint $\beta$ decay energies. 
On the other hand, $^9$Li has a non-negligible yield (around $1.9 \times 10^{-7}~\mu^{-1}\cdot$g$^{-1}\cdot$cm$^2$~\cite{bib:liandbeacom1}), a medium half-life of 
about $0.18$~sec, and decays into a $\beta$$+$$n$ pair with a branching ratio of 50.8\%. The associated $\beta$ spectrum is shown in Fig.~\ref{fig:li9spec}, and extends up to $13.5$~MeV reconstructed energy when accounting for resolution effects. The $\beta$+$n$ decay of $^9$Li is hence similar to the IBD of a DSNB neutrino, except that the $\beta$ from
${\rm ^{9}Li}$ is an electron and the neutron energy is higher than that from IBDs.
Since SK does not distinguish electrons from positrons and does not allow to measure the neutron energy, we model $^9$Li decay using the IBD Monte-Carlo simulation, renormalizing events to reproduce the $^9$Li $\beta$+$n$ spectrum. Finally, we estimate the total expected number of $^9$Li $\beta$+$n$ decays in the DSNB search window by using an earlier SK measurement~\cite{bib:skli9} that estimated the $^9$Li production rate to be $0.86 \pm 0.12({\rm stat.}) \pm 0.15({\rm syst.})~{\rm kton^{-1}}$$\cdot$${\rm day^{-1}}$. Note that the difference in neutron energies with respect to IBD processes leads to a systematic uncertainty. However, this uncertainty is taken into account when calibrating neutron tagging efficiencies, as neutrons from the calibration source have similar energies to the $^9$Li decay products. More details about this calibration procedure are given in Section~\ref{subsec:ntag}.

%
%
%

\subsection{Reactor neutrinos}
\label{subsec:reactor}
\begin{figure}[htbp]
    \centering
    \includegraphics[width=\linewidth]{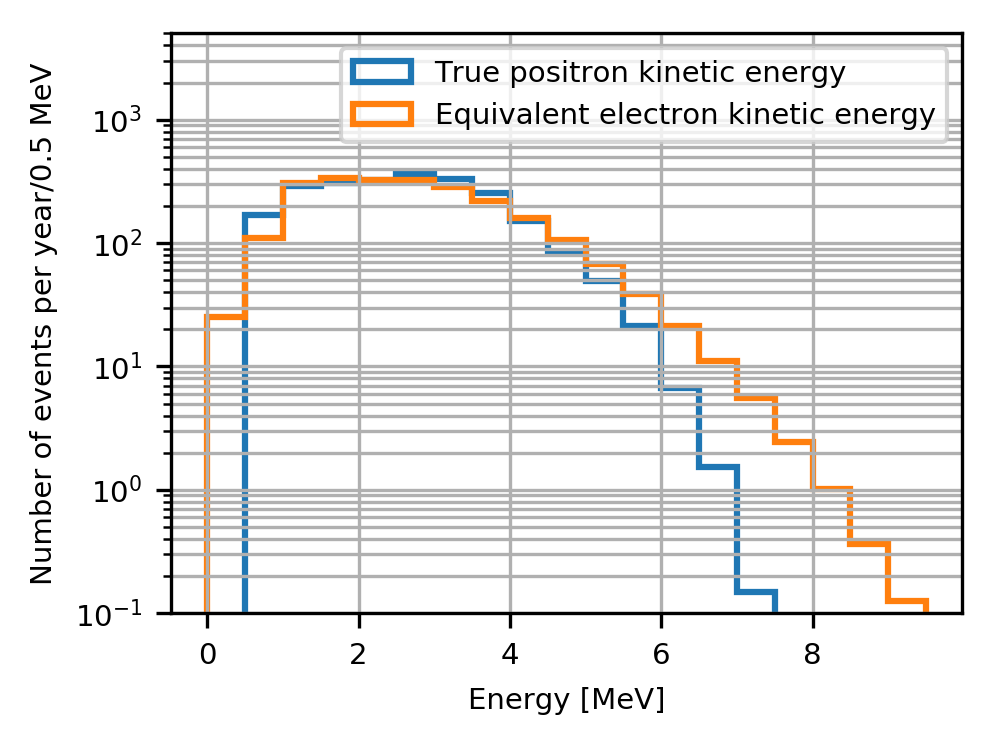}
    \caption{True positron kinetic energy (blue) and reconstructed kinetic energy (orange) spectra from reactor neutrino IBD interactions. 
    The y-axis shown the average number of events per year at SK-IV.
    The event rate is not constant with time due to reactors being powered on during the data taking period.
    }
    \label{fig:reacspec}
\end{figure}

Antineutrinos emitted from nearby nuclear reactors can undergo IBDs in SK and therefore be mistaken as the DSNB signal at low energies. We estimate the reactor antineutrino flux using the \texttt{SKReact} code~\cite{bib:skreact} developed for the SK reactor neutrino analysis. This code uses the reactor neutrino model from Ref.~\cite{bib:reactornumodel} based on IAEA records~\cite{international2005iaea} to evaluate the electron antineutrino fluxes for any reactor in the world, and simulates oscillation effects. The resulting flux prediction also accounts for the time dependence of the reactor antineutrino flux due to changes in reactor activity. The expected total reactor neutrino flux at SK can thus be predicted up to $E_\nu^\text{max} = 9.2$~MeV. The expected reactor neutrino spectrum is obtained from this predicted flux using the IBD simulation, and is shown in Fig.~\ref{fig:reacspec}. Above 6~MeV, as can be seen in this figure, the reconstructed spectrum is primarily shaped by resolution effects. Since the reactor antineutrino spectrum has been constrained up to 12~MeV by Daya Bay~\cite{bib:dayabay}, possible antineutrino contributions above $E_\nu^\text{max}$ can be safely neglected in this study.
In the 7.5$-$9.5~MeV reconstructed kinetic electron energy range, the reactor antineutrino background is about 6 times higher than the DSNB signal predicted by the Horiuchi$+$09 6~MeV model~\cite{bib:horiuchi09}. Reactor antineutrino backgrounds thus effectively set a lower energy threshold for the DSNB search.
%

\subsection{Solar neutrinos}
\label{subsec:solar}
Electron neutrinos from the Sun cause elastic interactions with electrons in SK. In the DSNB search window, the $^8$B and {\it hep} solar neutrino fluxes represent an important background up to $\sim$20~MeV. This background can be drastically reduced by neutron tagging, although a small fraction of electrons from solar neutrinos will be accidentally paired with background fluctuations or radioactive decays, like for the spallation backgrounds. In the absence of neutron tagging, a solar neutrino interaction can still be identified by comparing the reconstructed directions of the scattered electron to the direction of the Sun at the detection time. The angle between these two directions, called the opening angle, is expected to be close to zero for solar electron neutrinos scattering elastically. It is however smeared by reconstruction effects that need to be modeled accurately. In this study, we use the simulation designed for the latest SK-IV solar neutrino analysis~\cite{bib:sksolar4} to model solar neutrino spectra and evaluate the impact of opening angle cuts.

Finally, note that in this analysis we do not study the impact of solar antineutrino production, since the rate predicted by the Standard Model (SM) is negligible. Antineutrino production via beyond the SM processes would yield a signal that can only be distinguished from the DSNB signature by spectral shape studies. So far, only upper limits on the solar antineutrino flux have been derived at SK, notably in Ref.~\cite{bib:sksolarantinu}.

\section{Data reduction}
\label{sec:reduc}

We apply reduction cuts to the data taken by the SHE trigger in SK-IV, corresponding to an exposure of $22.5\times2970.1$~kton$\cdot$days. 
The SHE trigger efficiency is close to $100$\% over the entire analysis window. 
The following sections describe the four reduction steps used in this analysis: noise reduction in Section~\ref{subsec:noise}, spallation reduction in Section~\ref{subsec:spallreduc}, DSNB positron candidate selection in Section~\ref{subsec:posreduc}, and neutron tagging in Section~\ref{subsec:ntag}.

\subsection{Search energy range and noise reduction}
\label{subsec:noise}
The lower energy threshold for this analysis is the SHE trigger threshold, which corresponds to $E_{\rm rec} >$ 7.5 or 9.5~MeV depending on the observation time (see Table~\ref{tab:runsummary}). 
The upper bound of the DSNB analysis window is set to $29.5$~MeV for the model-independent analysis and $79.5$~MeV for the spectral analysis. 
%

We first select SHE-triggered events with $E_{\rm rec}$ below $79.5$~MeV and apply a set of cuts aimed at removing PMT noise, calibration events, radioactivity from the surrounding rock and the detector wall, and cosmic ray activity. In particular, we require candidate events to be located at least $2$~m away from the ID wall, thus defining a fiducial volume (FV) of $22.5$~kton. Additionally, we remove events associated with an OD trigger, or within $50~\mu$s of another event with reconstructed electron kinetic energy larger than $6$~MeV, in order to remove electrons produced by the decay of low energy cosmic ray muons stopping in the detector. In addition, we apply fit quality cuts to eliminate noise-like events. Inside the FV these noise reduction cuts have a signal efficiency larger than $99\%$ as confirmed by the IBD Monte-Carlo simulation.
Finally, we require all SHE events to be followed by an AFT trigger to allow for neutron tagging. Due to occasional deadtime induced by trigger software failure, this step is associated with a $\sim$94\% signal efficiency. Events passing these criteria will be subsequently referred to as ``DSNB candidates''.

\subsection{Spallation reduction}
\label{subsec:spallreduc}
Spallation backgrounds largely dominate over putative DSNB signals for reconstructed electron kinetic energies lower than $23.5$~MeV. Here, we present a set of dedicated cuts that will allow for a substantial reduction of these backgrounds, even in the absence of neutron tagging.
\subsubsection{Spallation preselection} 
\label{sec:spallpresel}

We apply two sets of preselection cuts to reduce contributions from the most energetic muons and some long-lived spallation isotopes such as $^{16}$N, with minimum harm to the signal efficiency. 

\begin{figure}[htbp]
  \begin{center}
   \includegraphics[clip,width=8.0cm]{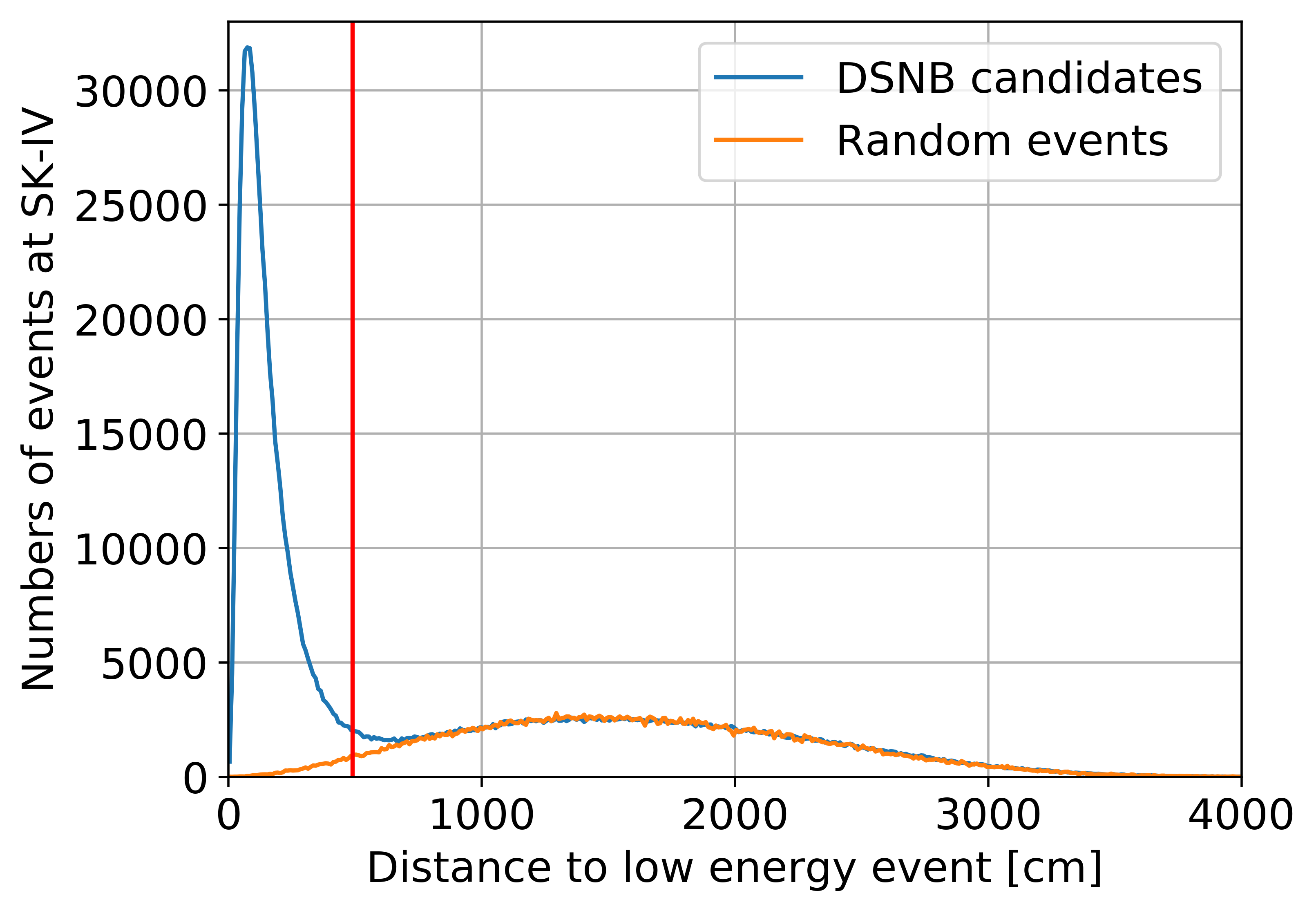}
  \end{center}
  \vspace{-10truept}
  \caption{Minimal distance between a given DSNB candidate (blue) or SK low energy event with a random vertex (orange) and a well-reconstructed event with $E_{\rm rec} > 5.5$~MeV observed within $60$~sec of it. The DSNB candidates are defined as all the SK-IV events with $E_{\rm rec} > 7.5$~MeV passing the noise cuts. Here we also show the $4.9$~m separation required in this analysis with a red vertical line. 
  }
  \label{fig:multispa}
  \end{figure}

First, since energetic muons could induce the production of multiple radioactive isotopes, we remove all DSNB candidates observed within 60~sec and 4.9~m from at least one other low energy event with $E_{\rm rec}$ in the 5.5-24.5~MeV range~\cite{LockeCoffani}. The 60~sec time window has been chosen to include decays of abundant and long-lived isotopes such as $^{16}$N while the 4.9-m distance cut provides optimal signal over background separation. We estimate the signal efficiency of this so-called ``multiple spallation cut'' using a sample of low energy events from radioactive decays at SK ($4 < E_{\rm rec} < 5$~MeV), whose reconstructed vertices have been replaced by random vertices inside the ID. The cut parameters introduced above allow to remove about $45\%$ of the background with a $98\%$ signal efficiency, as can be seen in Fig.~\ref{fig:multispa}. Since this cut removes low energy events clustered in space and time, it also removes low energy muons that are misclassified as electrons with tens of MeV energies. These muons are not targeted by the other spallation cuts applied in this study, which assume that the prompt event is an electron or a positron. Consequently, in this study, we will apply this cut over the whole energy range for each of our analyses, including the spallation-free $E_{\rm rec}>23.5$~MeV region.

\begin{figure}[htbp]
  \begin{center}
   \includegraphics[clip,width=8.0cm]{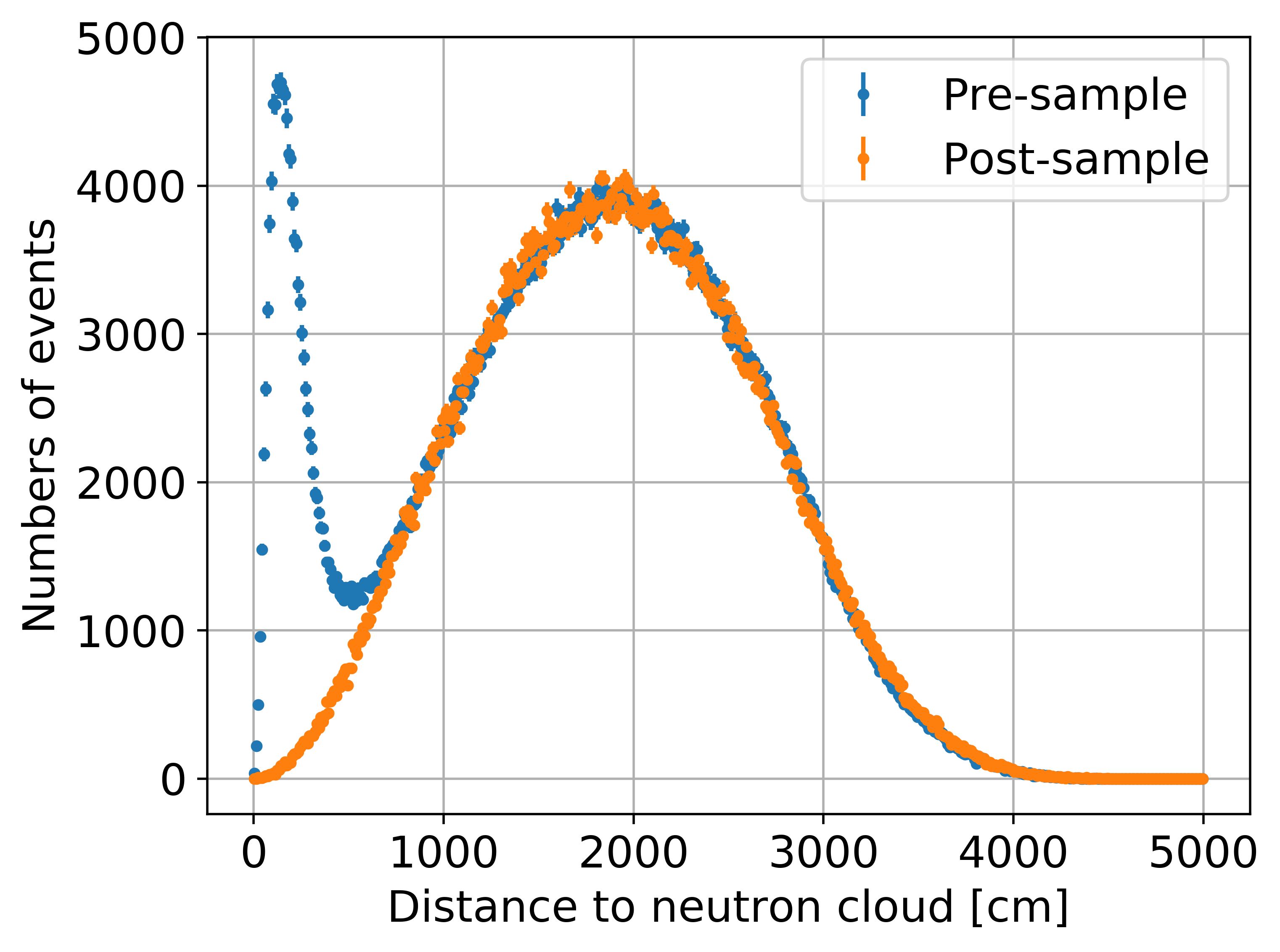}
  \end{center}
  \vspace{-10truept}
  \caption{Distributions of distance between DSNB candidates and neutron clouds for pre- and post-samples, where the selected neutron clouds have been observed up to $60$~sec before and after the DSNB candidates respectively. The distribution for the post-sample has been rescaled by assuming that contributions from spallation pairs beyond 20~m are negligible. Contributions from pairs formed by spallation isotope decays and neutron clouds associated with their parent muons appear as a peak in the pre-sample below 10~m. The associated distribution can be extracted by subtracting the rescaled post-sample from the pre-sample.}
  \label{fig:ncloud}
  \end{figure}

In addition to removing multiple spallation events, we locate muon-induced showers by identifying neutron captures observed less than $500~\mu$s after muons. These neutron clusters, called ``neutron clouds'', are often produced in muon-induced hadronic showers, that are the birthplaces of spallation isotopes. 
Neutron capture events following muons are collected separately from DSNB candidates, using a specific trigger called the Wideband Intelligent Trigger (WIT)~\cite{wit2011,wit2015,wit2017}. This trigger, in addition to setting a threshold on the number of PMT hits, applies cuts on the event quality and distance of the reconstructed vertex to the ID wall. Events triggered by WIT less than $500~\mu$s after a given muon are considered as neutron captures if their reconstructed vertex is within $5$~m of the reconstructed muon track. If multiple neutron captures are observed after a muon event, we define a neutron cloud using the criteria from Ref.~\cite{LockeCoffani}. We then remove DSNB candidates that are found close in time and space to these neutron clouds,
following the procedure introduced in Ref.~\cite{LockeCoffani}. More details about the cut criteria are given in Appendix~\ref{sec:appendixspall}. 
%
To estimate the signal and background efficiencies of the neutron cloud cut, we pair muons with DSNB candidates in data, defining two samples: a ``pre-sample'', including muons found up to $60$~sec \emph{before} a DSNB candidate, and a ``post-sample'', with muons found up to $30$~sec \emph{after} a candidate. While the pre-sample contains a mixture of correlated and uncorrelated pairs, the post-sample contains only uncorrelated pairs. This sample can hence be used to both readily estimate signal efficiencies and characterize the properties of the pairs formed by each isotope decay and their parent muon in the pre-sample. An example of how to extract the distribution of the distances between spallation events and the neutron clouds of their parent muons is shown in Fig.~\ref{fig:ncloud}. 
We optimize the neutron cloud cut by estimating the shape of the neutron clouds from data and maximizing the statistical significance of the signal over the spallation background. 
The performance of this cut is however currently limited by the weakness of the neutron capture signal and the fact that the WIT trigger has only been available only during the last 388 live days of SK-IV. While the signal efficiency is larger than $99$\% this cut only removes about $10$\% of the spallation background in this analysis. Over the 388 days of the WIT trigger livetime, however, it removes about 40\% of spallation isotope decays. 
%

Finally, it should be noted that the multiple spallation and the neutron cloud cuts described above are expected to overlap since muons associated with large hadronic showers are also likely to lead to the production of multiple isotopes. Accounting for this overlap, these preselection cuts allow to remove about 55\% of the spallation background when neutron cloud data are available.

\subsubsection{Searching for parent muons}
\label{sec:spaloglike}

Since preselection cuts remove only about half of the spallation background, a more in-depth study of the correlations between muons and DSNB candidates is necessary. In particular, in order to identify the decays of spallation isotopes, it is crucial to associate each isotope with its parent muon.
Due to the long half-lives of isotopes such as $^{11}$Be and $^{16}$N, we need to investigate all possible pairings between DSNB candidates and muons observed up to 30~sec \emph{before} them. This window size allows to accommodate most decay products, without processing a prohibitively high number of muons. As with neutron cloud cut optimization, we define pre- and post-samples in order to extract the observable distributions associated with spallation pairs and estimate signal efficiencies.
%
%
%

We define candidate muons by selecting events associated with more than 31 PMT hits (34 before May 2015) in 200~ns and depositing more than 500~$p.e.$ in the ID. Additionally, since the SK trigger windows can contain multiple events, we also look for muons around the main activity peak in these windows. For the same reason, calibration trigger windows are also thoroughly investigated. The properties of the muon candidates, such as the charge deposited in the detector, the entry point, and the direction of the track, are then extracted using a dedicated fitter~\cite{bib:zconner,bib:sdesai,bib:sksrn123}.
This fitter classifies muons into five categories: misfits (1.0\%), single through-going muons (82.2\%), stopping muons (4.9\%), multiple-track muons (7.6\%), and corner clippers (4.3\%); the fractions of the different categories in the SK-IV data are shown in parentheses. Misfit muons with a charge lower than $1000$~$p.e.$ are removed in order to reject non-muonic high energy events. This cut does not affect removing spallation background because the track length of these low charge muons through the ID is typically less than 50~cm. 
%

After having paired DSNB candidates with the muons selected above, we extract the following observables, related to the intrinsic properties of the muons and their correlations with DSNB candidates. 

\begin{figure}[htbp]
  \begin{center}
    \includegraphics[clip,width=8.0cm]{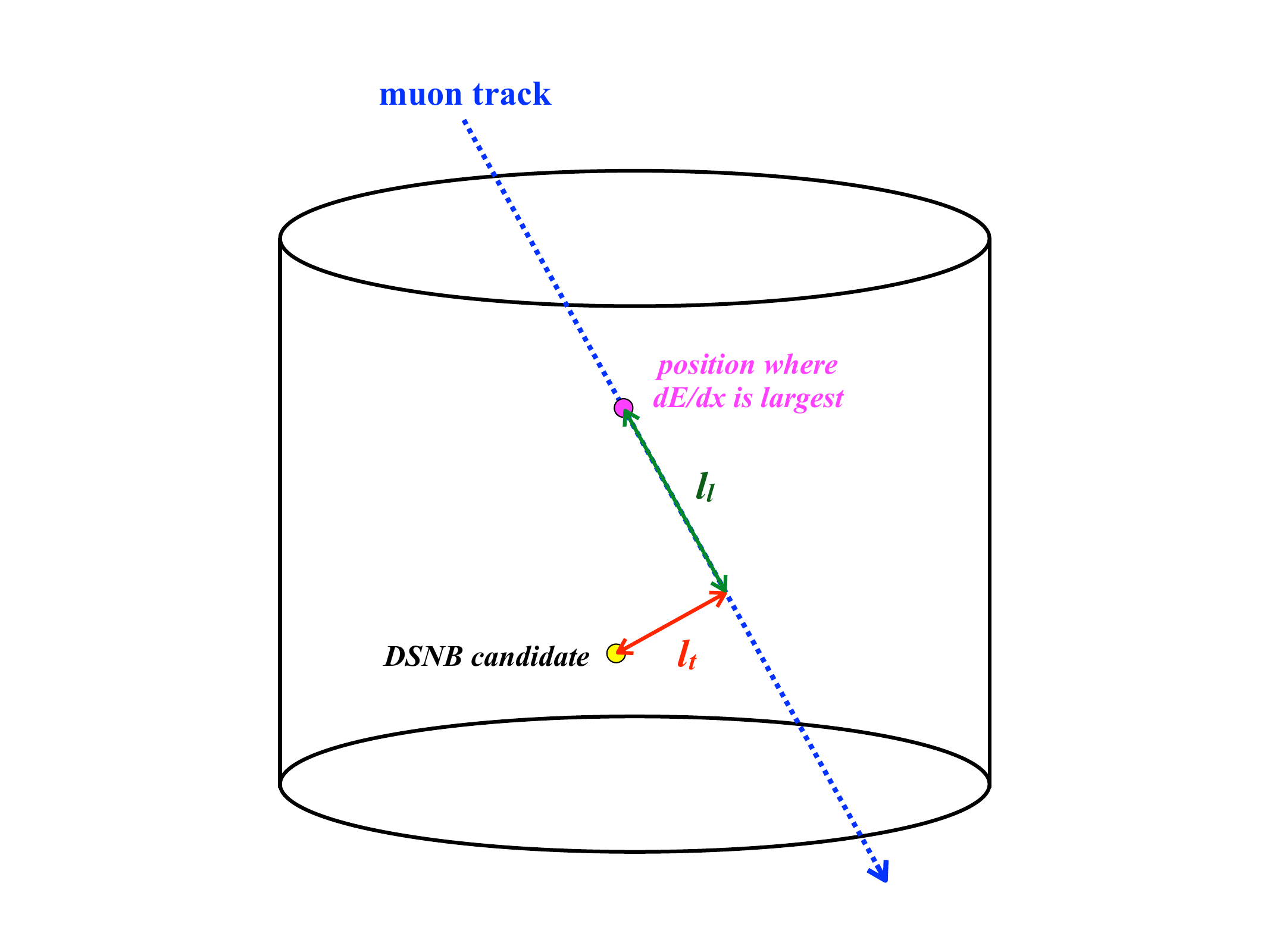}
  \end{center}
  \vspace{-15truept}
  \caption{Schematic of the spallation observables $\ell_t$ and $\ell_l$.
           The cylindrical shape represents the SK ID.}
  \label{fig:spavariable}
  \end{figure}

  \begin{itemize}
    \item $dt$: 
    time difference between a muon and the DSNB candidate. 
    For spallation events, this observable reflects the half-lives of the produced isotopes.
    %
    \item $\ell_t$: 
    transverse distance of the DSNB candidate to the muon track (see Fig.~\ref{fig:spavariable}). 
    This observable is related to the path lengths of the secondary particles from muon showers. For well-fitted single through-going muons paired with their associated spallation isotope decays, $\ell_t$ is typically no larger than a few meters.
    \item $\ell_l$: 
    longitudinal distance to the point of the muon track associated with the maximal energy deposition 
    (see Fig.~\ref{fig:spavariable}). The energy deposition along the muon track is determined by projecting back the light seen by each PMT to a point on the muon track, following the procedure described in Ref.~\cite{bib:sksrn123}.
    So $\ell_l$ provides an estimate of the distance between the low energy event and the origin of the particle shower and, if the shower position is correctly identified, should also not be larger than a few meters for spallation pairs.
    \item $Q_\mu$: 
    total charge deposited by muons in the detector. Higher values of this observable are expected when a shower is produced. While this observable also includes contributions from minimum ionization along the muon track ---and hence strongly depends on the muon track length, it provides the most robust shower predictor for poorly-fitted muons and muon bundles. Conversely, for well-fitted single through-going muons, the shower can be more accurately characterized by the $Q_{\rm res}$ observable described below.
    %
    \item $Q_{\rm res}$: 
    residual charge deposited by a muon compared to the value expected from the minimum ionization. 
    This observable can be expressed as: $$Q_{\rm res} = Q_\mu - q_{\rm MI} \cdot L,$$
    where $q_{\rm MI}$ and $L$ are the number of $p.e.$ per centimeter expected from the minimum ionization and the track length, respectively. Here $q_{\rm MI}$ depends on detector properties, such as water transparency, that vary with time and is therefore recomputed for each run.
    For muons with multiple tracks, $L$ is the length of the first track. 
    This observable allows to determine how likely a given muon is to induce a shower.
  \end{itemize}

\noindent
An example set of these spallation variables for single through-going muons is shown in Appendix~\ref{sec:appendixspall}.
The distributions associated with corner clipping muons show no evidence for spallation. Moreover, as discussed in Ref.~\cite{LockeCoffani}, spallation background events associated with corner clipping muons are expected to represent less than $10^{-5}$ of the total number of spallation events. In the rest of this study, we will therefore neglect contributions from these muons to spallation backgrounds.

Showers induced by muon spallation can be extremely energetic and involve up to thousands of particles, notably neutrons and pions. 
Such neutrons cause nuclear reactions and possibly produce deexcitation $\gamma$ rays,
and are later captured at a timescale up to $\sim$500~$\mu$s. 
The MeV-scale $\gamma$ rays or $\beta$ particles ---produced by the initial neutron, secondary nuclear reactions, or the decay product of a short-lived isotope, followed by a 2.2~MeV $\gamma$ ray from neutron capture can mimic an IBD signature. To prevent these from leaking into the sample, we require DSNB candidates to be more than $1$~ms away from any muon. The impact of this cut on the signal efficiency is negligible. Additionally, we apply a set of cuts on $dt$ and $\ell_t$ for different energy ranges, exploiting the dependence of the isotopes' half-lives in their endpoint energies shown in Fig.~\ref{fig:spall_products}. Detailed descriptions of these rectangular cuts are given in Appendix~\ref{sec:appendixspall}. These cuts are particularly efficient above 15.5~MeV where short-lived isotopes dominate. In 15.5$-$19.5~MeV, notably, they allow to remove about 85\% of the spallation background while keeping 88\% of the signal events. Finally, in order to further eliminate spallation backgrounds, we use distributions of the different observables considered here to define log-likelihood ratios. First, we prepare two probability density functions (PDFs): the spallation PDF 
(${\rm PDF}_{\rm spall}^i$) and the random PDF (${\rm PDF}_{\rm random}^i$), for 
each variable $i = dt, \ \ell_t, \ \ell_l, \ Q_\mu, \ Q_{\rm res}$.
We isolate contributions from spallation events by subtracting the post-sample distributions from the corresponding pre-sample distributions and obtain ${\rm PDF}_{\rm spall}^i$ after area normalization of these subtracted distributions. For ${\rm PDF}_{\rm random}^i$ we normalize post-sample distributions by their areas. We repeat this procedure for each category of muons. For single through-going and multiple track muons, that generate most of the spallation background, we tune the PDFs in $dt$ and $\ell_t$ bins. These bins, adjusted by considering the isotopes' typical half-lives and the $\ell_t$ distributions, account for the correlations between these observables. For each set of PDFs, we then define log-likelihood ratios as:
  \begin{align}
   \mathcal{L}_\text{spall} = \log \left( \prod_{i} \frac{{\rm PDF}_{\rm spall}^{i}}{{\rm PDF}_{\rm random}^{i}} \right). 
  \label{eq:spalike}
  \end{align}

\noindent
Separate likelihoods are defined for each muon category, except for corner clippers, whose contributions to spallation backgrounds are negligible as mentioned above. Note that for misfit muons only $dt$ is used to calculate log-likelihood ratios, as the other observables are not reliable. 
An example distribution of the log-likelihood ratio is given as Fig.~\ref{fig:spaloglike} in Appendix~\ref{sec:appendixspall}. 
We finally determine cut conditions for each likelihood ratio, accounting for complex correlations between spallation observables, geometrical and muon reconstruction effects. 

The signal efficiencies and background rejection rates of the resulting cuts are estimated using the random sample introduced in this section, as well as spallation-dominated data samples. These samples as well as our methodology to estimate the spallation cut performance are described in detail in Appendix~\ref{sec:appendixspall}. Note that the estimates of the spallation remaining rate presented there are used only for cut optimization and not for the final background predictions shown in Sections~\ref{sec:mianalysis} and \ref{sec:spectral}.
The current cuts achieve significant improvement over the previous cuts used in Refs.~\cite{bib:sksrn123,bib:sksrn4}, especially at low energies; for the same spallation remaining rate, the signal efficiency is increased by up to 60\% for $E_{\rm rec} < 11.5$~MeV, 20\% for $11.5 < E_{\rm rec} < 13.5$~MeV, and comparable for higher energies.
%
%

\subsection{DSNB positron candidate selection}
\label{subsec:posreduc}

In the region above $\sim$20~MeV, spallation backgrounds can be reduced to negligible levels using a series of spallation cuts while keeping most of the signal. However, significant backgrounds from atmospheric neutrino interactions and radioactive decays remain. To identify them, we define the following discriminating observables, aimed at characterizing the prompt event.

\subsubsection{Incoming event cut}

\begin{figure}[htbp]
    \centering
    \includegraphics[width=\linewidth]{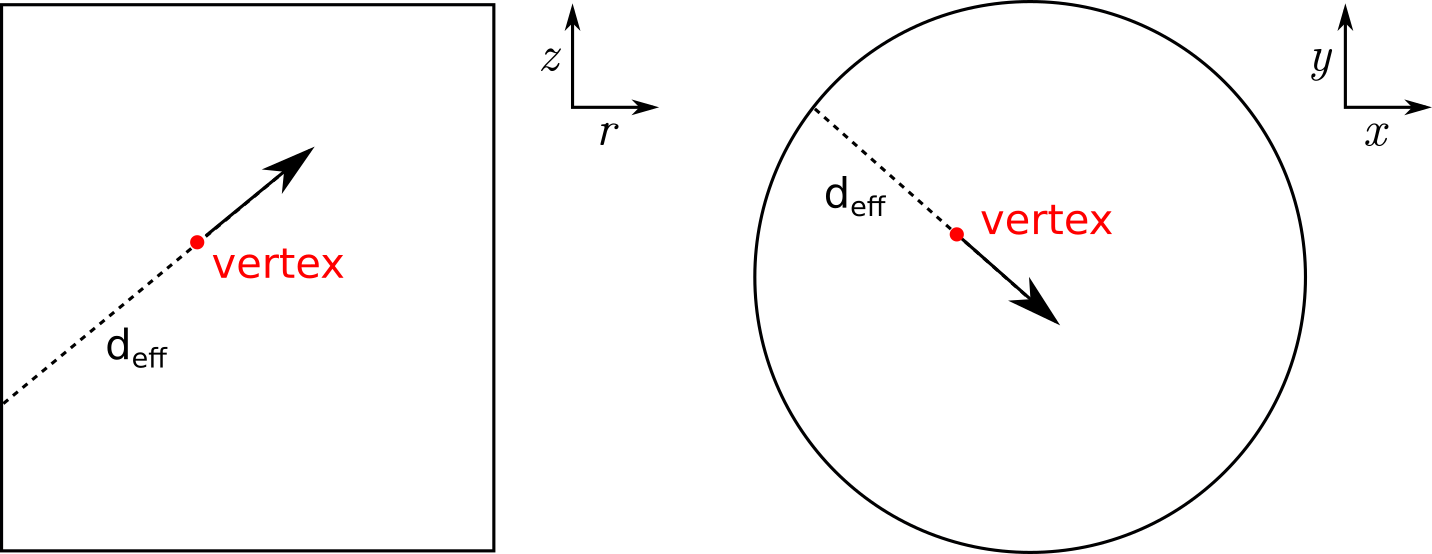}
    \caption{Views of the ID, showing the effective distance of a reconstructed event to the detector wall. The $d_\text{eff}$ is the distance that a particle emitted near the ID wall ---e.~g. from by a radioactive decay--- needs to travel to give the observed signal.}
    \label{fig:deff_schema}
  \end{figure}

\begin{figure*}[htbp]
    \centering
    \begin{minipage}{0.45\hsize}
     \begin{center}
        \includegraphics[width=8.cm]{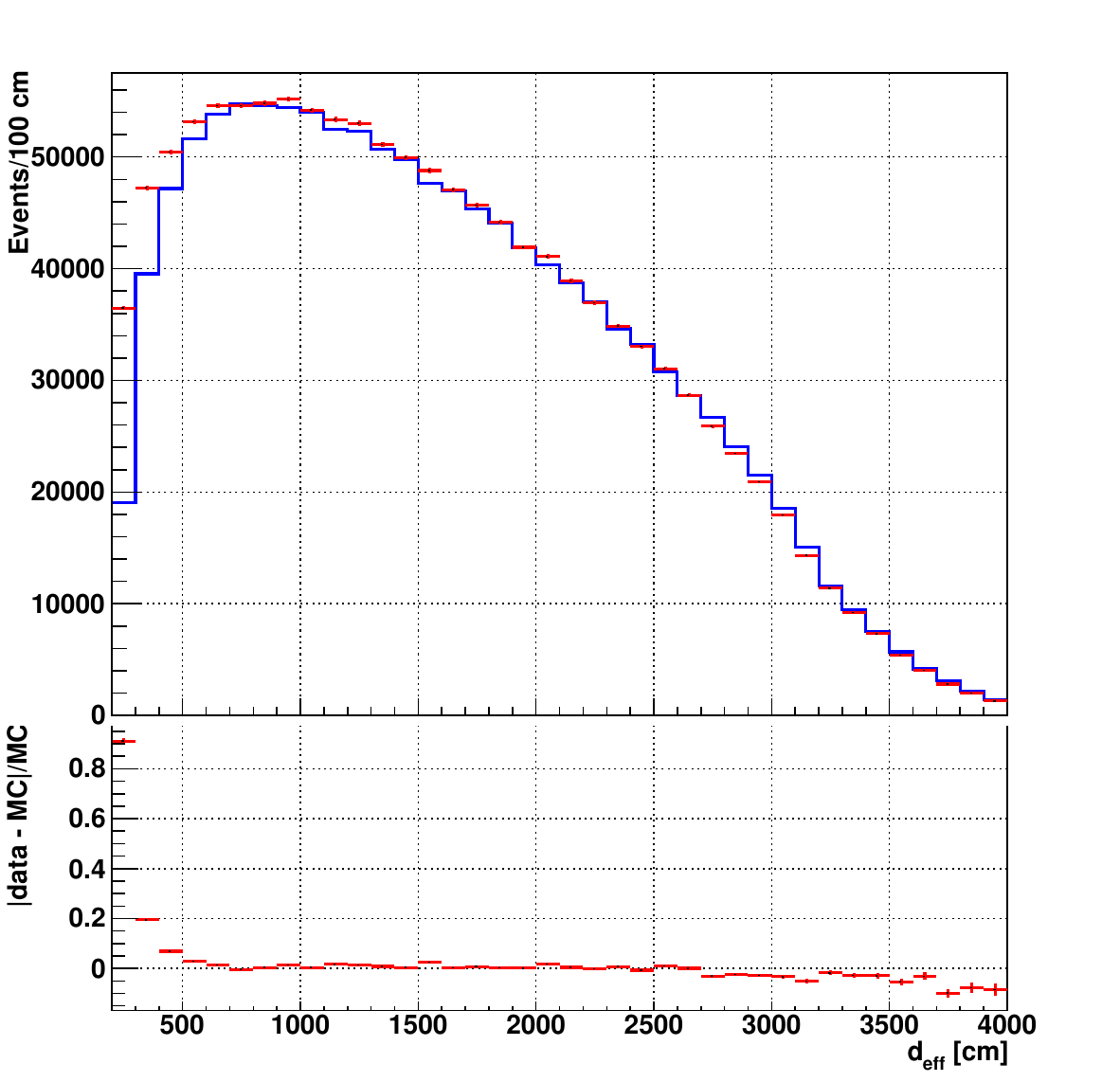}
     \end{center}
    \end{minipage}
        \begin{minipage}{0.45\hsize}
     \begin{center}
        \includegraphics[width=8.cm]{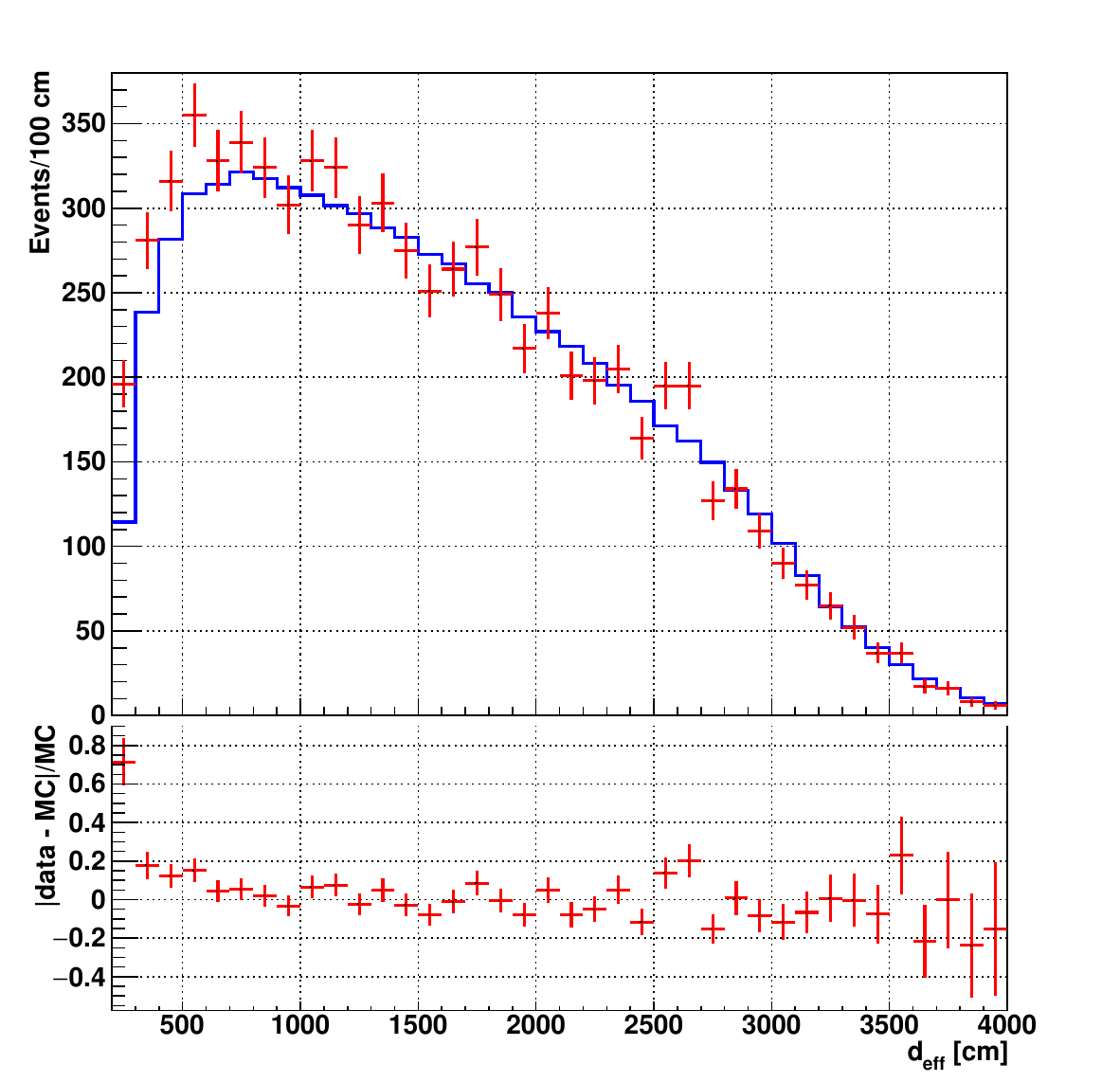}
     \end{center}
    \end{minipage}
    \caption{The $d_\text{eff}$ distributions for the observed data and IBD Monte-Carlo simulation, for events with $E_{\rm rec} \in [7.5, 29.5]$~MeV (left) and $E_{\rm rec} \in [15.5, 79.5]$~MeV (right). The rate of IBD Monte-Carlo simulation events has been scaled to match the number of events observed in the data for $d_\text{eff} > 1000$~cm. The lower panels show the relative difference between the numbers of data and IBD simulation events. We consider only events reconstructed in the FV and passing the noise reduction cuts. The assumed DSNB spectrum is taken from the Horiuchi$+$09 6~MeV model~\cite{bib:horiuchi09}.}
    \label{fig:deff}
  \end{figure*}

Radioactivity near the detector wall as well as muon spallation in the rock surrounding the detector, can lead to electrons or $\gamma$ rays entering FV. 
Instead of tightening the FV cut, we consider the effective distance of each event to the ID wall $d_\text{eff}$~\cite{bib:sksrn123,bib:sksolar1}. This observable is computed by following the reconstructed direction of each event backwards from its reconstructed vertex to the ID wall, and can be interpreted as the minimal distance needed for a radioactive particle produced near the wall to travel to the event vertex, as shown in Fig.~\ref{fig:deff_schema}. 
The $d_\text{eff}$ distributions for the DSNB signal and for the events selected using the cuts outlined in Section~\ref{subsec:noise} are shown in Fig.~\ref{fig:deff}, for the reconstructed energy ranges corresponding to the model-independent analysis, $7.5$ to $29.5$~MeV, and the spectral analysis, $15.5$ to $79.5$~MeV. Comparing the $d_\text{eff}$ distributions from the data and the Monte-Carlo simulation allows to estimate contributions from radioactivity near the wall, and determine a suitable $d_\text{eff}$ cut.
In this study we impose lower thresholds on $d_\text{eff}$ ranging from $3$ to $5$~m depending on the reconstructed energy: 

\begin{align}
    d_\text{eff} > \text{max}\left\{ 300, \ 500 - \frac{E_{\rm rec} - 15.5~\text{MeV}}{1~\text{MeV}}\times 50\right\}~\text{cm}. \nonumber
\end{align}

\subsubsection{Pre- and post-activity cuts}

\begin{figure*}[htbp]
    \centering
    \begin{minipage}{0.45\hsize}
        \begin{center}
            \includegraphics[clip,width=8.cm]{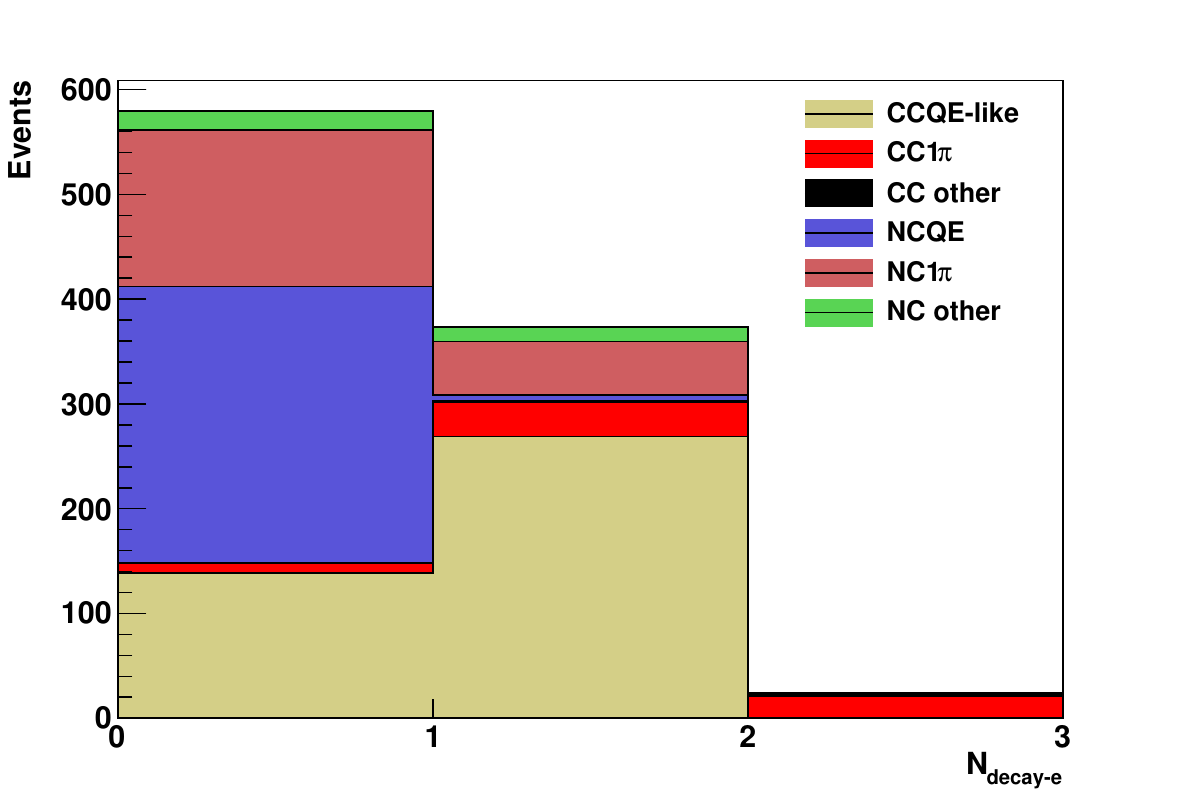}
        \end{center}
    \end{minipage}
    \begin{minipage}{0.45\hsize}
        \begin{center}
            \includegraphics[clip,width=8.cm]{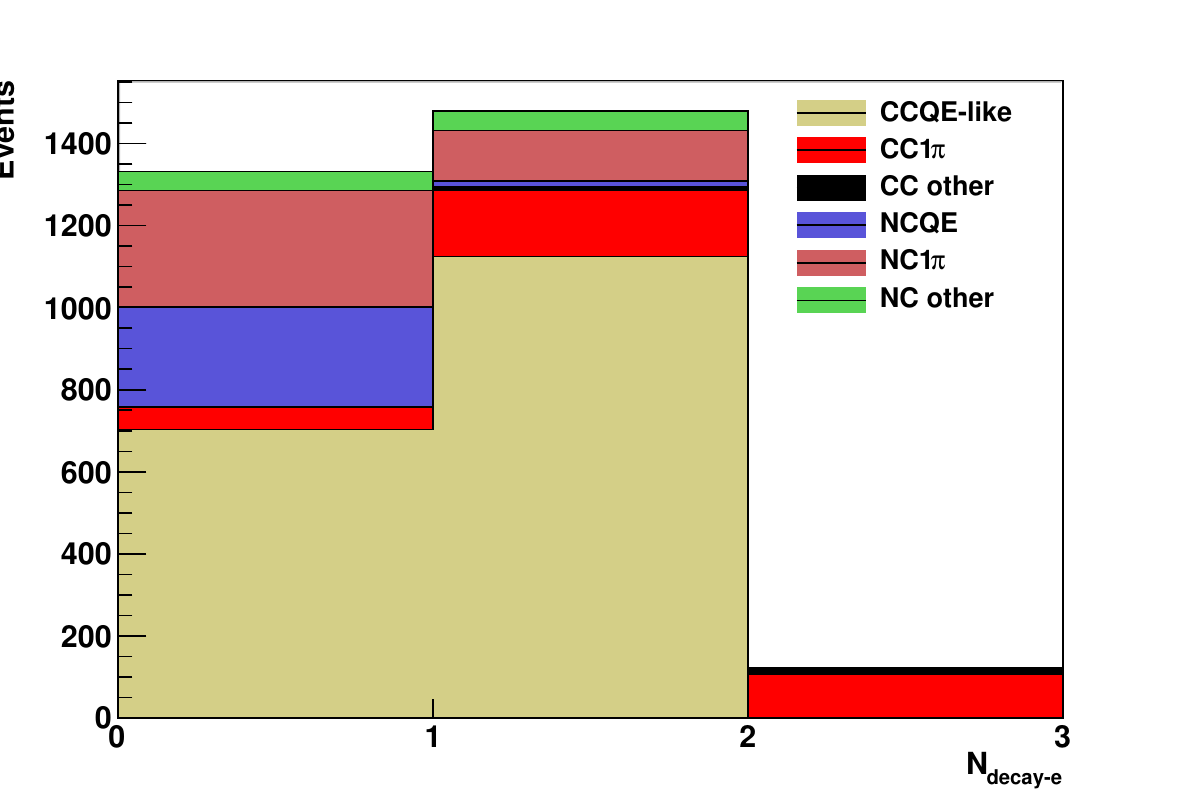}
        \end{center}
    \end{minipage}
    \caption{The $N_{{\rm decay}\mathchar`-e}$ distributions for the atmospheric neutrino Monte-Carlo simulation, for events with $E_{\rm rec} \in [7.5, 29.5]$~MeV (left) and $E_{\rm rec} \in [15.5, 79.5]$~MeV (right). We consider only events reconstructed in the FV and passing the noise reduction cuts. }
    \label{fig:post}
\end{figure*}

Atmospheric neutrino interactions can produce both prompt signals from photons, electrons, or energetic heavy particles, and delayed signals from decay of muons and pions to electrons. These signals will be typically separated by a few microseconds and can therefore share the same trigger window. Depending on which particle deposits most light, unusually high activity can be noticed either before or after the main peak. For pre-activity we compute the maximal number of hits, $N_{\rm pre}^{\rm max}$, in a $15$-ns time-of-flight subtracted window between $5~\mu$s and $12$~ns before the main peak. We apply a similar procedure to post-activity, using an algorithm developed for other analyses in SK and T2K that computes the number of electrons from muon and pion decay $N_{{\rm decay}\mathchar`-e}$~\cite{bib:t2k2016oa}. The $N_{{\rm decay}\mathchar`-e}$ distributions for low and high energy events are shown in Fig.~\ref{fig:post}. In the analyses discussed here we require $N_{\rm pre}^{\rm max} < 12$ and $N_{{\rm decay}\mathchar`-e} = 0$. 

\subsubsection{Cherenkov angle}

\begin{figure*}[htbp]
    \centering
    \begin{minipage}{0.45\hsize}
        \begin{center}
            \includegraphics[clip,width=8.cm]{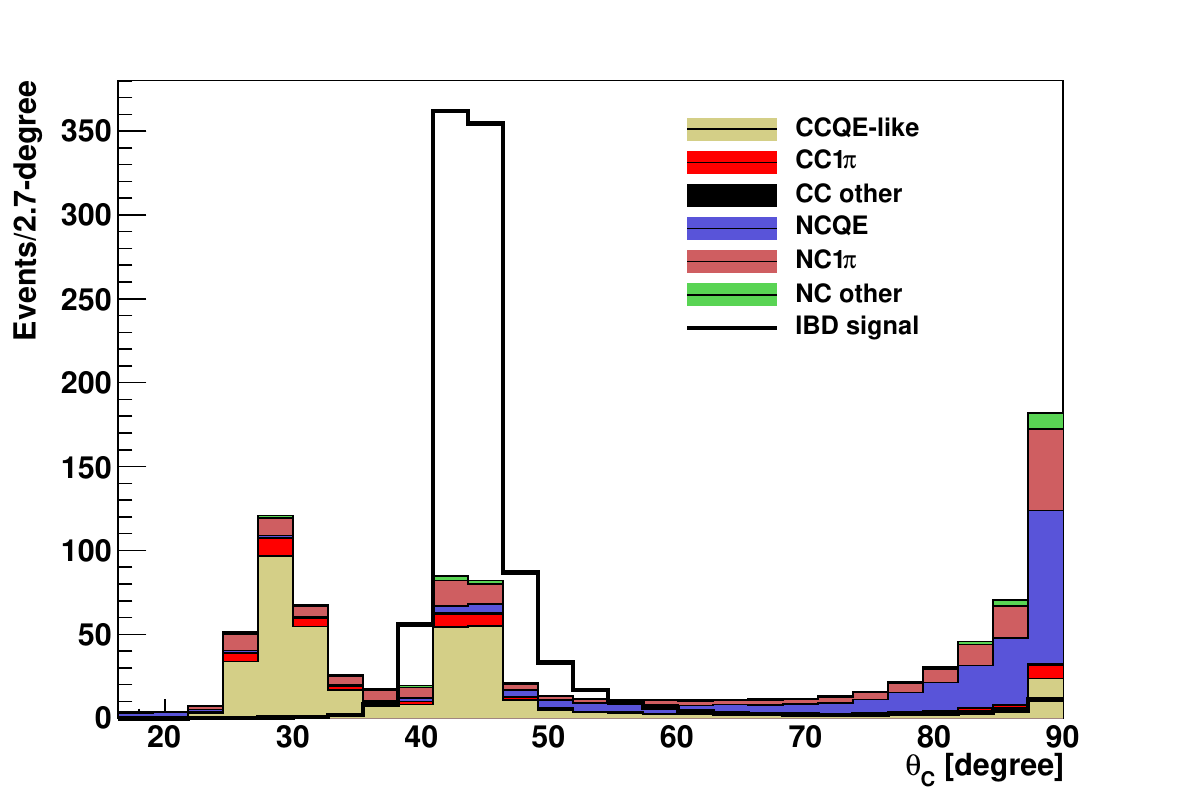}
        \end{center}
    \end{minipage}
    \begin{minipage}{0.45\hsize}
        \begin{center}
            \includegraphics[clip,width=8.cm]{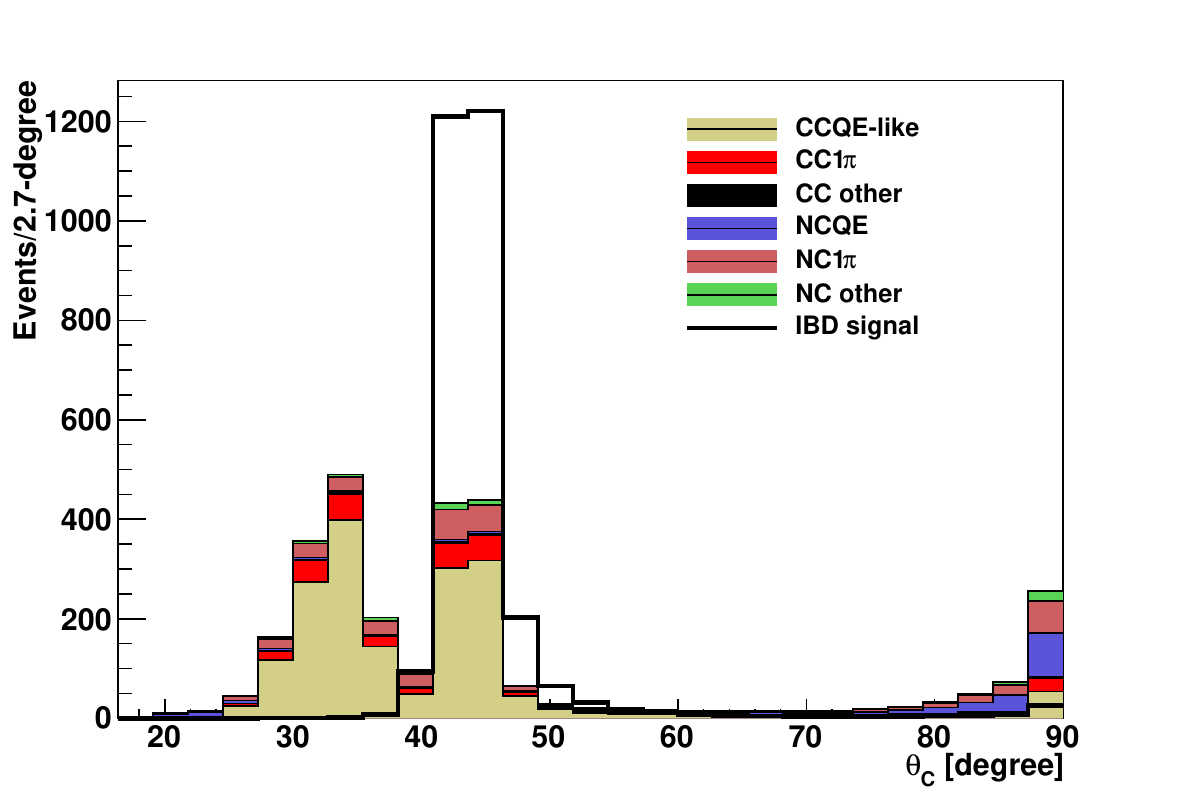}
        \end{center}
    \end{minipage}
    \caption{The $\theta_{\rm C}$ distributions for the atmospheric neutrino and IBD Monte-Carlo simulations, for events with $E_{\rm rec} \in [7.5, 29.5]$~MeV (left) and $E_{\rm rec} \in [15.5, 79.5]$~MeV (right). We consider only events reconstructed in the FV and passing the noise reduction cuts. The assumed DSNB spectrum is taken from the Horiuchi$+$09 6~MeV model~\cite{bib:horiuchi09}. The signal and background distributions are normalized to the same area.}
    \label{fig:cherenkov}
\end{figure*}

The opening angle of the Cherenkov light cone ($\theta_{\rm C}$) emitted by highly relativistic particles, electrons and positrons in the current context, in pure water is around $42^\circ$. Conversely, at the $\mathcal{O}(10)$~MeV energies considered here, heavier particles like muons and pions will be typically observed near the Cherenkov threshold, leading to cones with smaller angles in the current analysis range. In addition, NCQE interactions with multiple photon emission can produce multiple overlapping Cherenkov cones, that will be reconstructed as a single cone with a particularly large opening angle~\cite{bib:t2kncqe1to3,bib:t2kncqe1to9}. Consequently, $\theta_{\rm C}$ is one of the most powerful observables for reducing both NCQE and $\mu/\pi$-producing atmospheric backgrounds. The Cherenkov angle distributions for the signal and atmospheric neutrino events are shown in Fig.~\ref{fig:cherenkov} for the energy ranges of the model-independent analysis and the spectral analysis. In what follows we will require the Cherenkov angle in the signal regions to be in $[38, 50]^\circ$. In the spectral analysis, we will also use this angle to define sidebands to evaluate atmospheric backgrounds. 

\subsubsection{Ring clearness}

Electrons and positrons do not follow a straight trajectory in SK due to scattering and bremsstrahlung, which leads to fuzzy Cherenkov rings. Pions, on the other hand, lead to well-delineated ring patterns. In order to characterize this property we consider a $15$-ns time-of-flight subtracted window around the main activity peak and compute the opening angles of the cones formed by the directions of all possible 3-hit combinations. We then identify the peak of this opening angle distribution $\theta_0$ and estimate the ``clearness'' of the ring by computing: 

\begin{align}
   L_\texttt{clear} = \frac{N_\text{triplets}(\theta_0\pm 3^\circ)}{N_\text{triplets}(\theta_0\pm 10^\circ)}.
\end{align}

\begin{figure*}[htbp]
    \centering
    \begin{minipage}{0.45\hsize}
        \begin{center}
            \includegraphics[clip,width=8.cm]{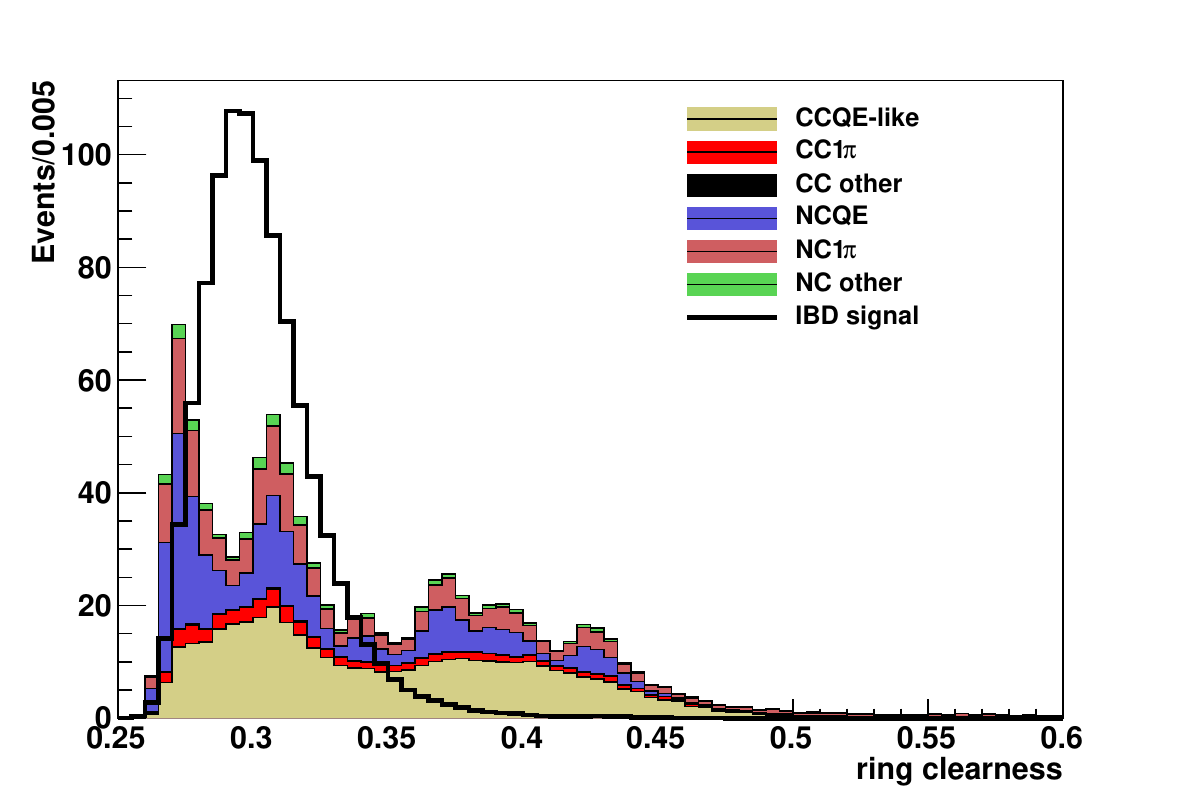}
        \end{center}
    \end{minipage}
    \begin{minipage}{0.45\hsize}
        \begin{center}
            \includegraphics[clip,width=8.cm]{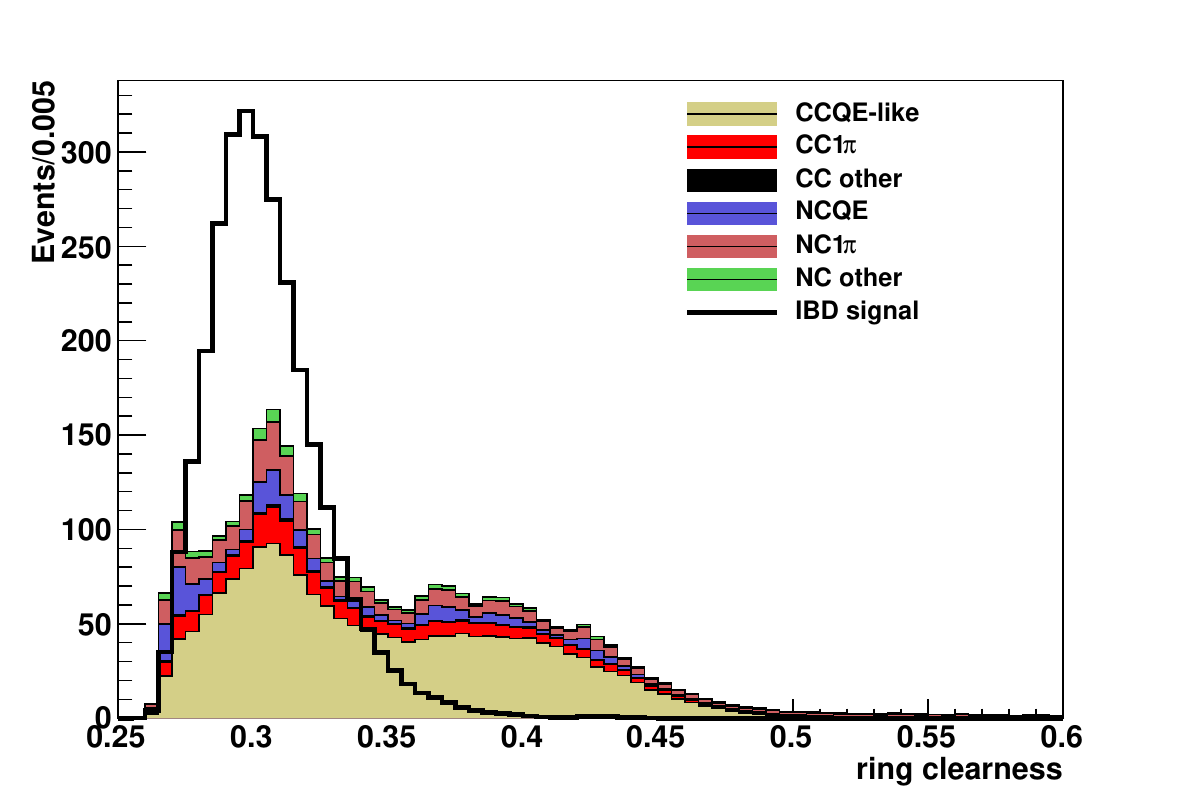}
        \end{center}
     \end{minipage}
    \caption{The $L_\texttt{clear}$ distributions for the atmospheric neutrino and IBD Monte-Carlo simulations, for events with $E_{\rm rec} \in [7.5, 29.5]$~MeV (left) and $E_{\rm rec} \in [15.5, 79.5]$~MeV (right). We consider only events reconstructed in the FV and passing the noise reduction cuts. The assumed DSNB spectrum is taken from the Horiuchi$+$09 6~MeV model~\cite{bib:horiuchi09}. The signal and background distributions are normalized to the same area.}
    \label{fig:pilike}
\end{figure*}

\noindent 
The $L_\texttt{clear}$ distributions from the atmospheric and IBD Monte-Carlo simulations are shown in Fig.~\ref{fig:pilike}. 
Events with lower $L_\texttt{clear}$ have fuzzier rings. For the analyses presented here we require $L_\texttt{clear} < 0.36$.

\subsubsection{Average charge deposit}
\begin{figure*}[htbp]
    \centering
    \begin{minipage}{0.45\hsize}
        \begin{center}
            \includegraphics[clip,width=8.cm]{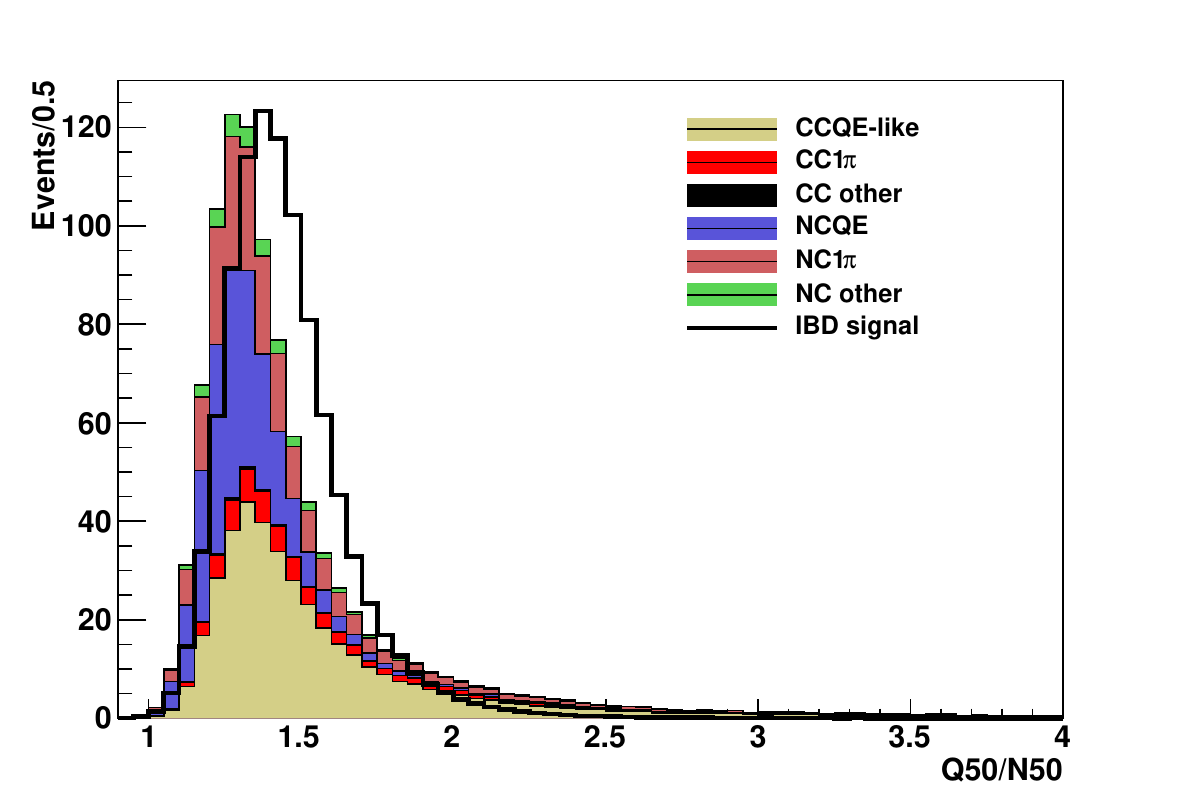}
        \end{center}
    \end{minipage}
    \begin{minipage}{0.45\hsize}
        \begin{center}
            \includegraphics[clip,width=8.cm]{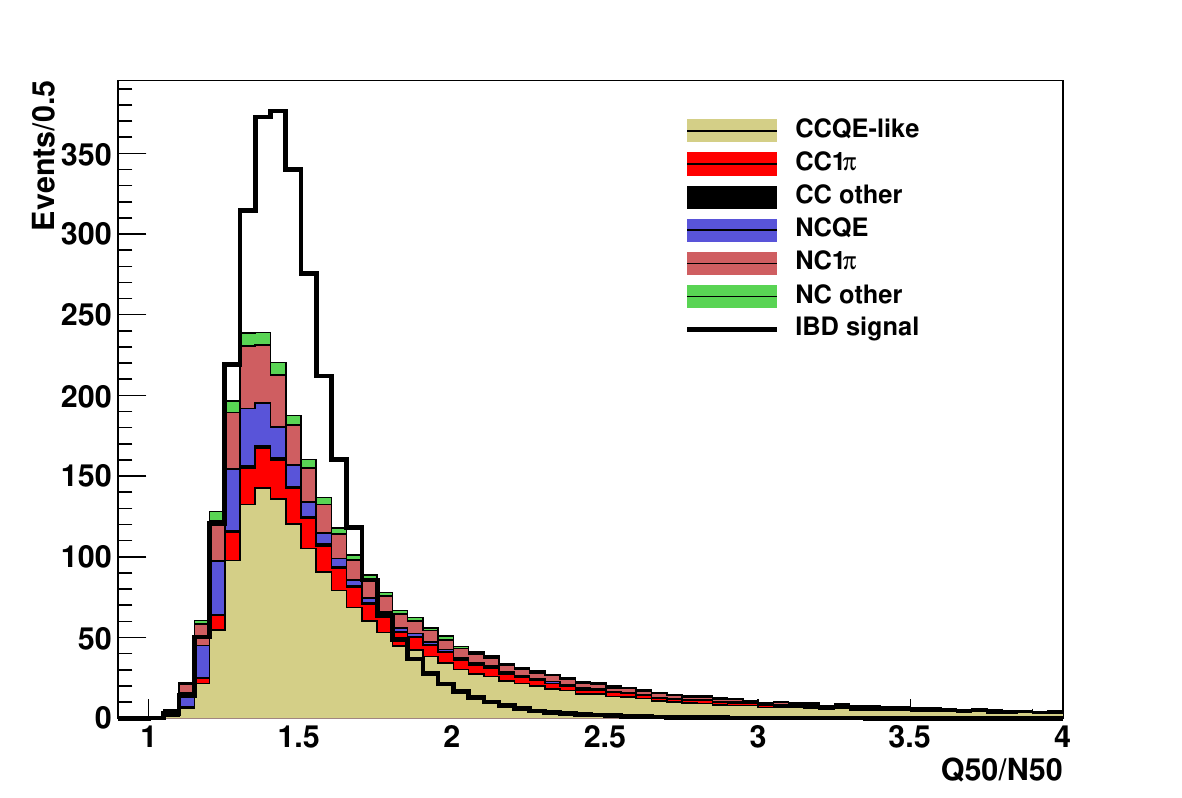}
        \end{center}
    \end{minipage}
    \caption{The $q_{50}/n_{50}$ distributions for the atmospheric neutrino and IBD Monte-Carlo simulations, for events with $E_{\rm rec} \in [7.5, 29.5]$~MeV (left) and $E_{\rm rec} \in [15.5, 79.5]$~MeV (right). We consider only events reconstructed in the FV and passing the noise reduction cuts. The assumed DSNB spectrum is taken from the Horiuchi$+$09 6~MeV model~\cite{bib:horiuchi09}. The signal and background distributions are normalized to the same area.}
    \label{fig:q50n50}
\end{figure*}

Energetic muons scatter less than electrons and hence often deposit more charge in individual PMTs than $\mathcal{O}(10)$~MeV electrons and positrons. We exploit this feature by considering a $50$-ns time-of-flight subtracted window around the main activity peak and calculate the average charge deposited per PMT hit ($q_{50}/n_{50}$) in this window. The distributions are given in Fig.~\ref{fig:q50n50}. We require $q_{50}/n_{50}$ to be less than $2$ in the analysis.

\subsubsection{Cut efficiencies and systematic uncertainties}

We estimate the signal efficiencies of most of the cuts detailed above using the IBD Monte-Carlo simulation. Hence, the main source of systematic uncertainties on the cut efficiencies stems from the modeling of the SK detector and particle propagation in water. We estimate these uncertainties using calibration data taken by injecting a monoenergetic electron beam produced by a LINAC at different positions inside the detector~\cite{bib:sklinac}. For a given cut, we compare the measured and predicted efficiencies using a dedicated Monte-Carlo simulation, and define the systematic uncertainty as the maximum discrepancy over all calibration tests. LINAC calibration thus allows to estimate uncertainties for ring clearness, the average charge, and Cherenkov angle cuts. Conversely, for the incoming event cut ---that requires considering uniformly distributed events--- we use a strategy developed for the SK solar neutrino analyses~\cite{bib:sksolar1,bib:sksolar2,bib:sksolar3,bib:sksolar4}: we measure the variation of the $d_\text{eff}$ cut efficiency after shifting the event directions and vertices in the IBD Monte-Carlo simulation. The vertex and direction shifts used for these estimates are obtained from calibration studies using a nickel source~\cite{bib:skcalib}. Finally, efficiencies for the pre- and post-activity cuts can be directly estimated using a spallation-dominated sample of low energy events, with negligible systematic uncertainty. Overall, positron candidate selection cuts allow to remove a large fraction of atmospheric backgrounds while keeping up to 85\% of the signal. The total systematic uncertainty computed for positron candidate selection cuts using these procedures are of a few percent.

\subsection{Neutron tagging}
\label{subsec:ntag}

The introduction of a new trigger scheme at SK-IV has made characterizing IBDs via neutron tagging possible, by requiring the prompt and delayed signals be detected within 500~$\mu$s of each other. In what follows, we will devise a neutron tagging algorithm tailored to the DSNB analysis, inspired by previous SK studies from Refs.~\cite{bib:hzhang,bib:yzhang}.

\subsubsection{DSNB candidate selection and neutron preselection}
\label{sec:ntagpresel}

As explained in Section~\ref{sec:sk}, most SHE triggered events are followed by a 500~$\mu$s AFT trigger window in order to record the neutron capture signal.
Before attempting to identify neutron candidates in this combined trigger window, we apply a cut on the maximum hits inside a $200$-ns window: $N_{200}<50$ to eliminate SHE+AFT trigger windows containing muons. This cut is associated with a signal efficiency larger than $99$\%. 
After applying the $N_{200}$ cut we scan the SHE+AFT trigger windows of the remaining events to select hit clusters that could be associated with neutron captures. Here, we take advantage of the fact that a typical neutron capture vertex is within a few tens of centimeters of the well-reconstructed IBD vertex. Since the SK resolution is about 50~cm for $\mathcal{O}(10)$~MeV events, this proximity allows to reliably estimate the required photon emission time for each PMT hit to be the result of the neutron capture. A neutron candidate is therefore defined as a group of hits that is clustered in emission time. We then look for clusters maximizing $N_{10}$, the number of hits in a $10$-ns window.

In previous SK-IV searches, neutron candidates were defined as clusters of hits satisfying $N_{10} \geq 7$~\cite{bib:hzhang}. This preselection cut has an efficiency of around $35\%$ and was motivated by the energy range of these searches, where spallation backgrounds were largely dominating. In this study, we modify this preselection step as follows. First, we apply a cut to reduce the continuous PMT dark noise, likely due to scintillation in the PMT glass, that causes a single PMT to flash more than once in rapid succession with a timescale of $\sim$10~$\mu$s. To do so, for each PMT we remove hit pairs separated by less than 6 (12)~$\mu$s for $N_{10} > 6$ ($N_{10} \leq 6$). After applying this cut, we define neutron candidates as hit clusters verifying $N_{10} \geq 6$. This preselection cut will allow loosening of the neutron tagging cut in energy regions where spallation no longer dominates, in particular for the spectral analysis. Finally, since the continuous dark noise cut relies on investigating PMT hit pairs, its efficiency sharply increases near the edge of the SHE+AFT trigger window. We restrict the neutron search window to [$14, 523$]~$\mu$s ([$14, 373$]~$\mu$s for events with a 350-$\mu$s AFT trigger window). After these different steps, neutron candidate preselection is associated with an efficiency of $44.7\%$.


\subsubsection{Boosted Decision Tree}
\label{sec:ntagbdt}

\begin{figure}[htbp]
    \centering
    \includegraphics[width=\columnwidth]{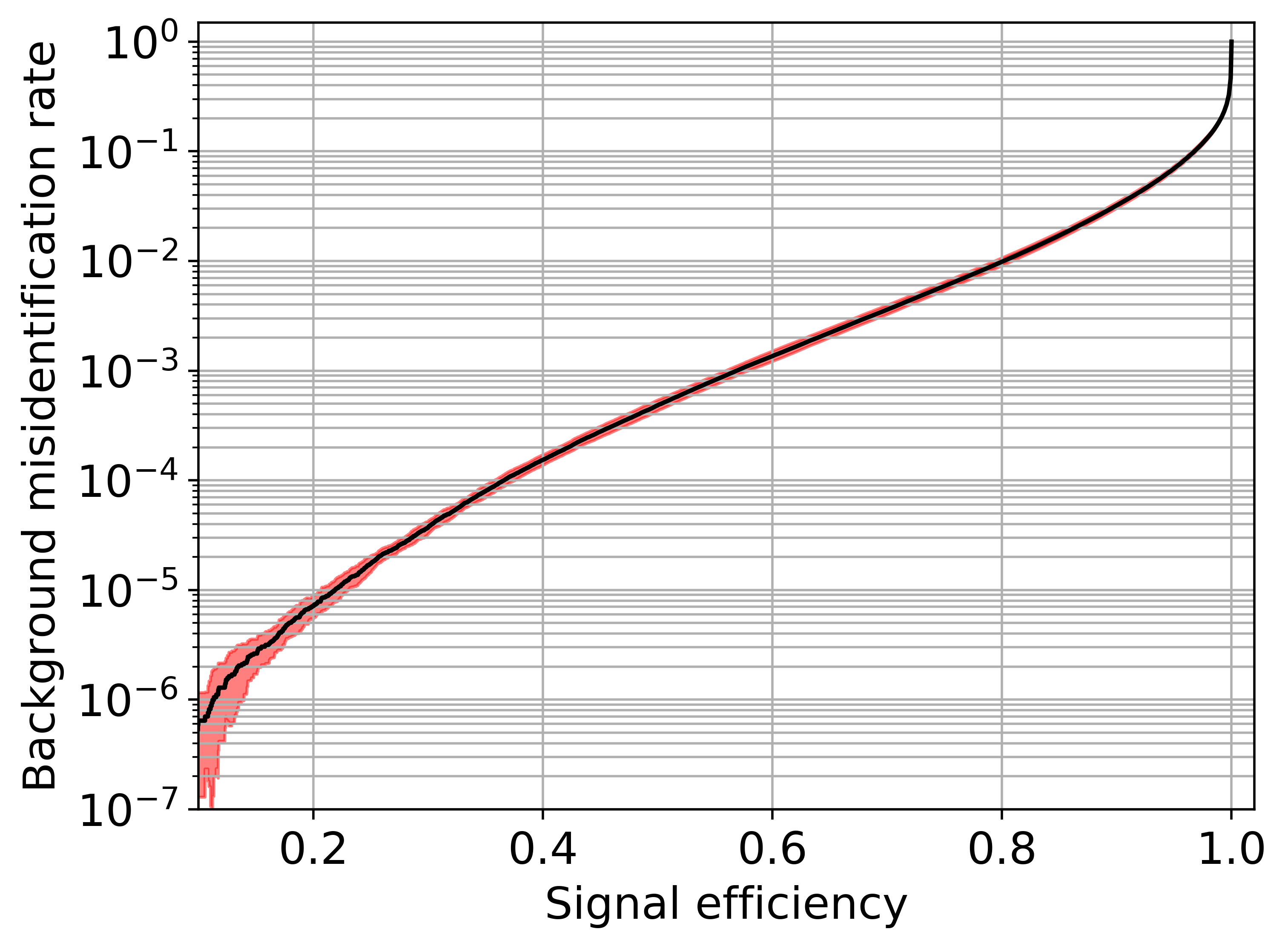}
    \caption{Neutron identification efficiency and mistag rate of BDT used for the current analysis. The red band depicts the uncertainty associated with the time-dependence of the noise in SK, as described in Section~\ref{subsubsec:acc}. The fraction of remaining IBD neutrons and the number of accidental coincidences per event after preselection, averaged over the entire SK-IV period, are $44.7\%$ and 7, respectively. These values depend linearly on the PMT noise~\cite{bib:rakutsu}, and hence vary by about 15\% over the SK-IV period.}
    \label{fig:bdtroc}
\end{figure}

After preselection, we are still left with a large background from e.~g.~dark noise fluctuations and radon decays~\cite{Nakano:2019bnr}, totaling about 7 neutron candidates per event. To further reduce this background, for each neutron candidate, 22 discriminating variables are calculated. These variables are related to specific features of noise events (such as clustered PMT hits), the Cherenkov light pattern of the neutron candidate, and its position with respect to the primary event vertex. 
These $22$ observables are then used in a Boosted Decision Tree (BDT) implemented using the xgboost Python library \cite{Chen:2016}. The BDT is trained using a data set of $2.8\times 10^8$ neutron candidates. In this sample, about $2\times 10^6$ candidates are neutron captures simulated using the procedure described in Section~\ref{subsec:signalsimu} and the rest is composed of low energy accidental coincidences from random trigger events. Table~\ref{tab:bdthyp2} summarizes the BDT training parameters. The final BDT performance is shown in Fig.~\ref{fig:bdtroc}. Even accounting for preselection cuts, this neutron tagging algorithm allows reduction the background rate by a factor of $7$ compared to the analysis presented in Ref.~\cite{bib:yzhang} for the same signal efficiency. 

\begin{table}[htbp]
    \centering
    \caption{Parameters used for the training of the neutron tagging BDT in the \texttt{xgboost} Python library framework.}
    \vspace{+3truept}
    \begin{tabular}{rr}
        \toprule
        Parameter & Value\\
        \hline
        {learning rate} & 0.02522 \\
        {subsample} & 0.97 \\
        {max depth} & 10 \\
        {tree method} & \texttt{approx} \\
        {max iterations} & 1500 \\
        {best iteration} & 1170 \\
        \botrule
    \end{tabular}
    \label{tab:bdthyp2}
\end{table}


\subsubsection{Systematic uncertainty on neutron tagging efficiency}
\label{sec:ntag_ambe}
\begin{figure}[htbp]
  \begin{center}
   \includegraphics[width=9.5cm]{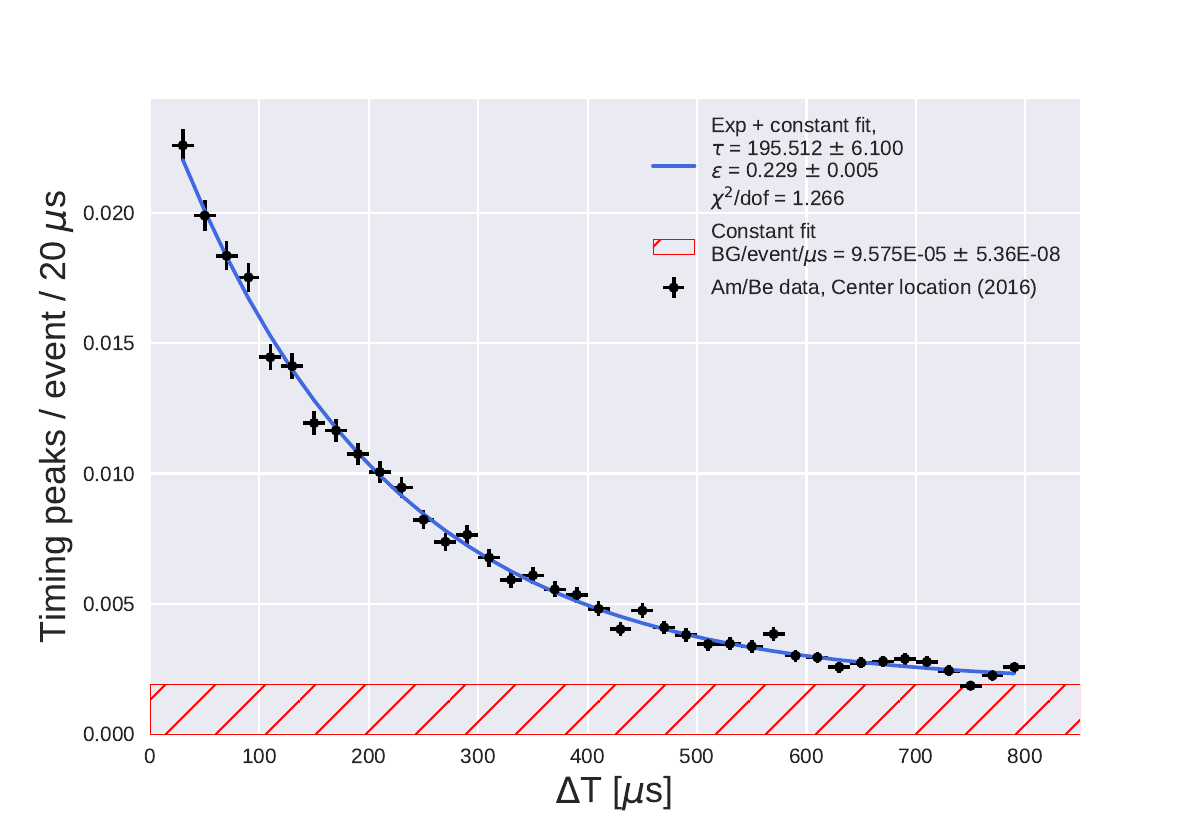}
  \end{center}
  \caption{Exponential $+$ constant fit of neutron candidate timings in the 2016 $^{241}$Am/Be sample collected at the center of the SK tank. The resulting time constant, $195.5\pm6.1$~$\mu$s, is consistent with prior measurements.}
  \label{fig:ambe_exp2}
  \end{figure}

To estimate the systematic uncertainty on the BDT efficiency associated with mismodeling of neutron propagation in water and the $2.2$~MeV photon emission, we compare the efficiencies predicted by the IBD Monte-Carlo simulation to measurements performed using an americium-beryllium ($^{241}$Am/Be) radioactive source embedded in a BGO scintillator. A detailed description of this procedure is given in Ref.~\cite{bib:skntagfirst}.
To calculate the neutron identification efficiency for data, we follow the methodology described in Ref.~\cite{bib:hayatontag} and fit the timing distribution of tagged neutrons by a decaying exponential plus a constant term as shown in Fig.~\ref{fig:ambe_exp2}. True neutron capture event counts decay exponentially in time from the detection time of the prompt event, while uncorrelated backgrounds are constant in time. By comparing the efficiency in data and simulation for BDT cut points relevant in our analysis, as described in Appendix~\ref{sec:appendixntag}, we assign a relative uncertainty of $12.5\%$ for all neutron tagging cuts.

\subsection{Solar neutrino cut}
\label{subsec:solarcuts}
\begin{figure}[htbp]
        \centering
        \includegraphics[width=\linewidth]{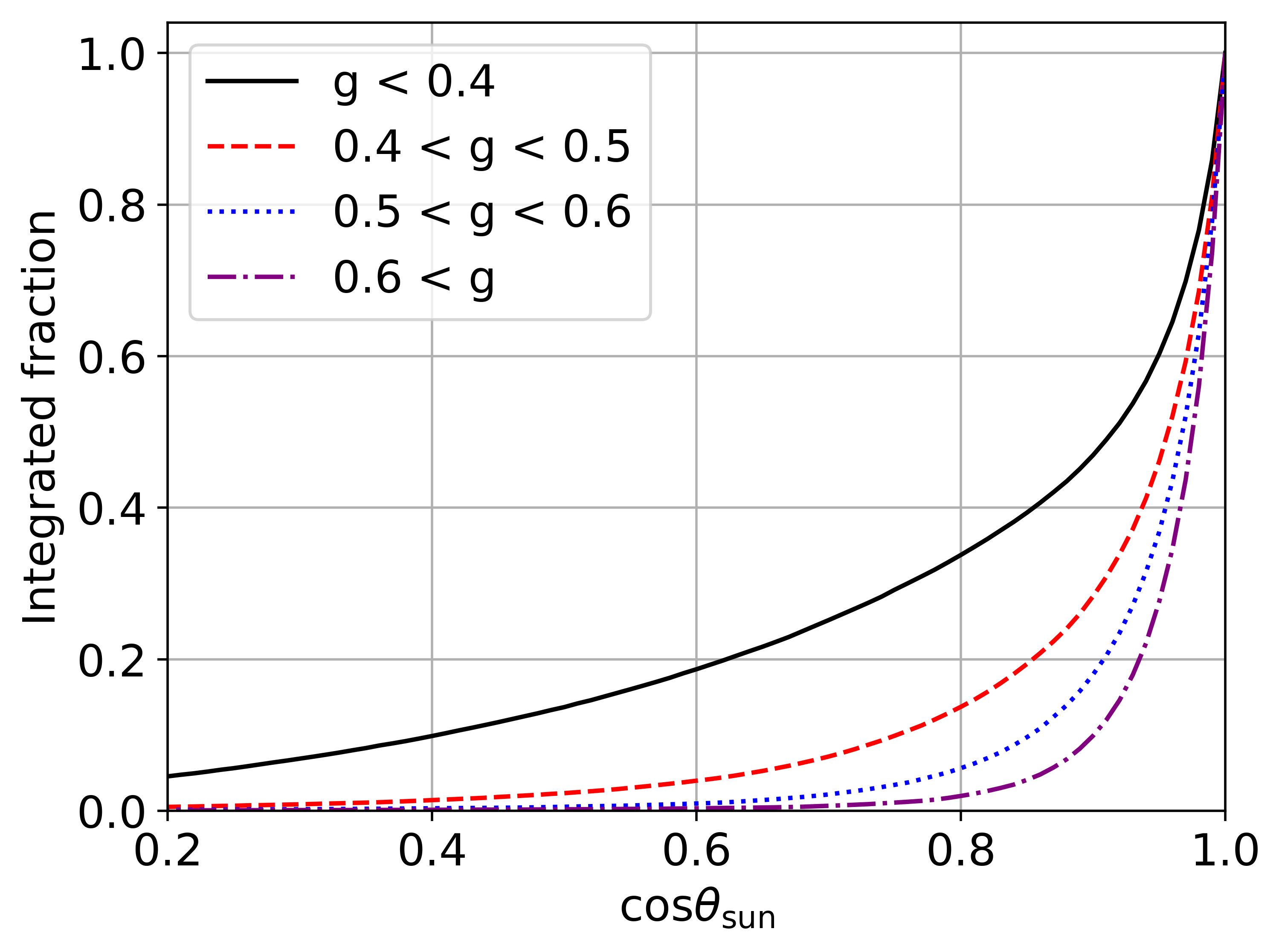}
        \caption{Opening angle distributions for simulated solar neutrino events for the different MSG bins used in this analysis. Events with low MSG have wider opening angle distributions due to direction reconstruction issues.}
        \label{fig:solarmsg}
    \end{figure}

Since even mild neutron tagging cuts can reduce solar neutrino backgrounds to negligible levels, dedicated solar neutrino cuts are not implemented for the DSNB model-independent analysis. For the spectral analysis, however, events with no tagged neutrons are also considered in order to maximize the effective exposure. When considering these events, we hence apply additional cuts specifically targeting solar neutrino backgrounds.
These cuts are based on the opening angle $\theta_\text{sun}$ between the reconstructed direction of candidate events and the direction of the Sun. The cosine of this angle is peaked at 1 for solar neutrino interaction products, and is smeared by kinematic and reconstruction effects. To estimate the impact of the latter, an specific observable, the Multiple Scattering Goodness (MSG) $g$ has been designed for the solar neutrino analysis~\cite{bib:sksolar1,bib:sksolar2,bib:sksolar3,bib:sksolar4}. Simulated opening angle distributions for solar neutrino events in different MSG bins are shown in Fig.~\ref{fig:solarmsg}.

In this analysis we use the method described in Ref.~\cite{bib:sksrn123} and apply $\cos\theta_{\rm sun}$ cuts in the four MSG bins shown in Fig.~\ref{fig:solarmsg}. We estimate the impact of these cuts on DSNB signal events by assuming that these events have a uniformly distributed $\theta_{\rm sun}$, and we obtain the associated MSG distributions using the IBD  Monte-Carlo simulation. 
Conversely, to estimate the number of solar neutrino events above 15.5~MeV, we evaluate the number of solar events in 13.5-15.5~MeV from data and extrapolate the result to higher energies using the solar neutrino MC simulation from Ref.~\cite{newsolarpaper}, a procedure already used in the DSNB search at SK-I,II,III~\cite{bib:sksrn123}. Here, we can safely assume the $\cos\theta_\text{sun}$ distribution for non-solar event to be uniform~\cite{bib:sksolar4}, and therefore approximate the number of solar events in 13.5-15.5~MeV by:

\begin{align}
    N^\text{solar}_\mathrm{low} = N_\mathrm{low}(\cos\theta_\text{sun} > 0) - N_\mathrm{low}(\cos\theta_\text{sun} < 0),
\end{align}

\noindent
using the data passing the noise, spallation, and positron selection cuts described in Secs~\ref{subsec:noise}, \ref{fig:spall_products}, and \ref{subsec:posreduc}. Using the solar neutrino MC simulation from Ref.~\cite{newsolarpaper}, the predicted number of solar neutrino events above 15.5~MeV will then be: 
\begin{align}
    N^\text{solar} \approx 0.12N^\text{solar}_\mathrm{low}.
\end{align}
The impact of solar and neutron tagging cuts can then be assessed by rescaling $N^\text{solar}$ by the corresponding efficiencies.

Finally, we evaluate the systematic uncertainties on the signal efficiency with the solar event cut using the LINAC calibration results. Specifically, we compute the scaling factor that allows to minimize the difference between the predicted and observed MSG distributions for LINAC events. We then evaluate the impact of this rescaling on the final efficiency to be of about 1\%.

\subsection{Cut optimization}
\label{subsec:cutoptspec}

In this study, we perform two types of analyses. 
Here we adopt a common cut optimization procedure for both analyses.

\subsubsection{DSNB positron candidate selection cuts} 

We determined the positron candidate selection cuts by comparing the IBD Monte-Carlo simulation with the atmospheric neutrino simulation for the ring clearness, average charge deposit, and Cherenkov angle, and the observed data for the effective distance $d_\text{eff}$. Since the impact of the positron candidate selection cuts depends weakly on the DSNB model considered, we assumed the Horiuchi$+$09 spectrum~\cite{bib:horiuchi09}. As discussed in Section~\ref{subsec:posreduc}, we estimate systematic uncertainties using results from the solar neutrino analysis~\cite{bib:sksolar1,bib:sksolar2,bib:sksolar3,bib:sksolar4} and the LINAC calibration~\cite{bib:sklinac}. The final cuts are the following:

\begin{align*}
    &d_\text{eff} > \text{max}\left\{ 300, \ 500 - \frac{E_{\rm rec} - 15.5~\text{MeV}}{1~\text{MeV}}\times 50\right\}~\text{cm}, \\ \vspace{10pt}
    &\qquad\qquad\qquad\qquad N_{\rm pre}^{\rm max} < 12, \\ \vspace{10pt}
    &\qquad\qquad\qquad\qquad N_{{\rm decay}\mathchar`-e} = 0, \\ \vspace{10pt}
    &\qquad\qquad\qquad\qquad L_\texttt{clear} < 0.36, \\ \vspace{10pt}
    &\qquad\qquad\qquad\qquad q_{50}/n_{50} < 2, \\ \vspace{10pt}
    &\qquad\qquad\qquad\qquad 38^\circ < \theta_{\rm C} < 50^\circ. 
\end{align*}

\noindent 
The efficiencies of these cuts are shown in Fig.~\ref{fig:signal_cuts} as a function of $E_{\rm rec}$ and vary between 70 and 85\%. Overall, positron candidate selection cuts allow to remove $80\%$ of the atmospheric neutrino backgrounds, and bring contamination from natural radioactivity to negligible levels.

\subsubsection{Spallation and neutron tagging cuts} 

Unlike positron candidate selection cuts, spallation and neutron tagging cuts need to be optimized in multiple energy regions to account for the important variations of the spallation and atmospheric background rates in the analysis windows. While only events with one tagged neutron are used in the DSNB model-independent analysis, the spectral analysis also considers events with zero or more than one tagged neutron. Here, we first optimize spallation and neutron tagging cuts for the signal region where exactly one tagged neutron is required. Using the thus optimized neutron tagging cuts, we then derive optimal spallation cuts for events with $\neq$ 1 tagged neutron.

\paragraph{Events with one tagged neutron: } We devise a cut optimization scheme common to both the model-independent and spectral analyses by simultaneously adjusting spallation and neutron tagging cuts in $E_{\rm rec}$ bins. 
%
For each energy bin, we select events with exactly one tagged neutron and find the spallation and neutron tagging cuts that yield optimal sensitivity under the background-only hypothesis. We define the sensitivity as the predicted number of signal events after cuts over the 90\% C.L. upper limit on the number of signal events, computed using the Rolke method~\cite{Rolke:2004mj}. To simplify the interpretation of the results presented in this paper, we apply the same cuts for both the model-independent and spectral analyses in the 15.5$-$29.5~MeV reconstructed energy range, where the two analyses overlap. In particular, we require spallation and solar neutrino backgrounds to be negligible above $15.5$~MeV.
The remaining rates of these backgrounds can be reliably evaluated by rescaling the number of events observed after noise reduction, spallation, and positron candidate selection cuts by the neutron mistag rate. 

\begin{figure}[htbp]
  \begin{minipage}{1.0\hsize}
   \begin{center}
    \includegraphics[clip,width=9.0cm]{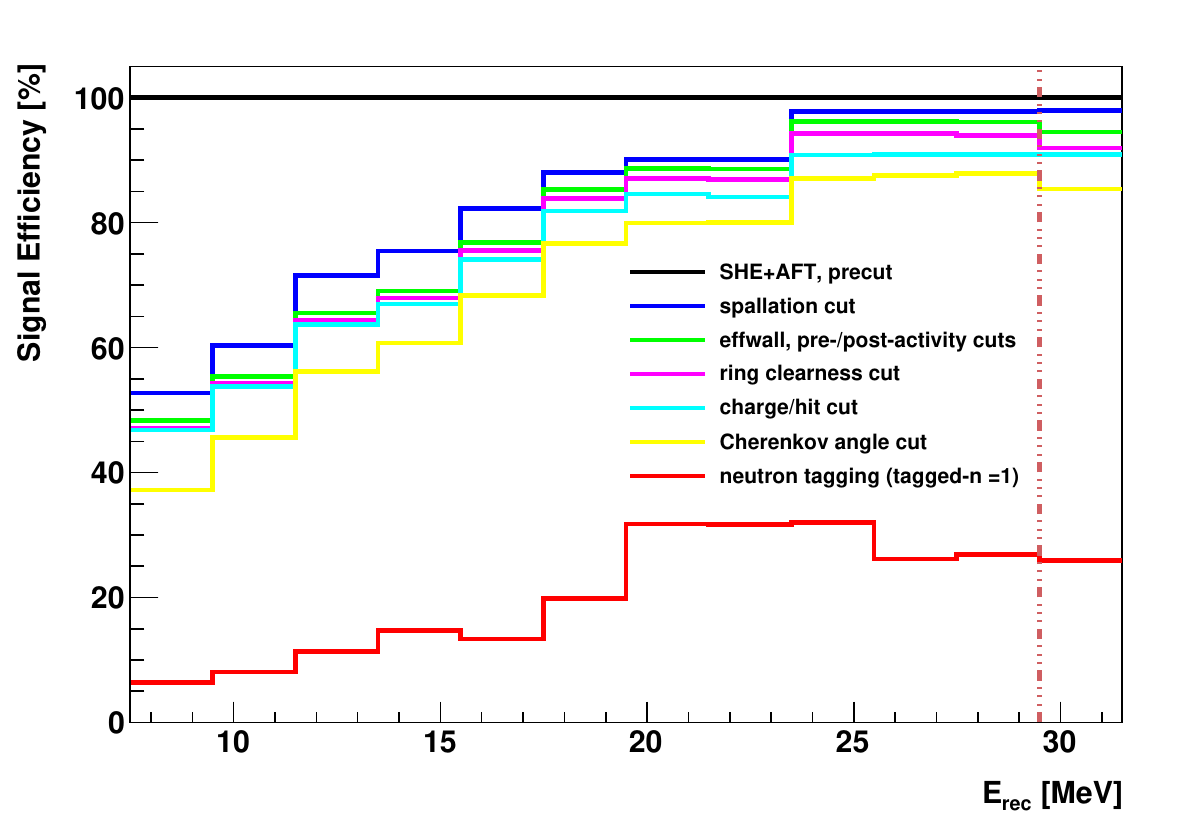}
   \end{center}
  \end{minipage}
  \begin{minipage}{1.0\hsize}
   \begin{center}
    \includegraphics[clip,width=9.0cm]{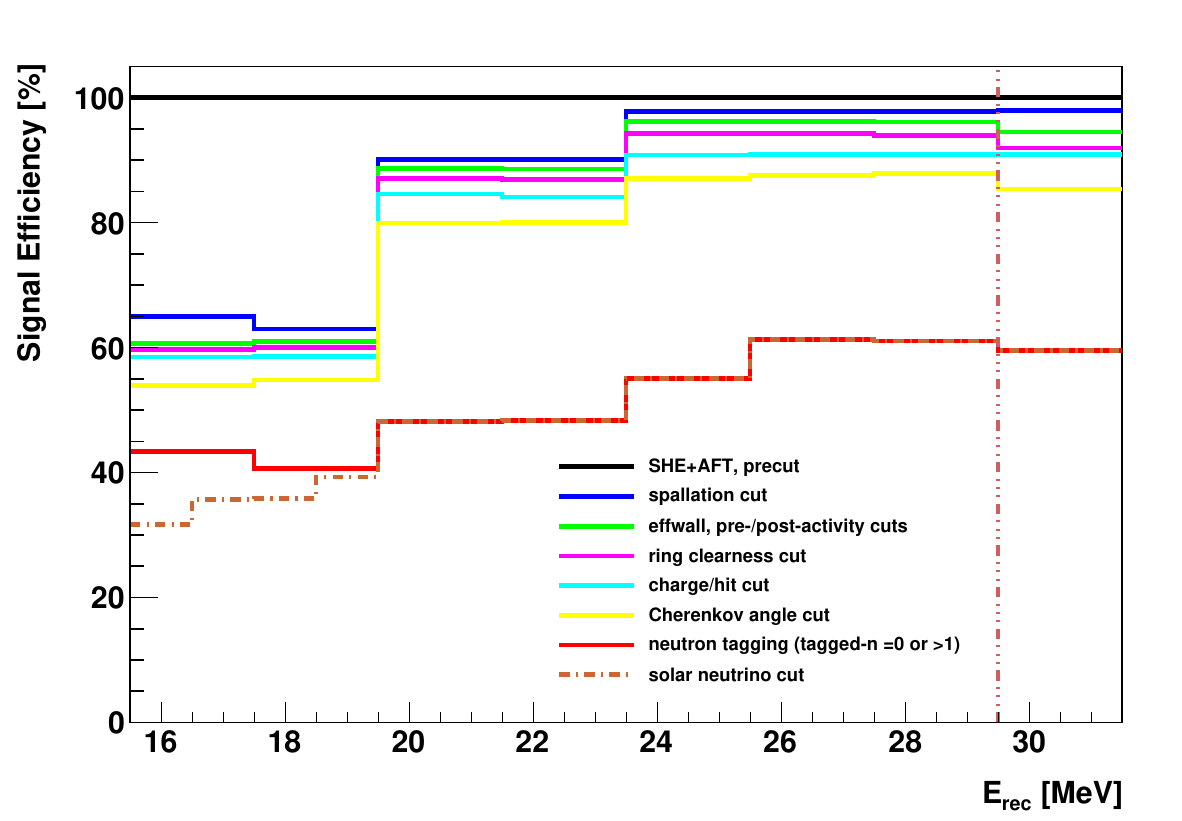}
   \end{center}
  \end{minipage}
  \vspace{+5truept}
  \caption{DSNB signal efficiencies for the signal regions where one tagged neutron (top) and $\neq$ 1 tagged neutrons (bottom) are required, as started after the SHE$+$AFT trigger requirement and the noise reduction cuts. While the one tagged neutron region is common to both the model-independent and spectral analyses above 15.5~MeV, the $\neq$ 1 tagged neutron region is used only in the spectral analysis. Here, we applied noise reduction, spallation, third reduction, neutron tagging, and solar cuts sequentially and show cumulative efficiencies at each stage. The result in the final bin above 29.5~MeV corresponds to the efficiency for the range through 29.5 to 79.5~MeV, as calculated assuming the Horiuchi$+$09 6~MeV spectrum~\cite{bib:horiuchi09}. Note also that efficiencies for the solar neutrino cut in the bottom are shown in 1-MeV bins.}
  \label{fig:signal_cuts}
  \end{figure}

The signal efficiencies for our choice of cuts are shown in the top panel of Fig.~\ref{fig:signal_cuts} for each energy region. 
For each set of cuts we also present the associated systematic uncertainties, computed as described in Sections~\ref{subsec:spallreduc}, \ref{subsec:posreduc}, and~\ref{subsec:ntag}, in Table~\ref{tab:cut_systematics}. 

    \begin{table}[htbp]
    \centering
    \caption{Systematic uncertainties for the reduction cuts, computed as described in Sections~\ref{subsec:spallreduc}, \ref{subsec:posreduc}, and~\ref{subsec:ntag}. The systematic uncertainties for the AFT trigger, noise reduction, pre- and post-activity cuts are negligible. The solar neutrino cut is used only for the spectral analysis, when events have $\neq$1 tagged neutrons.}
    \vspace{+3truept}
    \begin{tabular}{l c}
    \toprule
        Cut & Relative systematic error \\
         \hline
    Spallation cut & $0.1\%$\\ 
    $d_{\rm eff}$ cut & $0.1\%$\\
    $L_\texttt{clear}$ cut & $0.2\%$\\
    $q_{50}/n_{50}$ cut & $1.2\%$\\
    $\theta_{\rm C}$ cut & $0.7\%$\\
    Neutron tagging & $12.5\%$\\
    Solar neutrino cut & $1.0$\%\\
    \botrule
    \end{tabular}
    \label{tab:cut_systematics}
    \end{table}

\paragraph{Events with zero or more than one tagged neutrons: } As previously discussed, the events rejected by the neutron tagging cuts obtained above can still be considered in the spectral analysis. In the 23.5$-$79.5~MeV region, that covers most of the spectral analysis window, spallation and solar neutrino backgrounds are negligible and only noise reduction and positron candidate selection cuts need to be applied to these events. In the 19.5$-$23.5~MeV region, solar neutrino backgrounds are still negligible and spallation backgrounds are mostly linked to short-lived isotopes, that are easy to reject using the cuts from Section~\ref{subsec:spallreduc}. For these energies, we therefore apply the spallation cut derived above, whose efficiency is close to 90\%. In the 15.5$-$19.5~MeV range, however, both solar neutrino and spallation backgrounds largely dominate and need to be reduced using the targeted reduction strategies described in Sections~\ref{subsec:spallreduc} and \ref{subsec:solarcuts}. In this analysis, we remove the solar neutrino backgrounds using the cuts in Section~\ref{subsec:solarcuts} with the cut points from Ref.~\cite{bib:sksrn123}. The cut points and the corresponding signal efficiencies are shown in Table~\ref{tab:solarcuts} for the different multiple scattering goodness and reconstructed kinetic electron energy regions. The remaining number of solar neutrino events after all cuts is less than 2 for the entire energy region. 

    \begin{table}[htbp]
    \centering
    \caption{Cuts on $\cos\theta_{\rm sun}$ (upper limits), and signal efficiencies for SK-IV. These cuts are only applied to events with $\neq 1$ tagged neutron, in the spectral analysis. They depend on the value of the multiple scattering goodness, $g$, and on the reconstructed kinetic electron energy. They are the same as the ones applied for SK-I and III in Ref.~\cite{bib:sksrn123}. The signal efficiency for SK-IV is within 1\% of the one obtained in Ref.~\cite{bib:sksrn123} for SK-I and III, which is similar to the associated systematic uncertainty.}
    \vspace{+3truept}
    \begin{tabular}{ccccc}
    \toprule
    $E_{\rm rec}$ [MeV] & 15.5$-$16.5 & 16.5$-$17.5 & 17.5$-$18.5 & 18.5$-$19.5\\
    \hline
        $g < 0.4$ & 0.05 & 0.35 & 0.45 &0.93 \\
        $0.4<g<0.5$& 0.39& 0.61& 0.77 & 0.93\\
        $0.5<g<0.6$& 0.59& 0.73& 0.81 & 0.93\\
        $g>0.6$& 0.73& 0.79& 0.91 & 0.93\\ \hline
        Efficiency & 73.1\% & 82.2\% & 88.3\% & 96.6\%\\
        \botrule
    \end{tabular}
    \label{tab:solarcuts}
    \end{table}

In contrast, the impact of spallation cuts is difficult to evaluate in the 15.5$-$19.5~MeV range . 
To optimize spallation cuts for the events considered here, we therefore proceed as follows. First, we bring contributions from short-lived isotopes ---with $\mathcal{O}(0.01)$~sec half-lives--- to a negligible level, using the cuts shown. 
in Appendix~\ref{sec:appendixspall}.
As can be seen in Fig.~\ref{fig:spall_products}, the remaining event sample will be dominated by decays from $^8$B and $^8$Li, that have a half-life close to 1~sec. We then optimize the spallation likelihood cuts 
using the method described in Appendix~\ref{sec:spacutperf} to compute the remaining amount of $^8$B and $^8$Li decays, and
maximize $S/\sqrt{B}$ where $B$ and $S$ are the numbers of spallation and non-spallation events respectively. 
The efficiencies of the optimal cuts are of 65\% and 63\% in 15.5$-$17.5 and 17.5$-$19.5~MeV, for a $^8$B+$^8$Li remaining fraction of about 2.5\%. The corresponding remaining number of spallation events, as well as the performance of this cut will be discussed and compared to results from the previous SK DSNB analyses in Section~\ref{subsec:specresults}.

\subsection{Selected events in the final data samples}
\label{sec:eventquality}

\begin{figure}[htbp]
    \centering
    \includegraphics[width=9.5cm]{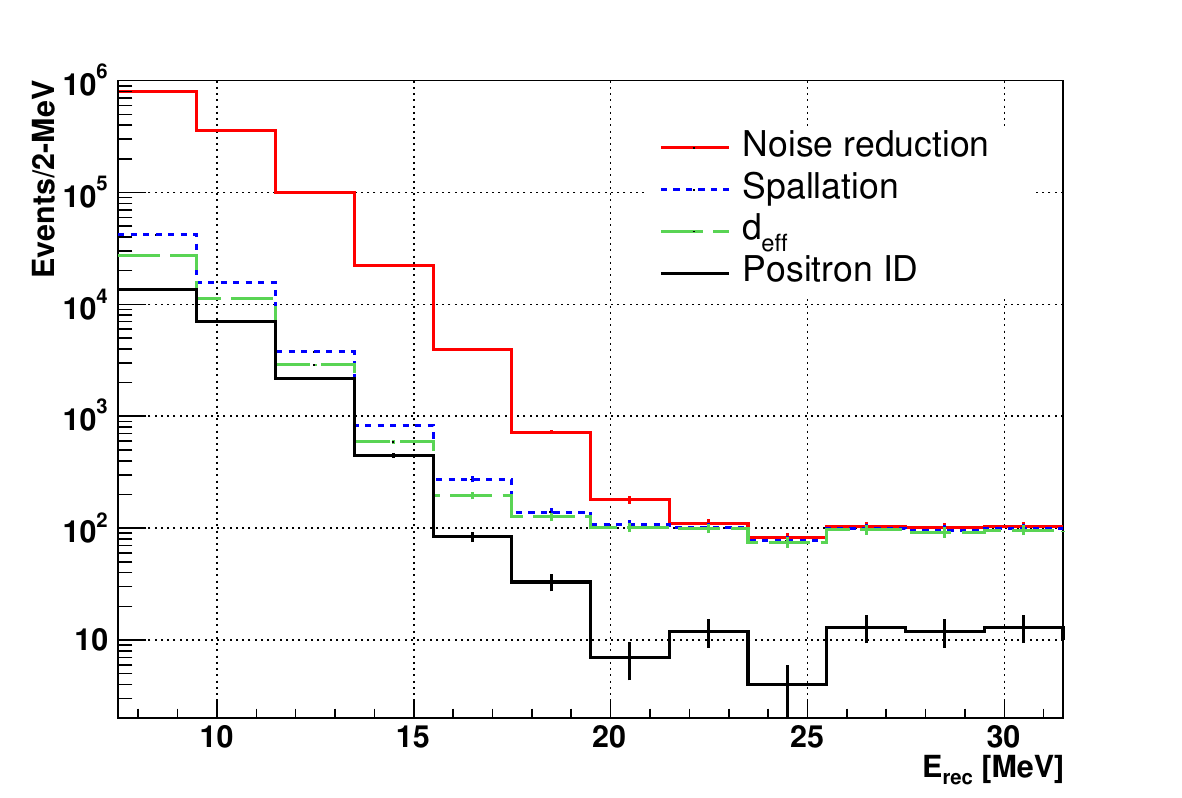}
    \caption{Data spectrum after noise reduction, spallation, $d_\mathrm{eff}$, and all other positron candidate selection cuts. Here the cuts are applied in the order shown above.}
    \label{fig:datacutflow}
\end{figure}

\begin{figure}[htbp]
    \centering
    \includegraphics[width=9cm]{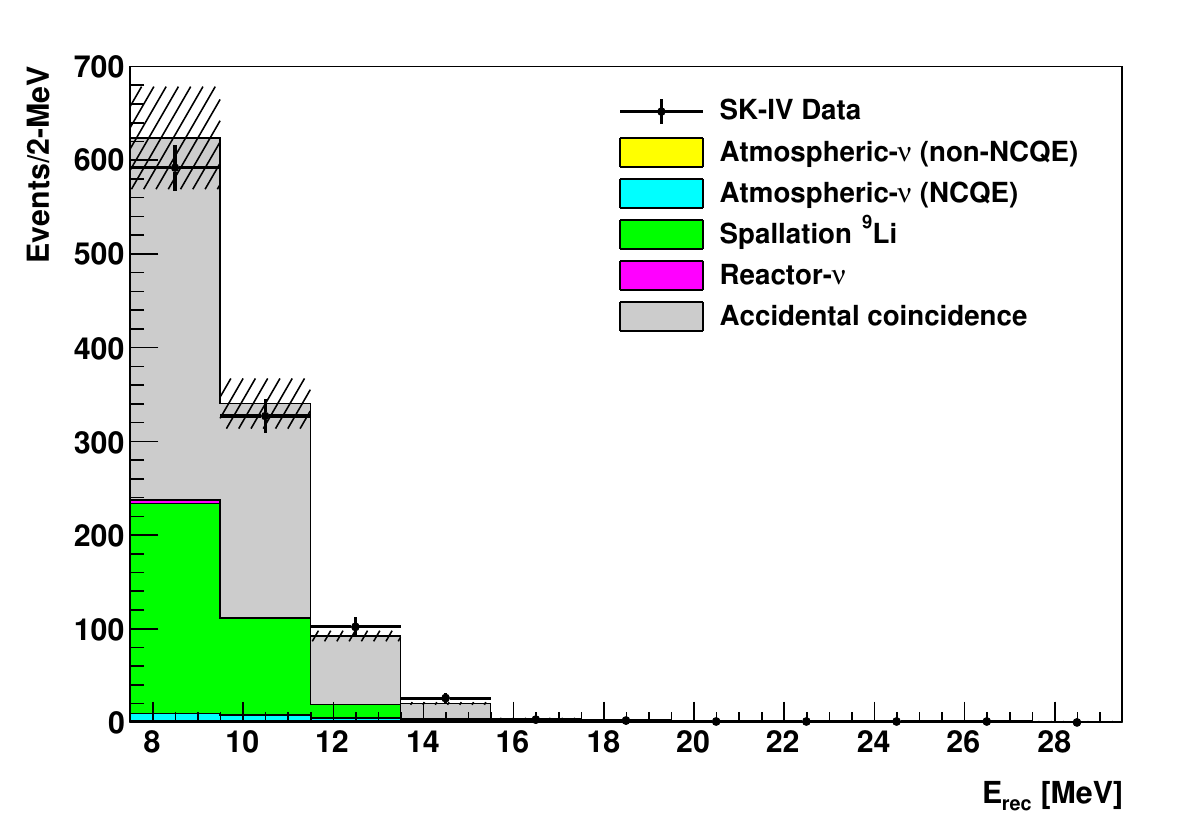}
    \caption{Predicted and observed spectra for events passing noise reduction, multiple spallation, DSNB positron candidate selection, and neutron tagging cuts. Here we apply a uniform neutron tagging cut with a 20\% signal efficiency for all energies. In the absence of most spallation cuts, the dominant backgrounds are accidental coincidences and $^9$Li decays. The systematic uncertainties are shown by the hatched lines.}
    \label{fig:nospacut}
\end{figure}

Here we discuss the behavior of the observed data after the cuts. In particular, the impact of spallation, $d_\mathrm{eff}$, and the other positron candidate selection cuts on data is shown in Fig.~\ref{fig:datacutflow}. As shown in this figure, for reconstructed energies below 20~MeV contributions from spallation and radioactivity near the wall are significant.
To validate our background modeling and neutron tagging procedure we also compare the observed data to the background expectations after the noise reduction, multiple spallation cuts, DSNB positron candidate selection, and neutron tagging cuts. To simplify the interpretation of the data, we impose the same neutron tagging cut, with a 20\% signal efficiency, for all energies. While most spallation cuts are removed in order to focus on neutron tagging, multiple spallation cuts are conserved in order to avoid contaminating the neutron detection window with isotope decays ---an effect not accounted in the mistag rate predictions. The predicted and observed event spectra are shown in Fig.~\ref{fig:nospacut} and agree within uncertainties. 

\begin{figure}[htbp]
    \centering
    \includegraphics[width=\linewidth]{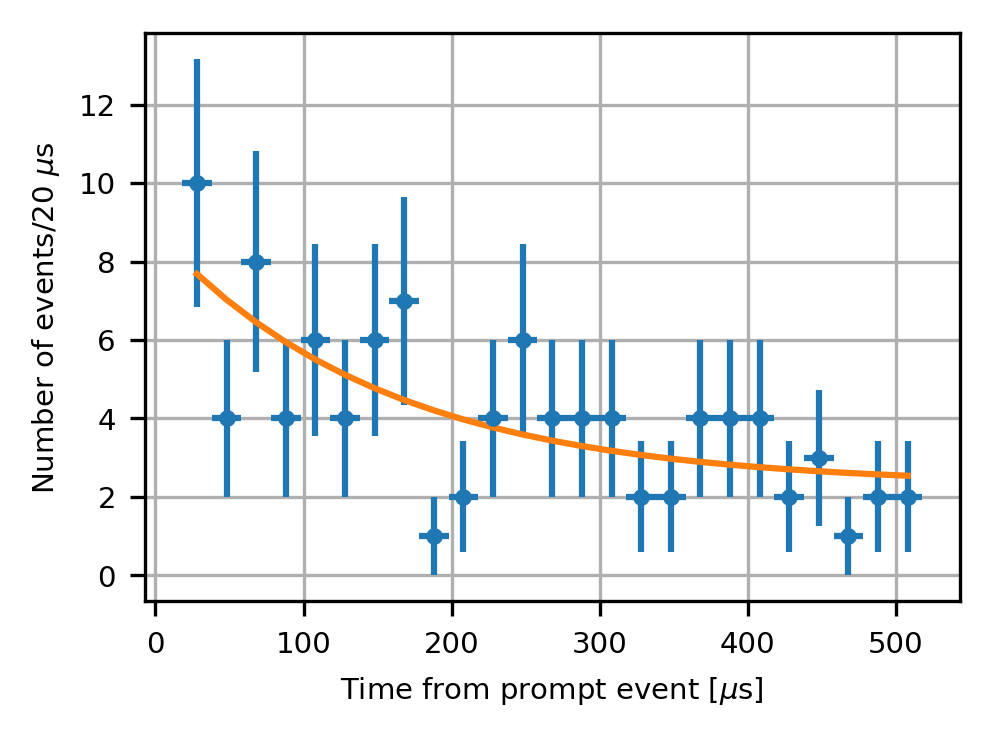}
    \caption{Time differences between reconstructed neutron captures and their corresponding prompt events (positron candidates). Here we show the data for one tagged neutron. The solid orange curve shows the fit result by an exponential plus a constant, giving a time constant of $\tau = 154 \pm 104~\mu$s.}
    \label{fig:neutrontime}
\end{figure}

\begin{figure}[htbp]
    \centering
    \includegraphics[width=\linewidth]{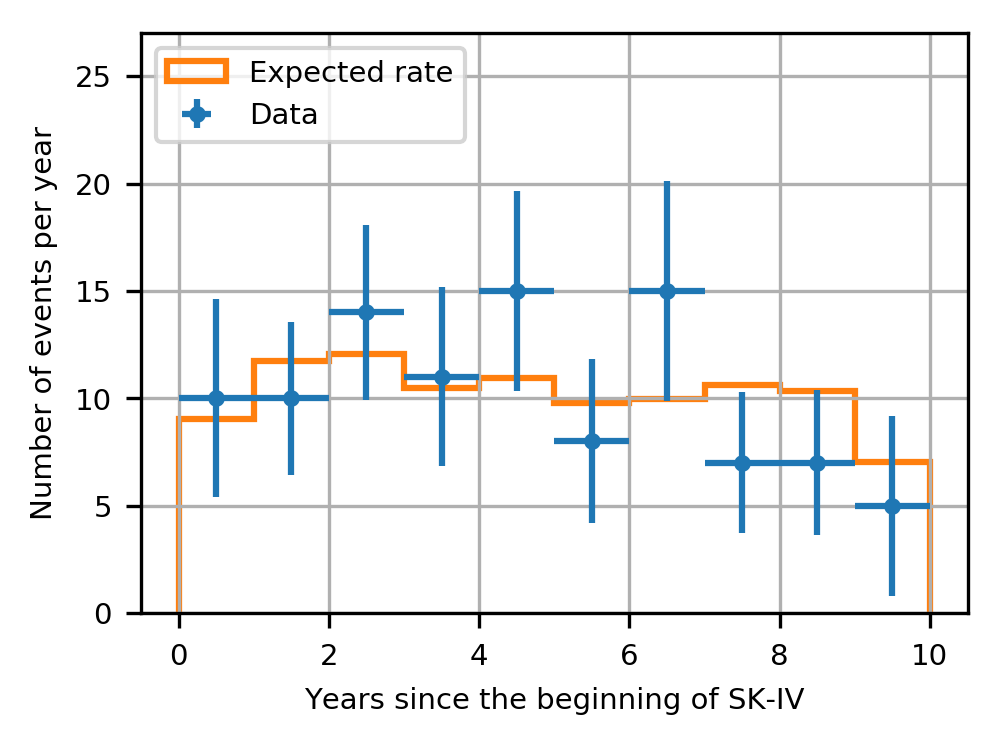}
    \includegraphics[width=\linewidth]{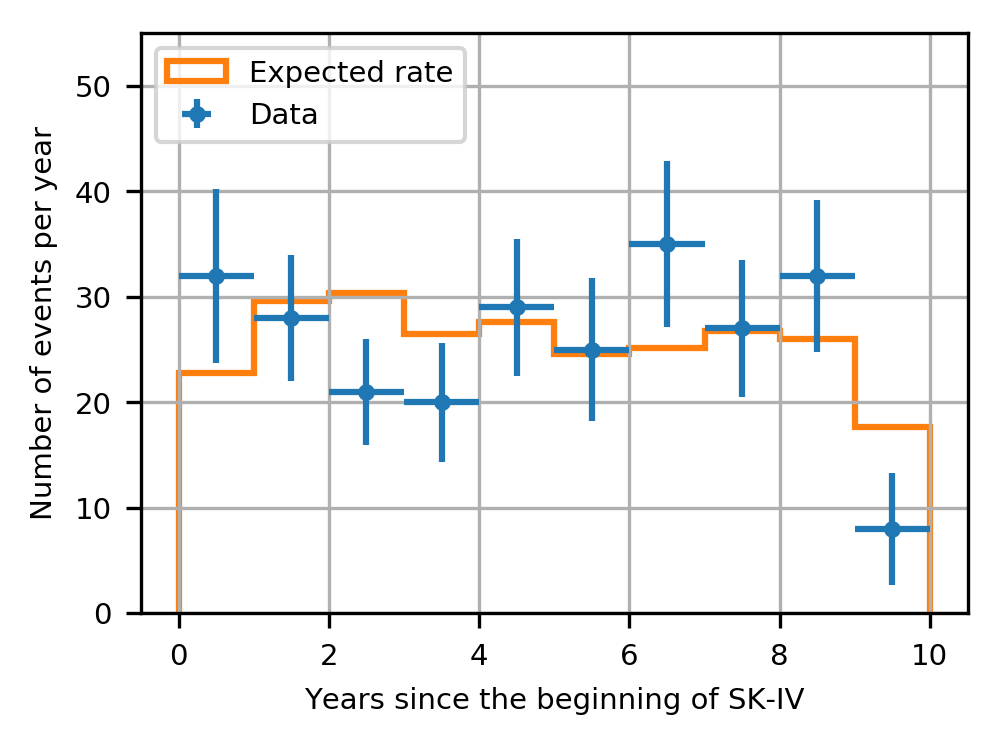}
    \caption{Observation times for events with one neutron (top), and with zero or more than one neutron (bottom) after the cuts. Overlaid is the expected time distribution, where variations of the data-taking rate over the SK-IV period are taken into account.}
    \label{fig:timeseries}
\end{figure}

\begin{figure*}[htbp]
    \centering
    \includegraphics[width=8cm]{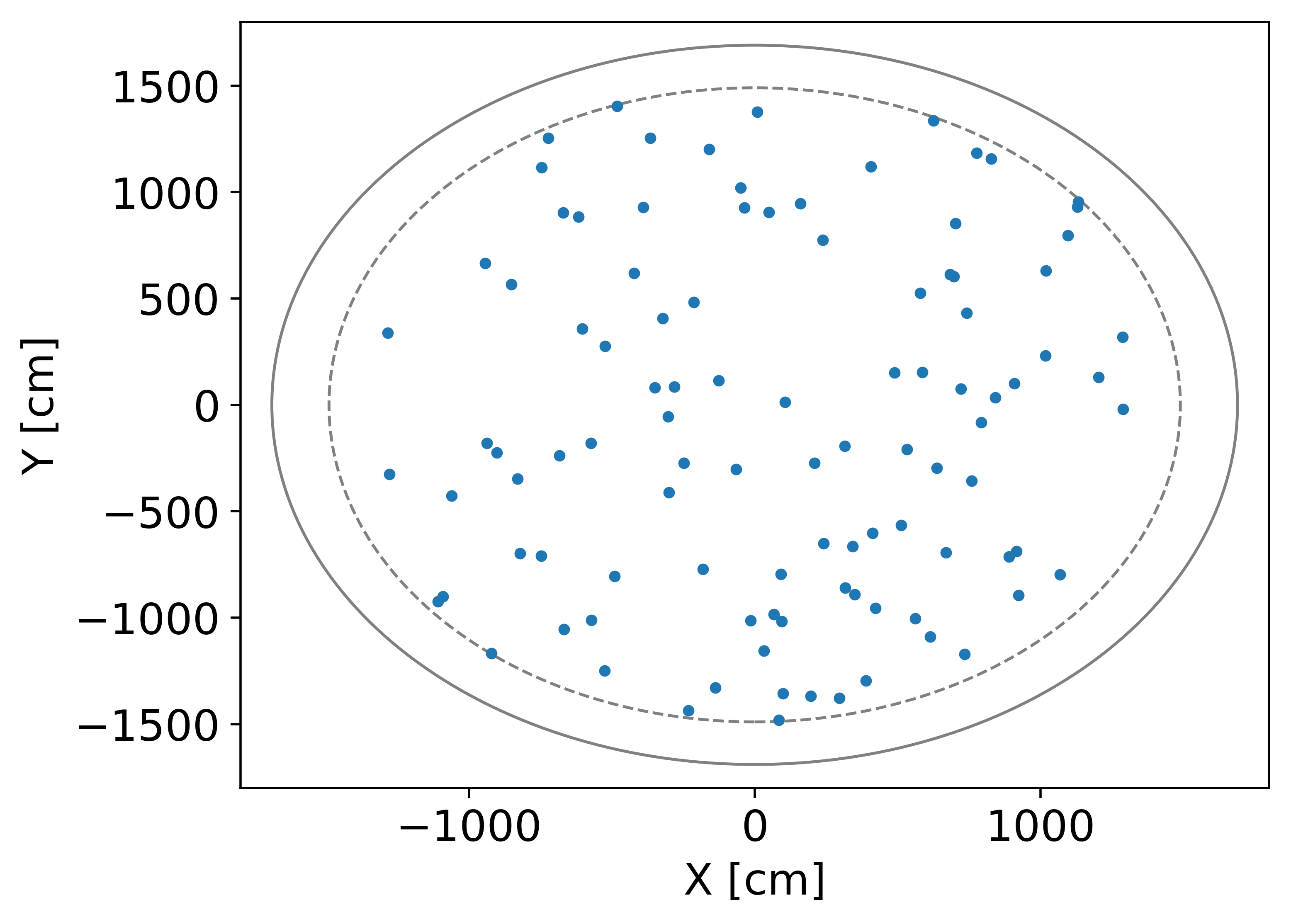}
    \includegraphics[width=8cm]{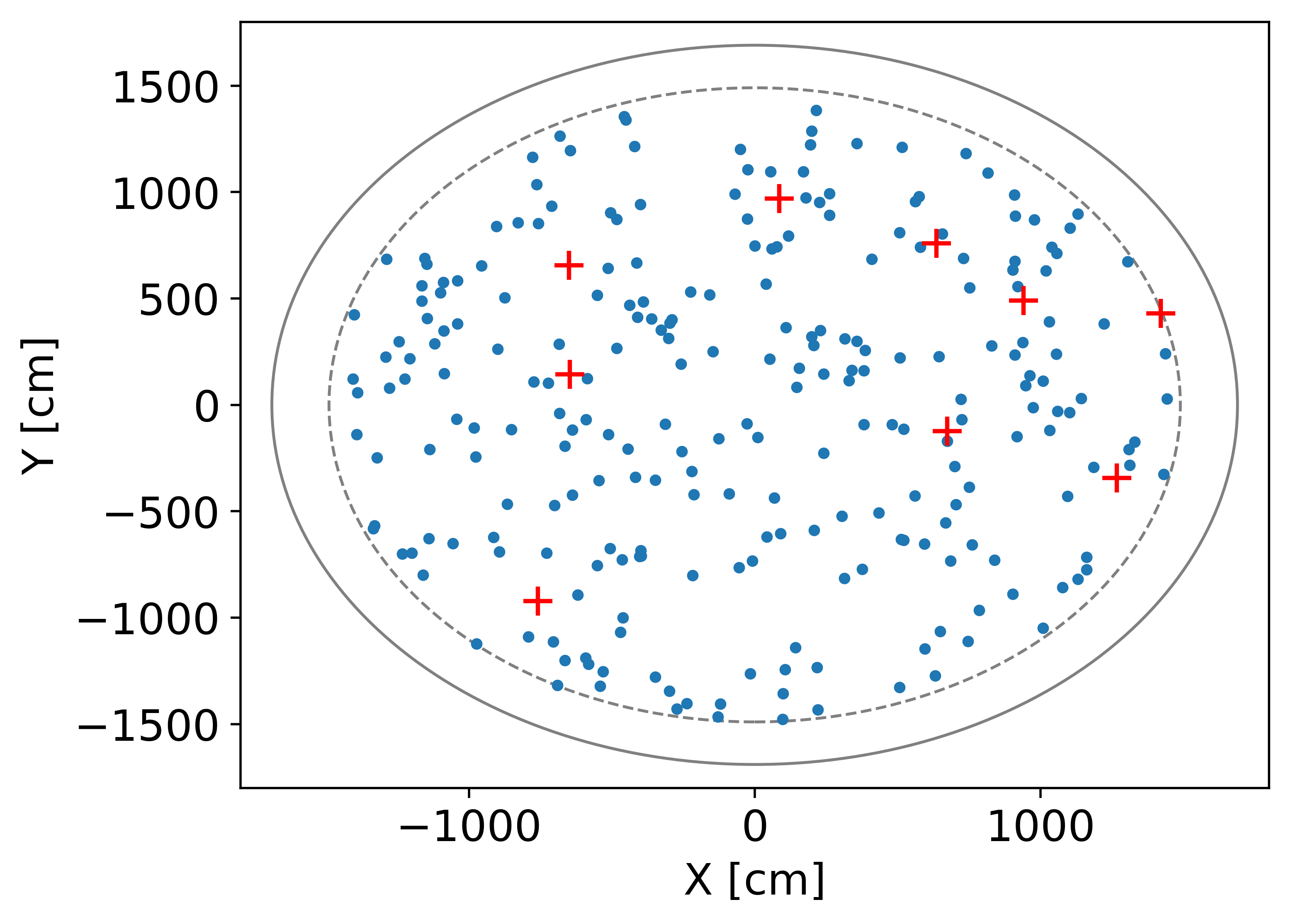}
    \includegraphics[width=8cm]{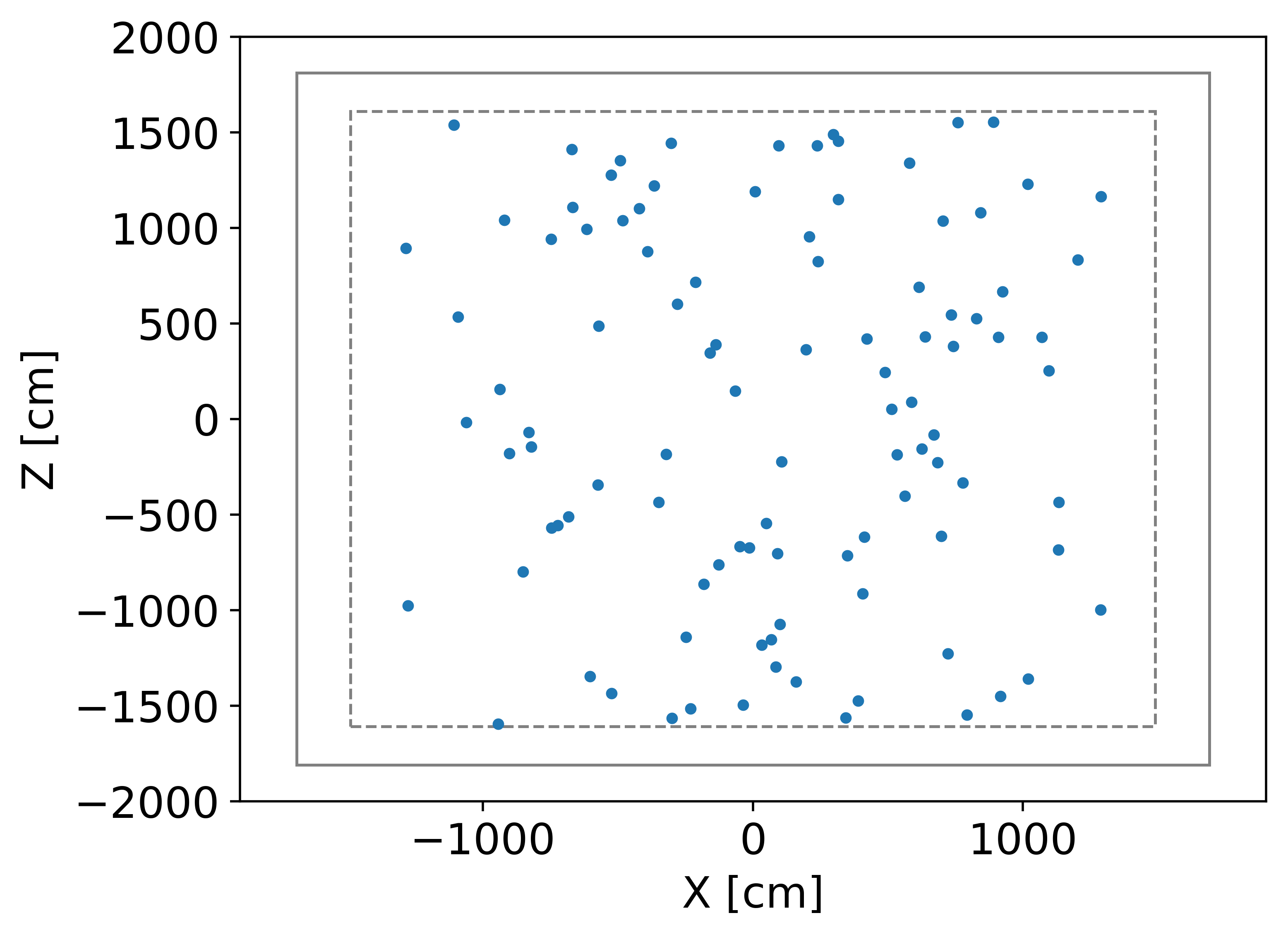}
    \includegraphics[width=8cm]{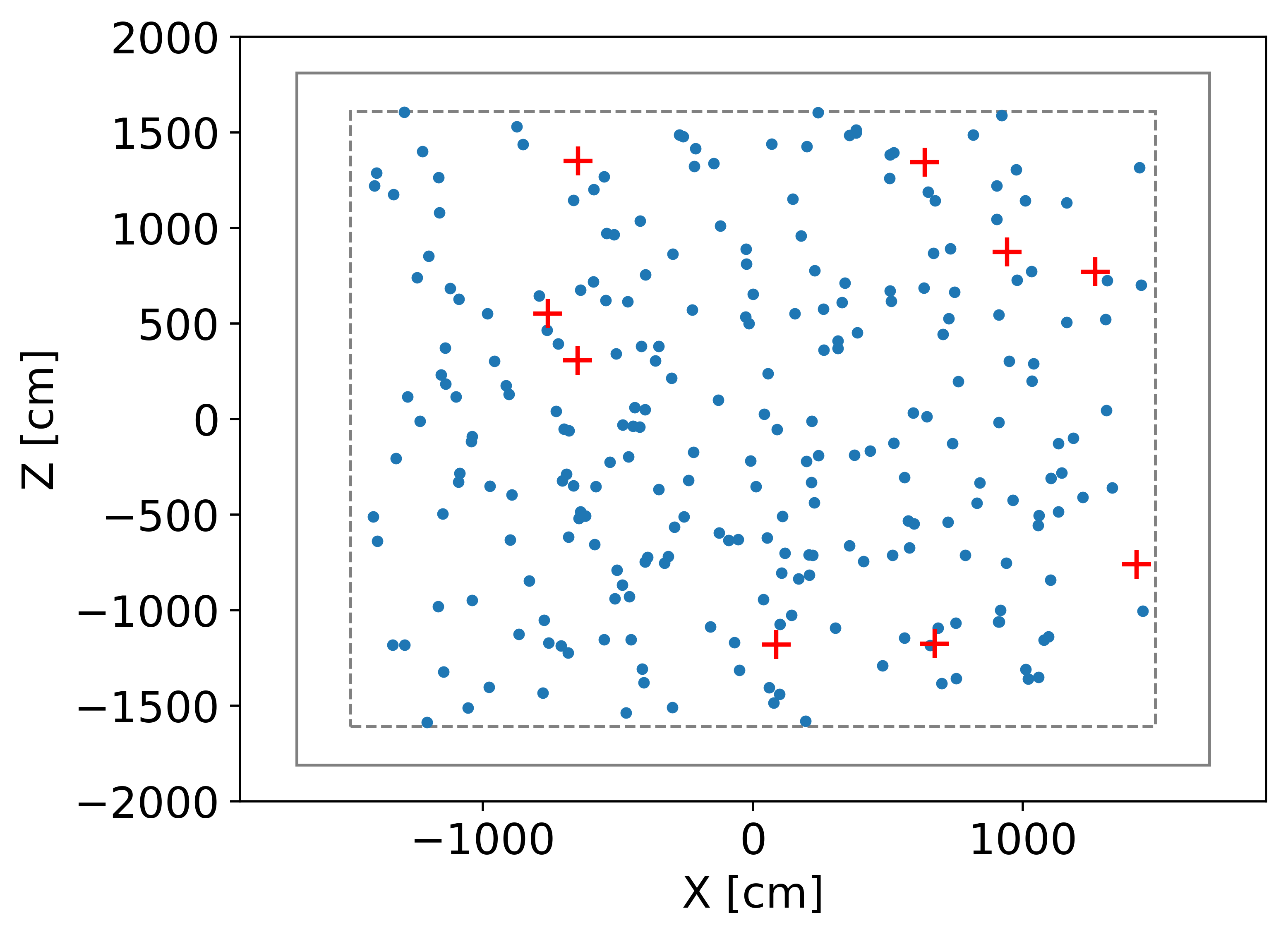}
    \caption{Top (top) and transverse (bottom) views of reconstructed vertices for events with one neutron (left), and with zero or more than one neutron (right) after the cuts. Events with more than one neutron are shown as red crosses. The solid and dashed gray lines show the ID and the FV, respectively.}
    \label{fig:eventpositions}
\end{figure*}

After the cuts derived in Section~\ref{subsec:cutoptspec} are applied, we find 102 events with exactly one neutron in 7.5$-$79.5~MeV. For these events, the distribution of the time difference between the neutron and the prompt event is shown in Fig.~\ref{fig:neutrontime}. Fitting this distribution by an exponential plus a constant, as is done in the $^{241}$Am/Be calibration, yields a time constant of $\tau = 154 \pm 104~\mu$s, which is compatible with the expected neutron capture time in water. In the 15.5$-$79.5~MeV region, that will be used for the spectral analysis, we also find 248 events with zero neutrons and 9 events with more than one neutron.  The observation times and reconstructed vertices for these events are shown in Figs.~\ref{fig:timeseries} and \ref{fig:eventpositions}. No event time cluster has been observed and the time-dependence of the event rate is consistent with the livetime variations over the SK-IV period. Events are also uniformly distributed all over the FV, irrespective of the number of neutrons. The space and time distributions of the selected events are therefore consistent with what is expected for DSNB candidates.

\section{DSNB Model-independent Analysis}
\label{sec:mianalysis}
In this section we perform a DSNB model-independent search.
Specifically, we divide the 7.5$-$29.5~MeV reconstructed energy range into $2$-MeV bins and, in each of them, search for DSNB IBDs.
%
In what follows, we describe how to estimate the various backgrounds encountered in this analysis, model their energy dependence, and estimate the corresponding uncertainties.



\subsection{Background estimation}
\label{subsec:bkgestimate}
\subsubsection{Atmospheric neutrinos}

In the high end of the analysis window ($E_{\rm rec} > 19.5$~MeV), the atmospheric neutrino spectrum is dominated by decay of
invisible muons or pions from CC or NC interactions, forming the well-known Michel spectrum. 
Systematic uncertainties in this region will therefore only affect the total number of predicted atmospheric neutrino events. 
Here, we estimate this number using the sideband region of $29.5 < E_{\rm rec} < 49.5$~MeV, that has a similar background composition.
The number of events in this sideband is $88.8\%$ of the atmospheric neutrino Monte-Carlo prediction, and we therefore use this factor to rescale the non-NCQE backgrounds. The associated uncertainty is due to statistical fluctuations in the number of events in the sideband and is around $19$\%. 
The resulting spectrum is labeled as ``non-NCQE'' in Fig.~\ref{fig:finalspectrumwide}. 

Below $19.5$~MeV, the dominant atmospheric neutrino background arises from NCQE interactions.
The cross section measurements from T2K~\cite{bib:t2kncqe1to9} are used to estimate this background. 
In particular, the Monte-Carlo simulation predictions of neutrino and antineutrino NCQE interactions are renormalized using the scaling factors from this measurement; $f_{\rm \nu\mathchar`-NCQE} = 0.80 \pm 0.08 ({\rm stat.}) ^{+ {\rm 0.24}}_{- {\rm 0.18}} ({\rm syst.})$ and $f_{\rm \bar{\nu}\mathchar`-NCQE} = 1.11 \pm 0.18 ({\rm stat.}) ^{+ {\rm 0.29}}_{- {\rm 0.22}} ({\rm syst.})$, respectively. 
Note that this rescaling provides an inclusive estimation of the NCQE and NC 2p2h interactions.
The errors of these factors are taken as cross section uncertainties.

Two additional flux uncertainties are considered for the NCQE background: the atmospheric neutrino flux uncertainty and the uncertainty due to the flux difference between the T2K beam and atmospheric neutrinos.
For the former, a $15$\% uncertainty is taken from Refs.~\cite{bib:hkkm2006,bib:hkkm2011}.
The latter uncertainty arises from the fact that the scaling factors above are measured for the T2K fluxes while the current focus is the atmospheric neutrino flux and then the effect of cross section model uncertainties is different.
To estimate this uncertainty, the ratios of flux-averaged events in the T2K beam and
atmospheric neutrino fluxes for different cross section models are calculated.
%
%
%
Here six models are considered: Spectral Function (SF)~\cite{bib:benhar1994,bib:benhar2005}, Relativistic Mean Field (RMF)~\cite{bib:horowitz1981,bib:maieron2003,bib:caballero2005,bib:gonzalez2013}, Superscaling (SuSA)~\cite{bib:amaro2005,bib:amaro2006}, Relativistic Green's Function (RGF)~\cite{bib:capuzzi1991,bib:gonzalez2013,bib:meucci2004,bib:meucci2014} with two functional forms for the potential (EDAI, Democratic), and Relativistic Plane Wave Impulse Approximation (RPWIA)~\cite{bib:gonzalez2013}.
The maximum differences are taken as systematic errors, 5\% for neutrinos and 7\% for antineutrinos.


There are two systematic error sources related to neutron tagging for the NCQE background: tagging efficiency and neutron multiplicity. 
For the uncertainty on the tagging efficiency, 12.5\% is employed in this analysis as the maximum difference between the measured and predicted efficiencies in the $^{241}$Am/Be calibration.
The neutron tagging efficiency will also depend on the distance between the neutron emission and capture vertices. Indeed, neutrons produced in atmospheric interactions can travel up to a few meters and are hence more difficult to identify than neutrons produced in DSNB IBDs. The effect of the uncertainties in modeling the neutron travel distance on the efficiency is of around $7$\%~\cite{bib:skncqe}.
In this analysis, the number of tagged neutrons is required to be one, therefore the model uncertainty affecting the neutron multiplicity has to be estimated.
Here the recent measurement of the neutron multiplicity after (mostly CC) interactions of neutrinos from the T2K beam inside the SK tank is used \cite{bib:rakutsu}.
This measurement allows to compare observations with Monte-Carlo simulation predictions for neutron multiplicity as a function of the reconstructed squared momentum transferred to the nucleus, $Q^2$, and could therefore be adjusted to study the NCQE background. 
We incorporate the maximum difference between observed and predicted neutron multiplicity in T2K for each $Q^2$ range, and renormalize the present Monte-Carlo simulation prediction to accommodate this gap. 
This procedure gives variances of approximately 40\% in the number of events with the tagged neutron being one, both for neutrinos and antineutrinos.    

The NCQE spectral shape is sensitive to the $\gamma$ emission model, affecting the current differential limit extraction. 
It is complicated to estimate the associated systematic uncertainty, as $\gamma$ rays are produced at every stage of the NCQE process, from the primary neutrino interaction to the secondary nuclear reaction. The modeling of these $\gamma$ emission is in particular a probable cause of the currently observed discrepancy between the observed and predicted Cherenkov angle distributions using neutrinos from the T2K beam ~\cite{bib:t2kncqe1to9}. One notable effect of this mismodeling is a smearing of the reconstructed energy analogous to a resolution effect. To evaluate the associated systematic uncertainty we model this smearing by convolving the original spectrum with a Gaussian distribution and using the energy-dependent ratio between the smeared and unsmeared spectra to renormalize the Monte-Carlo events.
We then compute the number of renormalized events in each energy bin after the reduction steps described in Section~\ref{sec:reduc}. The difference between this result and the nominal prediction is then taken as the systematic uncertainty associated to the $\gamma$ emission modeling. For this study, we set the smearing parameter $\sigma$ to 3~MeV for all energies; this value allows to cover the discrepancy observed between the predicted and observed $\theta_{\rm C}$ distributions in the T2K measurements mentioned above. The resulting uncertainty ranges from 30\% to 60\% in the 7.5$-$29.5~MeV range.

Finally, we combine all the systematic uncertainties described above by adding them in quadrature. 
The total systematic uncertainty on the NCQE background then varies from 60\% to 80\% for low to high energies. The predicted NCQE spectrum with the T2K scaling is shown in Fig.~\ref{fig:finalspectrumwide}.

\subsubsection{Spallation $^{9}$Li} 

The background rate of spallation $^9$Li can be estimated by the following formula: 

  \begin{align}
  \label{eq:li9estimate}
   B_{^9\text{Li}} = R_{^9\text{Li}} \times T_{\text{live}} \times 
                     22.5~\text{kton} \nonumber \\ 
                     \times \text{Br}[^9\text{Li}\rightarrow\beta+n] \times f_{\text{window}}
                     \times \epsilon_\text{reduc}, 
  \end{align}

\noindent 
where $R_{^9\text{Li}}$ is the production $^9$Li rate measured at SK ($0.86 \pm 0.12({\rm stat.}) \pm 0.15({\rm syst.})~{\rm kton^{-1}}$$\cdot$${\rm day^{-1}}$)~\cite{bib:skli9}, $T_{\text{live}}$ is the operational livetime (2075.3 and 2970.1~days below and above $E_{\rm rec} = 9.5$~MeV, respectively), $\text{Br}[^9\text{Li}\rightarrow\beta+n] = 0.508$ is the branching ratio for $\beta$+$n$ decay, $f_{\text{window}}$ is the fraction of the $^9$Li decay energy spectrum above the search energy threshold, and $\epsilon_\text{reduc}$ is the reduction efficiency. 
We estimate $f_{\text{window}}$ using the Monte-Carlo simulation elaborated for IBD events, renormalized to fit the $^9$Li spectrum. 
This same simulation can be used to estimate the efficiencies for the noise and positron reductions as well as the neutron tagging, that are the same as for the DSNB signal. 
The spallation cut efficiency is estimated using the procedure outlined in Section~\ref{sec:reduc}. 
The dominant systematic uncertainty on the number of $^9$Li events in each energy bin after cuts arises from estimating the impact of spallation cuts. Indeed, as discussed in Appendix~\ref{sec:spacutperf}, our method assumes that the time difference between an isotope decay and its parent muon is the only isotope-dependent spallation observable. In this study, we account for the impact of other spallation observables by defining a 50\% uncertainty on the number of $^9$Li events.
%
The uncertainty on the event rate is taken from the previous SK study~\cite{bib:skli9} as 22\%. 
Other uncertainties come from the reduction, especially from the spallation cut and neutron tagging, that is 20\% in total. 
The total uncertainty assigned for the $^9$Li background estimation is therefore 60\%.

\subsubsection{Reactor neutrinos} 

Reactor neutrino backgrounds are estimated by renormalizing the IBD Monte-Carlo simulation. 
They populates only in the lowest energy bin, as seen in Fig.~\ref{fig:finalspectrumwide}. 
In this analysis, a conservative 100\% uncertainty is assigned to their estimated rate.

\subsubsection{Accidental coincidences} 
\label{subsubsec:acc}
A prompt SHE signal accidentally paired with a subsequent dark noise fluctuation or a low energy radioactive decay can look like an IBD signature. 
Since accidental pairing can occur for both signal and background events, the resulting background can be readily estimated by computing the number of events left after noise reduction, spallation, and positron reduction cuts $N^\text{data}_\text{pre-ntag}$, and rescaling it by the neutron mistag rate: 

  \begin{align}
  \label{eq:accestimate}
   B_\text{acc} = \epsilon_\text{mis} \times N^\text{data}_\text{pre-ntag}. 
  \end{align}

\noindent 
The associated systematic errors are highly cut-dependent and arise from the time dependence of the PMT dark noise and natural radioactivity. In order to evaluate its effect on the background rejection, we separate the random trigger data mentioned in section~\ref{subsec:signalsimu} over all the SK-IV period into $10$ time bins of about eight months each. Then we evaluate the BDT mistag rate for each bin separately and, for a given signal efficiency, take the standard deviation from the average mistag rate as systematic error.  The typical size of uncertainty is a few to $30$\%, in the region of interest for the DSNB analyses, as shown in Fig.~\ref{fig:bdtroc}.

\subsection{Differential upper limit}
\label{subsec:upperlimit}

\begin{figure}[htbp]
  \begin{minipage}{1.0\hsize}
   \begin{center}
    \includegraphics[clip,width=9.0cm]{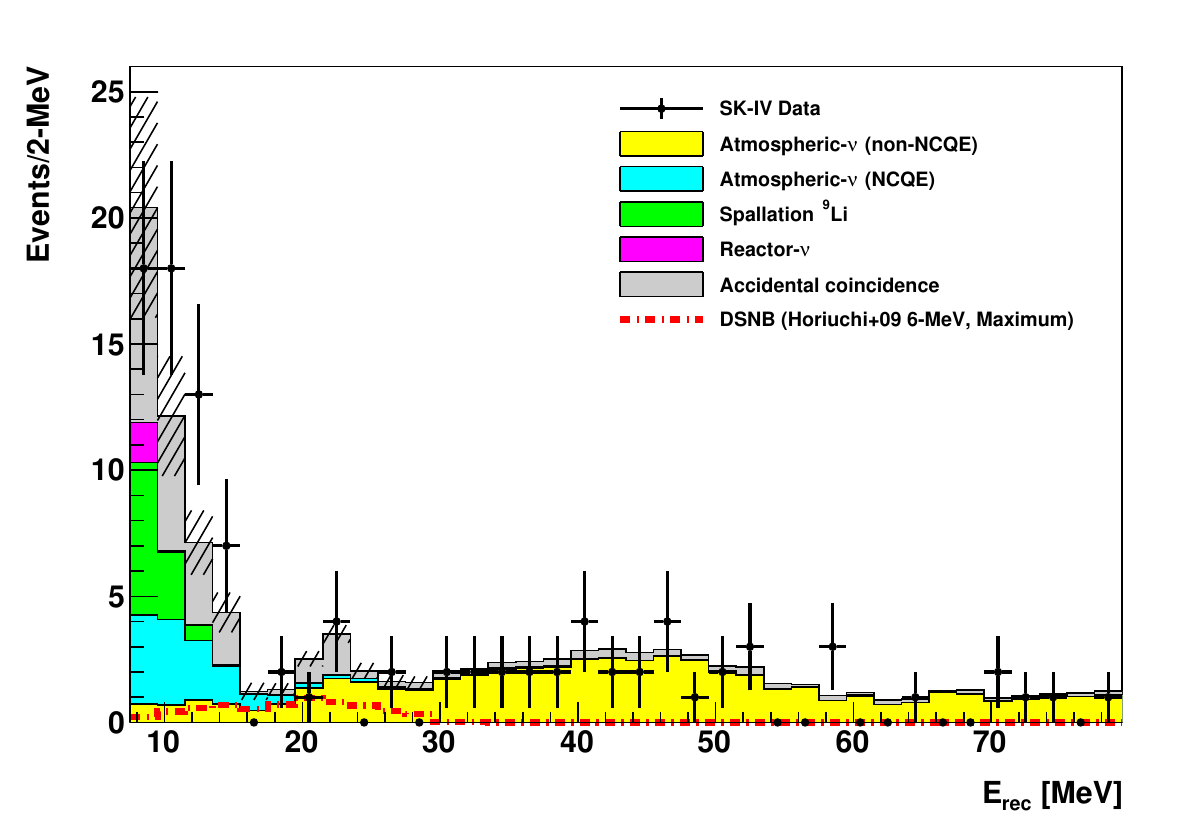}
   \end{center}
  \end{minipage}
  \begin{minipage}{1.0\hsize}
   \begin{center}
    \includegraphics[clip,width=9.0cm]{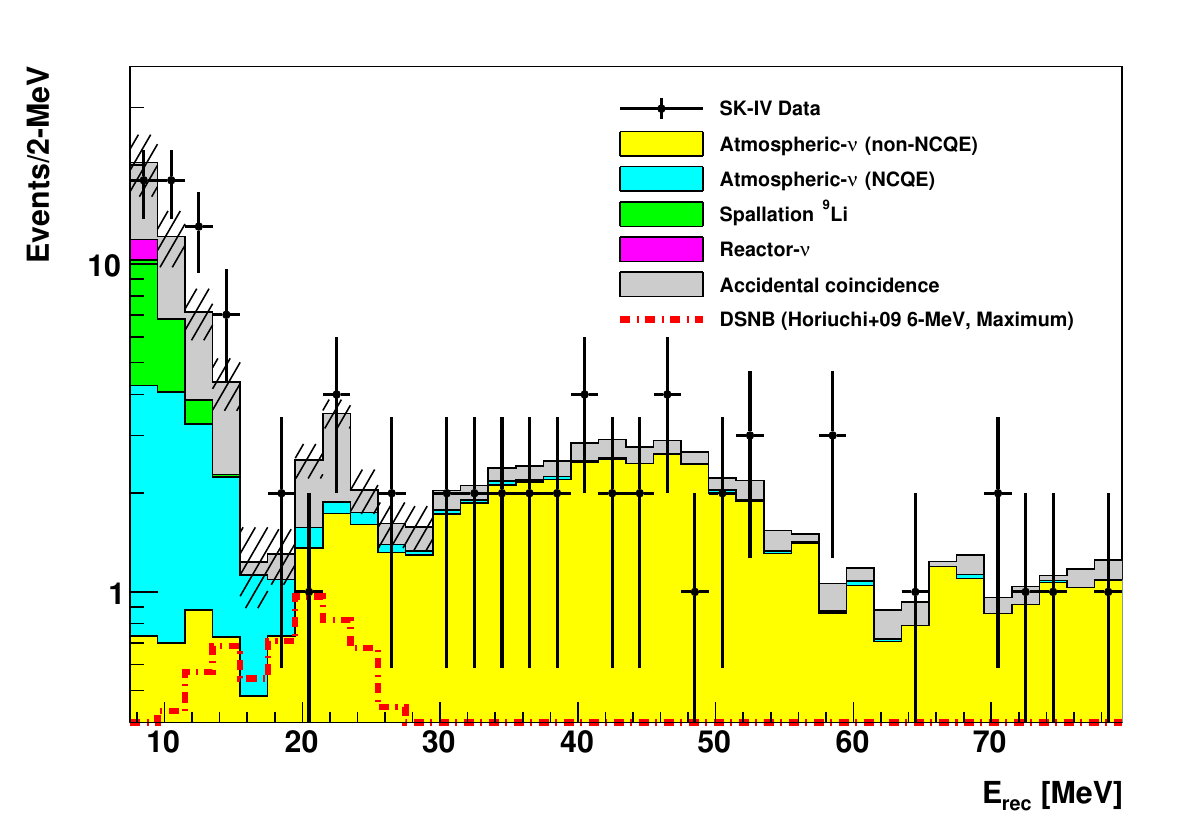}
   \end{center}
  \end{minipage}
  \vspace{+5truept}
  \caption{Reconstructed energy spectra after data reductions for the background 
           expectation and observation including the signal window and the sideband region
           with a linear (top) and a log (bottom) scale for the vertical axis. 
           Color-filled histograms correspond to each background source with 
           hatched lines in the signal window representing the total systematic uncertainty. 
           The red dashed line represents a DSNB signal expectation from the Horiuchi$+$09 
           model~\cite{bib:horiuchi09} shown only for the signal window.}
  \label{fig:finalspectrumwide}
  \end{figure}

\begin{figure*}[htbp]
  \begin{center}
   \includegraphics[clip,width=13.5cm]{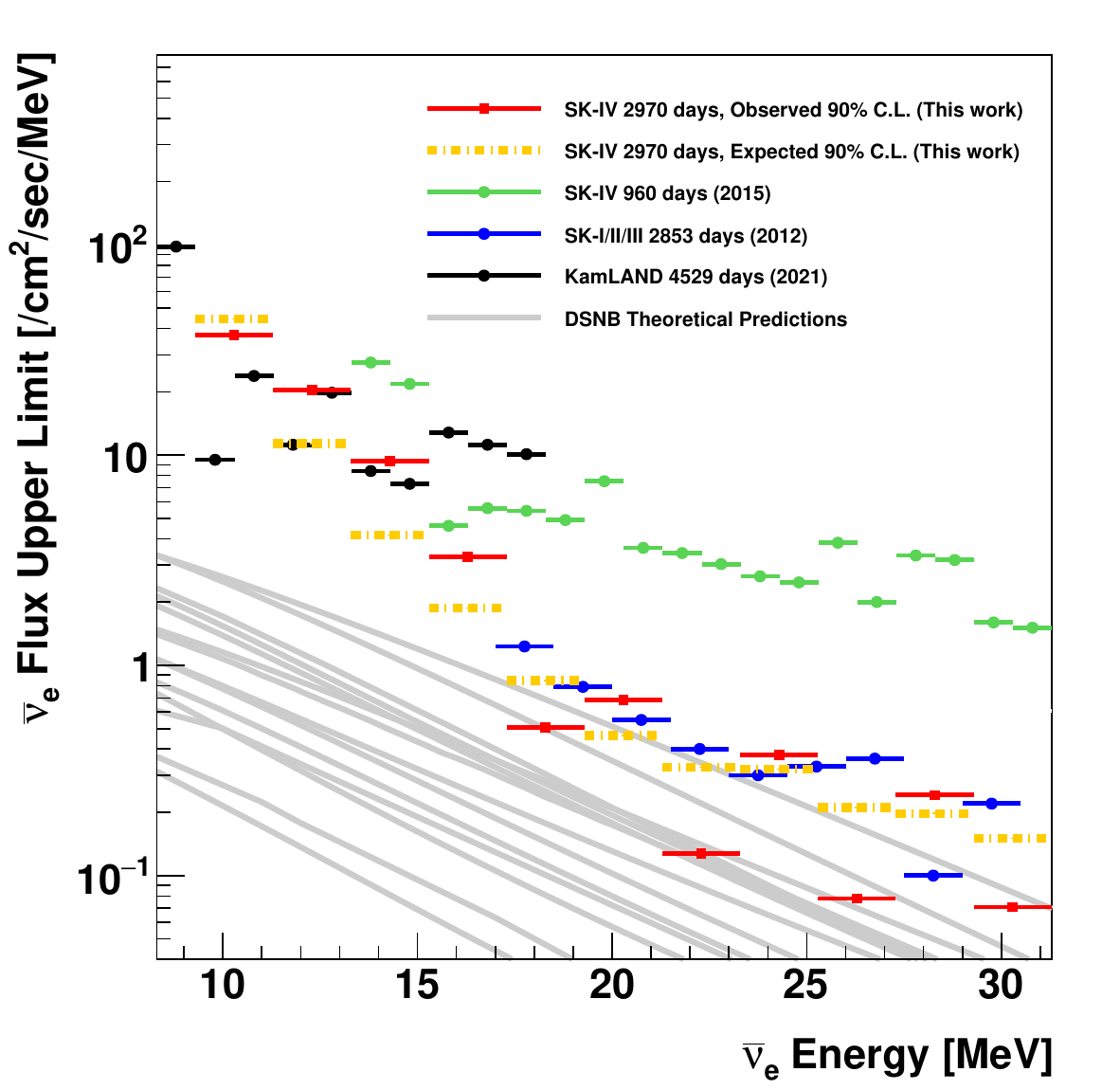}
  \end{center}
  \vspace{-10truept}
  \caption{The 90\% C.L. expected and observed upper limits on the extraterrestrial electron                antineutrino flux from the present work, in comparison with previously published 
           results from SK~\cite{bib:sksrn123,bib:sksrn4} and KamLAND~\cite{bib:kamlandsrnnew} and 
           DSNB theoretical predictions from Fig.~\ref{fig:srnmodels} (in gray). The upper limit from Ref.~\cite{bib:sksrn123} (blue) has been derived in Ref.~\cite{bib:sksrn4}.}
  \label{fig:srn_modelindep_limit}
  \end{figure*}

The observed data and expected background spectra are shown in Fig.~\ref{fig:finalspectrumwide}. The corresponding p-values are calculated for each energy bin by performing pseudo experiments as follows. We vary the number of expected background events randomly based on Gaussian distributions with their widths being the 1$\sigma$ systematic uncertainties from each background source. We then calculate p-values from the resulting distributions and the observed number of events in each bin, as shown in Table~\ref{tab:srn_modelindep_limit}. Here the most significant bins are found to be at 2$\sigma$ level. 
We therefore conclude that no significant excess is observed in the data over the background prediction in any energy bin, and place upper limits on the extraterrestrial $\bar{\nu}_e$ flux using the pseudo experiments above. Here we obtain the 90\% C.L. upper bound on the number of signal ($N_{90}^{\rm limit}$) as an excess of the observation over the background expectation, by varying the number of observed events in each energy bin by their statistical uncertainties as well as varying the number of expected background events by their systematic uncertainties from each source. The 90\% C.L. upper limit on the $\bar{\nu}_e$ flux is then calculated as:

  \begin{align}
   \label{eq:srnlimit}
    \phi_{90}^{\rm limit} = \frac{N_{90}^{\rm limit}}
                                  {t \cdot N_{p} \cdot \bar{\sigma}_{\rm IBD} 
                             \cdot \epsilon_{\rm sig}}, 
  \end{align}

\noindent 
where $t$ is the operational livetime [sec], $N_p$ is the number of free protons in the SK's FV,
$\bar{\sigma}_{\rm IBD}$ is the IBD cross section [${\rm 10^{-41}~cm^2}$] at a mean                
neutrino energy in the corresponding region ($\bar{E}_{\nu}$), and $\epsilon_{\rm sig}$ is the signal efficiency. Note that the neutrino energy is obtained as $E_{\nu} = E_{\rm rec} + 1.8$~MeV in IBD.
For the expected sensitivity, the same procedure is applied while the number of observed event is replaced with the number of nominal background prediction. Here statistical uncertainties on the backgrounds are considered instead. 
The expected sensitivities and observed upper limits from this work are summarized in Table~\ref{tab:srn_modelindep_limit}. These results are also compared with the previously published results and theoretical predictions in Fig.~\ref{fig:srn_modelindep_limit}. Note that the SK-I,II,III limits presented in this figure have been obtained using a different methodology, more similar to the one used for the spectral analysis in Section~\ref{sec:spectral}. In addition, the SK-I,II,III analysis not only used a higher energy threshold but also did not involve neutron tagging~\cite{bib:hzhang,bib:sksrn123}. Hence, we do not combine the SK-IV result with the SK-I,II,III one for the model-independent differential limit. 
The sensitivities obtained with this analysis are the world's tightest above neutrino energies of 11.3~MeV.
The current limit disfavors the Totani$+$95 model~\cite{bib:totani95} and the most optimistic predictions of the Kaplinghat$+$00 model~\cite{bib:kaplinghat00}, and is reaching close to several other model predictions. Due to not only a higher exposure but also higher cut efficiencies and more precise background estimation, this new analysis considerably improves on the previous SK-IV DSNB model-independent search~\cite{bib:hzhang}. 

  \begin{table*}[htbp]
    \centering
    \caption{Summary on the 90\% C.L. expected sensitivities and observed upper limits as well as the corresponding p-values in each electron antineutrino energy bin ($E_{\nu} = E_{\rm rec} + 1.8$~MeV).}
    \vspace{-5truept}
    \begin{tabular}{r l l c} \\ \hline\hline
      $E_{\nu}$ [MeV] & \ \ Expected [${\rm cm^{-2}sec^{-1}MeV^{-1}}$] & \ \ Observed [${\rm cm^{-2}sec^{-1}MeV^{-1}}$] & \ \ p-value \\ \hline  
      9.3$-$11.3 & \ \ \ $4.44\times10^{1}$  & \ \ \ $3.71\times10^{1}$  & \ \ $0.346$ \\
     11.3$-$13.3 & \ \ \ $1.14\times10^{1}$  & \ \ \ $2.04\times10^{1}$  & \ \ $0.886$ \\
     13.3$-$15.3 & \ \ \ $4.17\times10^{0}$  & \ \ \ $9.34\times10^{0}$  & \ \ $0.938$ \\
     15.3$-$17.3 & \ \ \ $1.87\times10^{0}$  & \ \ \ $3.29\times10^{0}$  & \ \ $0.830$ \\
     17.3$-$19.3 & \ \ \ $8.48\times10^{-1}$ & \ \ \ $5.08\times10^{-1}$ & \ \ $0.243$ \\
     19.3$-$21.3 & \ \ \ $4.64\times10^{-1}$ & \ \ \ $6.84\times10^{-1}$ & \ \ $0.686$ \\
     21.3$-$23.3 & \ \ \ $3.28\times10^{-1}$ & \ \ \ $1.27\times10^{-1}$ & \ \ $0.073$ \\
     23.3$-$25.3 & \ \ \ $2.11\times10^{-1}$ & \ \ \ $3.75\times10^{-1}$ & \ \ $0.597$ \\
     25.3$-$27.3 & \ \ \ $2.13\times10^{-1}$ & \ \ \ $7.77\times10^{-2}$ & \ \ $0.051$ \\
     27.3$-$29.3 & \ \ \ $1.98\times10^{-1}$ & \ \ \ $2.42\times10^{-1}$ & \ \ $0.605$ \\
     29.3$-$31.3 & \ \ \ $1.50\times10^{-1}$ & \ \ \ $7.09\times10^{-2}$ & \ \ $0.126$ \\ \hline\hline
    \end{tabular}
    \label{tab:srn_modelindep_limit}
  \end{table*}

\section{Spectral fitting}
\label{sec:spectral}
In this section, we derive model-dependent limits on the DSNB flux by fitting signal and background spectral shapes to the observed data in the 15.5$-$79.5~MeV range, and combine the results with the ones from previous SK phases~\cite{bib:sksrn123} to achieve a $22.5\times 5823$~kton$\cdot$day exposure. While atmospheric neutrino backgrounds, notably from decays of invisible muons and pions, will dominate over most of the analysis window, we also account for possible residual spallation in 15.5$-$19.5~MeV. 

\subsection{Signal and sideband regions}

In order to evaluate the number of atmospheric neutrino background events, we define three regions of parameter space based on the reconstructed Cherenkov angle: one signal region with $\theta_{\rm C}\in [38,50]^\circ$, that will contain most of the signal and the irreducible backgrounds, and two sidebands with $\theta_{\rm C}\in [20,38]^\circ$ and $\theta_{\rm C}\in [78,90]^\circ$. The low Cherenkov angle region will be populated with mostly atmospheric backgrounds involving visible muons and pions while the high angle region will be mostly populated by NCQE atmospheric neutrino events with multiple $\gamma$ rays. Finally, we separate events with exactly one identified neutron from the others, thus defining an ``IBD-like'' and a ``non IBD-like'' region. Note that due to the low efficiencies of the neutron tagging cuts the non IBD-like region is expected to contain a sizable amount of signal. Our analysis will hence involve six regions of parameter space: two signal regions with intermediate values of the Cherenkov angle, and four sidebands with low and high Cherenkov angle values, as summarized in Table~\ref{tab:sigregions}.

\begin{table}[htbp]
    \centering
    \caption{Overview of the regions used in the spectral analysis. We split the parameter space according to the reconstructed Cherenkov angle and the number of tagged neutrons. Regions with small and large Cherenkov angles are dominated by the interactions producing visible muons and pions and by NCQE interactions, respectively. We assign numbers for each region.}
    \vspace{3truept}
    \begin{tabular}{|c|ccc|}
    \hline
         \diagbox{$N_{{\rm tagged\mathchar`-}n}$}{$\theta_{\rm C}$}&
         20$-$38$^\circ$ & 38$-$50$^\circ$ & 78$-$90$^\circ$ \\
                  \hline
        1 & I & II & III \\
        $\neq$1 & IV & V & VI\\
        \hline
    \end{tabular}
    \label{tab:sigregions}
\end{table}

\subsection{Spectral shape fitting}
\label{subsec:shapefit}

We perform a simultaneous fitting of the signal and background spectra to the observed data in all six regions of parameter space defined in Table~\ref{tab:sigregions} using an extended maximum likelihood method. Performing this type of analysis requires knowing the shapes of the signal and background spectra in each of the regions. While the signal spectrum can be reliably predicted by the IBD Monte-Carlo simulation for a given DSNB model, the treatment of the atmospheric neutrino and spallation backgrounds is more complex. For this study, we follow the method described in Ref.~\cite{bib:sksrn123} and divide these backgrounds into the following five categories.
\paragraph{Invisible muons and pions:} this category regroups events with electrons produced by the decays of invisible muons and pions. The energy distribution of these electrons follows a Michel spectrum, whose shape is independent of the Cherenkov angle and the neutron multiplicity and will therefore be the same across all six regions. For this study, we estimate the shape of the Michel spectrum at SK directly from data, using a sample of electrons produced by cosmic ray muon decays (a similar sample for atmospheric and accelerator neutrino oscillation and proton decay analyses is described in Refs.~\cite{bib:ashie,bib:regis}). We then use the atmospheric neutrino Monte-Carlo simulation to compute the fractions of background events in the different signal and background regions. Since electrons cannot be distinguished from positrons at SK this background dominates in the signal regions II and V and are negligible everywhere else.
\paragraph{$\nu_e$ CC interactions:} in this category we find backgrounds arising from CC interactions of electron neutrinos and antineutrinos, with no visible muons and pions in the final state. Their contributions will dominate in the signal regions II and V above $50$~MeV. We estimate the associated spectral shapes in all regions using the atmospheric neutrino Monte-Carlo simulation. Similarly to Michel electrons, this background is negligible outside the regions II and V. 
\paragraph{$\mu/\pi$-producing interactions:} visible muons and pions will be associated with low Cherenkov angles, as these particles are significantly heavier than electrons. The associated background will therefore dominate in the low Cherenkov angle regions I and IV, and, after positron candidate selection cuts, will be negligible in the signal regions. We extract the associated spectral shapes by considering an atmospheric Monte-Carlo sample with only CC interactions, visible muons and pions, and no electrons.  
\paragraph{NCQE interactions:} unlike the other types of atmospheric neutrino events the NCQE spectrum peaks at low energies, similarly to the DSNB spectrum. Here we define spectral shapes using simulated NCQE events with no electrons. This background, often involving multiple $\gamma$ rays, will dominate in the high Cherenkov angle regions III and VI, but will also contribute to all other regions.
\paragraph{Spallation backgrounds:} these backgrounds are negligible in the one-tagged-neutron regions I, II, and III after the cuts. However, in the other regions, especially in signal region V, residual spallation backgrounds from $^8$Li, $^8$B, and $^9$C decays could remain. Since $^9$C has a shorter half-life and is produced closer to the muon track than the other two isotopes, it is expected to be subdominant. To model the spallation spectrum and its variability we parameterize this spectrum and vary its isotope composition as follows:

\begin{align}
    S_{\rm spall}(E) = &f_0 S_{^9{\rm C}}(E)\\
    \nonumber
    &+ (1 - f_0) \left(f_1 S_{^8{\rm B}}(E) + (1 - f_1) S_{^8{\rm Li}}(E)\right),\\
    \nonumber
    \text{with }f_0\in [&0,0.5]\quad\text{and}\quad f_1\in [0,1]
\end{align}

\noindent 
where $S_i(E)$ refers to the decay spectrum for a given isotope $i$, normalized to one in the spectral analysis energy window, and $f_0$ and $f_1$ are the isotope fractions. Here, we constrain $^9$C to always be less abundant than the other two isotopes after spallation cuts are applied. In what follows, we will take the nominal spallation spectrum to be the average between the steepest and the least steep spectra and parameterize possible deviations from this hypothesis as systematic uncertainties. We evaluate the $\beta$ spectra using external measurements~\cite{bib:winter06} for $^8$B decays and GEANT4 for $^8$Li and $^9$C. As with $^9$Li, we then use these spectra to renormalize events in the IBD Monte-Carlo simulation and thus model detector resolution effects. \bigskip\\

Although the first four background categories do not include all possible types of atmospheric neutrino interactions, their associated spectra can be linearly combined to model other background categories. Pion-producing NC backgrounds can notably be treated as a superposition of NCQE and decay electron backgrounds. The five categories described here hence allow to map all the background configurations relevant to this analysis.

For each background category we define probability distribution functions (PDFs) which model the corresponding energy spectra in each of the six Cherenkov angle and neutron regions. These PDFs are normalized to $1$ across all regions. We apply a similar procedure to obtain signal spectral PDFs associated to different DSNB models, using the IBD Monte-Carlo simulation.
For $N_\text{events}$ observed events with energies $\{E_1,...,E_{N_\text{events}}\}$ we then extract the most probable numbers of signal and background events by performing an extended unbinned maximum likelihood fit, using the following likelihood function:

\begin{align}
    \mathcal{L}(\{N_j\}) &= e^{-\sum_{j = 0}^5 N_j}\,\prod_{i = 1}^{N_\text{events}}  \sum_{j = 0}^5 N_j\,\text{PDF}_{j}^{(r)}(E_i).
    \label{eq:like}
\end{align}

\noindent 
Here, the index $j = 0,...,5$ refers to the signal and the five background categories described above. Then $\{N_j\}$ designates the numbers of events for the signal and the different background types, across all six Cherenkov angle and tagged neutron regions. For each event, $\text{PDF}_{j}^{(r)}(E_i)$ is the PDF for the signal or background category $j$ in the region $r$ (for the regions defined in Table~\ref{tab:sigregions}).

\subsection{Systematic uncertainties} 
\label{subsec:specsys}
One considerable advantage of using the likelihood defined in Equation~\ref{eq:like} is that the sizes of the different background contributions can be treated as nuisance parameters. This likelihood, however, does not account for the considerable uncertainties on the spallation and atmospheric background spectral shapes in the energy window of the DSNB analysis. In this section, we outline how to incorporate systematic uncertainties on these shapes as well as on detector modeling and predicted cut efficiencies.
We account for systematic uncertainties on the spallation and atmospheric neutrino background spectral shapes, energy reconstruction, and reduction efficiencies by convolving the likelihood introduced in Equation~\ref{eq:like} with suitable probability distributions. Here we neglect the systematic uncertainties associated with the $\mu/\pi$ background. The amounts of these background in the signal regions II and V are indeed constrained to be negligible by observations from the low angle regions I and IV, a result that is robust under large variations of the assumed spectral shapes in these regions. We also neglect the systematic uncertainties on the spectral shape of the background associated with invisible muon and pion decays since this shape has been directly extracted from a large sample of SK-IV data. We then consider five sources of systematic uncertainties: energy scale and resolution, spectral shape predictions for the spallation, NCQE, and $\nu_e$ CC backgrounds, and reduction cut efficiencies.

\subsubsection{Energy scale and energy resolution} 
To estimate the impact of energy scale and energy resolution uncertainties on the signal and atmospheric background spectra, we follow the procedure described in Ref.~\cite{bib:sksrn123}.
The energy scale and energy resolution uncertainties are obtained from the solar neutrino analysis and increased to 2\% and 3\% respectively to account for the higher energies studied here. We model the effect of these uncertainties by distorting the signal and atmospheric background PDFs in Equation~\ref{eq:like} using the energy resolution functions from Refs.~\cite{bib:sksolar1,bib:sksolar2,bib:sksolar3,bib:sksolar4}. We parameterize the amount of distorsion applied to these PDFs using the variable $\epsilon$ such that $\epsilon = 1$ represents a $1\sigma$ deviation in the energy scale and resolution. The final likelihood can then be expressed as:

\begin{align}
    \mathcal{L}'= \int_{-\infty}^{+\infty} e^{-\epsilon^2/2}\mathcal{L}(\epsilon)\,\mathrm{d}\epsilon,
\end{align}

\noindent 
where $\mathcal{L}(\epsilon)$ represents the likelihood from Equation~\ref{eq:like} with the PDFs distorted by an amount $\epsilon$. These uncertainties have a negligible impact on the final result and are therefore not taken into account in the final fit.

While energy scale and resolution uncertainties on signal and atmospheric background spectra can be neglected in this analysis, the spallation background spectrum is particularly steep and hence more sensitive to the modeling of energy reconstruction. In this analysis, we will therefore account for the associated modeling uncertainties when estimating spallation spectral shape systematic uncertainty. 

\subsubsection{Spallation background spectral shapes}
\label{sec:spallsys}
As discussed above, we consider two sources of systematic uncertainties here: energy scale and resolution uncertainties, and isotope composition. We estimate the latter by varying the $^8$B, $^8$Li, and $^9$C fractions as shown in Equation~\ref{eq:spalike}. Then, we distort the steepest and least steep spectra found using this method in order to account for the 2\% and 3\% energy scale and resolution uncertainties discussed above. We take these distorted spectra as our $\pm$1$\sigma$ variations from the nominal spallation spectrum. Finally, we parameterize the total spectral distorsion associated with the systematic uncertainty by using a third order polynomial $\mathcal{P}_3$ such that:

\begin{align}
    \text{PDF}_\text{new} &= \text{PDF}_\text{old}\times N\left[ 1 + \epsilon \mathcal{P}_3(E)\right],
    \label{eq:spallsyspoly}
\end{align}

\noindent 
where $\epsilon$ is expressed in units of $\sigma$. More details about these estimates are given in Appendix~\ref{sec:appendixspecfit}. We then convolve the likelihood $\mathcal{L}$ from Equation~\ref{eq:like} with a Gaussian prior on $\epsilon$:

\begin{align}
\label{eq:lgaus}
    \mathcal{L}' = \int_{-2}^{2} \mathcal{L}(\epsilon) e^{-\frac{\epsilon^2}{2}}\,\mathrm{d}\epsilon.
\end{align}


\subsubsection{Atmospheric $\nu_e$ CC spectral shapes} 
The shape of the $\nu_e$ CC background for electron energies below about $100$~MeV has not been studied in-depth to date, although Monte-Carlo simulations predict a linear increase in the number of events with the electron energy. In our analysis, we model this spectral shape uncertainty by allowing the slope of the $\nu_e$ CC spectrum to vary in the two signal regions. We then integrate the distorted PDF using pseudo-Gaussian weights. A detailed description of this procedure is provided in Ref.~\cite{bib:sksrn123}.



At this step of the analysis, we assume the spectral shape distortions in the $=$1 and $\neq$1 tagged neutron regions to be fully correlated and model the effect of neutron multiplicity uncertainties separately.

\subsubsection{Relative normalization for NCQE events} 
The Cherenkov angle distribution of atmospheric NCQE events is associated with large uncertainties stemming from the modeling of $\gamma$ ray emission. We incorporate this uncertainty in the current analysis by allowing the relative normalization of the NCQE PDFs between the different Cherenkov angle regions to vary. Following the procedure outlined in Ref.~\cite{bib:sksrn123}, we define a 1$\sigma$ variation of this normalization to correspond to a $100\%$ change of the number of NCQE events in the intermediate Cherenkov angle region. Additionally, in order for the overall normalization of the NCQE PDFs across regions to remain constant, we reweight these PDFs as follows:

\begin{align}
    \text{PDF'}^\text{NCQE}_\text{med, n}& = \text{PDF}^\text{NCQE}_\text{med, n} \times (1 + \epsilon), \\
    \text{PDF'}^\text{NCQE}_\text{high, n}& = \text{PDF}^\text{NCQE}_\text{high, n} \times (1 - f_\text{high}\epsilon), \\
    \text{PDF'}^\text{NCQE}_\text{low, n}& = \text{PDF}^\text{NCQE}_\text{low, n} \times (1 - f_\text{low}\epsilon), 
    \label{eq:ncqesys}
\end{align}

\noindent 
where the ``low'', ``med'', and ``high'' indices refer to the low (20$-$38$^\circ$), medium (38$-$50$^\circ$), and high (78$-$90$^\circ$) Cherenkov angle ranges from Table~\ref{tab:sigregions} and $n = 0,1$ for the tagged neutron regions respectively. 
$f_{\rm low}, f_{\rm med}, f_{\rm high}$ are scaling factors introduced to preserve the overall normalization of the PDFs. The $\epsilon$ parameterizes the PDF distortion, in units of $\sigma$. We finally inject these distorted PDFs into Equation~\ref{eq:like} and convolve the new likelihood with a weighted Gaussian prior, as with $\nu_e$ CC background uncertainties. 

\subsubsection{Uncertainty on the neutron multiplicity} 
\label{subsubsec:neutmultsys}
The fractions of signal and background events observed in the $=$1 and $\neq$1 tagged neutron regions depend on both the neutron tagging efficiency and the true neutron multiplicity of the observed events. As shown in the $^{241}$Am/Be calibration study, the efficiency of the neutron tagging cut for the DSNB spectral analysis is associated with a systematic uncertainty of around $6$\%. Conversely, the actual neutron multiplicity for atmospheric neutrino interactions is poorly known. As discussed in Section~\ref{subsec:bkgestimate}, neutron multiplicity uncertainties of up to $\sim$40\% have been measured for reconstructed energies above $100$~MeV using the T2K neutrino beam~\cite{bib:rakutsu}. 
In what follows, we therefore consider a $40\%$ systematic uncertainty on the neutron tagging. 

We model the effect of neutron multiplicity uncertainties by varying the relative normalization of the PDFs in the $=$1 and $\neq$1 tagged neutron regions for the $\nu_e$ CC, NCQE, and invisible muon and pion backgrounds. As mentioned at the beginning of this section we neglect the systematic uncertainty associated with $\mu/\pi$ backgrounds, which are associated with negligible contributions in the signal region. As discussed above, we set the 1$\sigma$ variation of the fraction of events with one tagged neutron to a conservative value of $40\%$. We then distort the PDFs in the $=$1 and $\neq$1 tagged neutron regions for each background category as follows:

\begin{align}
    \text{PDF}_{j,\text{new}}^\text{IBD} &= \text{PDF}_{j,\text{old}}^\text{IBD} \times (1 + \alpha_j \epsilon), \\
    \text{PDF}_{j,\text{new}}^\text{non-IBD} &= \text{PDF}_{j,\text{old}}^\text{non-IBD} \times (1 - \alpha_j f \epsilon), 
    \label{eq:nmultsys}
\end{align}

\noindent 
where $\alpha_j$ is the relative $1\sigma$ systematic uncertainty for background category $j$, $\epsilon$ measures the magnitude of the distortion in units of $\sigma$ and $f 
$ preserves the overall normalization of the PDFs. 
Finally, we incorporate this uncertainty into the final likelihood by integrating the initial likelihood similarly to Equation~\ref{eq:lgaus}. 

\subsubsection{Combining the background uncertainties}
The three categories of systematic uncertainties on backgrounds described above ---spallation and $\nu_e$ CC spectral shapes, NCQE Cherenkov angle distribution, and neutron multiplicity--- are incorporated into Equation~\ref{eq:like} using the modified likelihood defined in Equation~\ref{eq:lgaus} and the modified PDFs defined in Ref.~\cite{bib:sksrn123} as well as in Eqs~\ref{eq:ncqesys} and \ref{eq:nmultsys} with distorsion parameters $\epsilon_{\nu_e\text{CC}}$, $\epsilon_\text{NCQE}$, and $\epsilon_n$ respectively. The final likelihood $\mathcal{L}_\text{syst}$ can then be written as:

\begin{align}
\nonumber
    \mathcal{L}_\text{syst}(\{N_j\}) =& \int_{-1}^3 \mathrm{d}\epsilon_{\nu_e\text{CC}} \int_{-1}^3 \mathrm{d}\epsilon_\text{NCQE}\int_{-2}^3 \mathrm{d}\epsilon_{n}\int_{-2}^2 \mathrm{d}\epsilon_{\rm spall} \\
    \label{eq:likesys}
    &\times\mathcal{L}(\{N_j\}, \epsilon_{\nu_e\text{CC}}, \epsilon_\text{NCQE}, \epsilon_n, \epsilon_{\rm spall})\\
    \nonumber
    &\times G(\epsilon_{\nu_e\text{CC}}) \times G(\epsilon_\text{NCQE}) \times\frac{e^{-\frac{\epsilon_n^2}{2}}}{\sqrt{\pi}}\times\frac{e^{-\frac{\epsilon_{\rm spall}^2}{2}}}{\sqrt{\pi}}, 
\end{align}

\noindent 
where $\mathcal{L}(\{N_j\}, \epsilon_{\nu_e\text{CC}}, \epsilon_\text{NCQE}, \epsilon_n, \epsilon_{\rm spall})$ is the likelihood defined in Equation~\ref{eq:like} with distorted PDFs and $G$ is the weighted Gaussian introduced in Ref~\cite{bib:sksrn123}.

\subsubsection{Energy-independent efficiency systematic error}
Maximizing the likelihood $\mathcal{L}_\text{syst}$ defined in Equation~\ref{eq:likesys} over the numbers of events in all five background categories results in a marginalized likelihood $\mathcal{L}_\text{opt}(N_s)$ where $N_s$ is the number of signal events. Using this likelihood to constrain the total number of DSNB events that passed through the detector, $R$, requires rescaling $N_s$ by the average value of the signal efficiency over all analysis regions and their associated energy windows, $\epsilon_\text{sig}$. We therefore need to incorporate the systematic uncertainty on $\epsilon_\text{sig}$ into $\mathcal{L}_\text{opt}$, following a procedure similar to the treatment of the background systematic uncertainty. Since the neutron tagging efficiency only affects the relative normalization between regions, the main sources of uncertainties on $\epsilon_\text{sig}$ will arise from noise reduction and positron candidate selection cuts. These uncertainties are summarized in Table~\ref{tab:cut_systematics}. We also incorporate the uncertainties associated to the fiducial volume cut, inverse beta decay cross section, and livetime calculation, that have been taken from the solar neutrino analyses and are summarized in Table~\ref{tab:fv}. Adding all these uncertainties in quadrature, we then rewrite the likelihood as:

\begin{align}
    \mathcal{L}'(R) = \int_{0}^1 \mathcal{L}_\text{opt}(\epsilon R)e^{-\frac{(\epsilon - \epsilon_\text{sig})^2}{2\sigma_\text{sig}^2}}\,\mathrm{d}\epsilon, 
\end{align}

\noindent 
where  $\sigma_\text{sig}$ is the $1\sigma$ systematic uncertainty on $\epsilon_\text{sig}$. 

\begin{table}[htbp]
    \centering
    \caption{Uncertainties associated with the fiducial volume cut, the Strumia-Vissani IBD cross section calculation, and the livetime calculation for SK-I to IV. The uncertainties for SK-I,II,III have been taken from the SK solar neutrino analyses ~\cite{bib:sksolar1,bib:sksolar2,bib:sksolar3,bib:sksolar4}. For SK-IV the uncertainties have been obtained from the IBD Monte-Carlo simulation.}
    \vspace{3truept}
    \begin{tabular}{ccccc}
    \toprule
         SK phase & I & II & III & IV \\
         \hline
          Fiducial volume & 1.3\% & 1.1\% & 1.0\% & 1.5\%\\
          Cross section & 1.0\% & 1.0\% & 1.0\% & 1.0\%\\
          Livetime & 0.1\% & 0.1\% & 0.1\% & 0.1\%\\
          \botrule
    \end{tabular}
    \label{tab:fv}
\end{table}

Systematic uncertainties associated with reduction efficiencies weaken the final limits by around $1\%$. Similarly, while the uncertainties associated with spallation and atmospheric neutrino backgrounds are of order $100\%$, their overall effect on the final limits is of about $5\%$. The limited impact of spectral shape uncertainties is due to the large dominance of backgrounds associated with invisible muon and pion decays, whose spectrum is well-known. The final analysis will hence be largely dominated by statistical uncertainties, which will allow to easily combine SK-IV observations with results from previous SK phases.

\subsection{Combination with SK-I,II,III}
\label{subsec:combinationItoIII}
The methodology described above, based on an extended maximum likelihood fit over signal and sideband regions, is similar to the one described in Ref.~\cite{bib:sksrn123}, that performed a DSNB spectral analysis over the $22.5\times2853$~kton$\cdot$days of data corresponding to the first three SK phases. While this previous analysis did not include neutron tagging and used slightly different background categories and reduction cuts, it is also dominated by statistical uncertainties. Hence, the SK-I,II,III results can be readily combined with the SK-IV observations.

In order to study a wide variety of the discrete and parameterized models introduced in Section~\ref{sec:intro}, we build signal PDFs for all four phases of SK using Monte-Carlo simulation to model detector resolution effects over most of the energy range. For SK-IV, we renormalize the IBD Monte-Carlo simulation by the predicted DSNB spectrum for all energies. For SK-I,II,III, we renormalize the simulations used in Ref.~\cite{bib:sksrn123} ---with events generated in $E_{e^+}$ in 10$-$90~MeV--- and use the resolution functions from Refs.~\cite{bib:sksolar1,bib:sksolar2,bib:sksolar3} to extrapolate them to lower energies. For the Galais+10~\cite{bib:galais10} and the Nakazato+15~\cite{bib:nakazato15} models, for which DSNB fluxes are available only up to antineutrino energies of 50~MeV, we assume the predicted flux to be zero in the 50$-$81.3~MeV range. This assumption is not expected to affect the spectral fit, as backgrounds from $\nu_e$ CC interactions largely dominate at these energies.
Finally, for each DSNB model, we fit the data samples obtained in Ref.~\cite{bib:sksrn123} for the first three SK phases. Here, although we use the same cuts and search regions as in Ref.~\cite{bib:sksrn123}, we account for possible residual spallation by modeling the spectra and associated systematic uncertainty using the same procedure as at SK-IV. We then combine the four fit results by summing the corresponding likelihoods maximized over the nuisance parameters. 

\subsection{Results}
\label{subsec:specresults}
\begin{figure*}[htbp]
    \centering
    \includegraphics[width=16.cm]{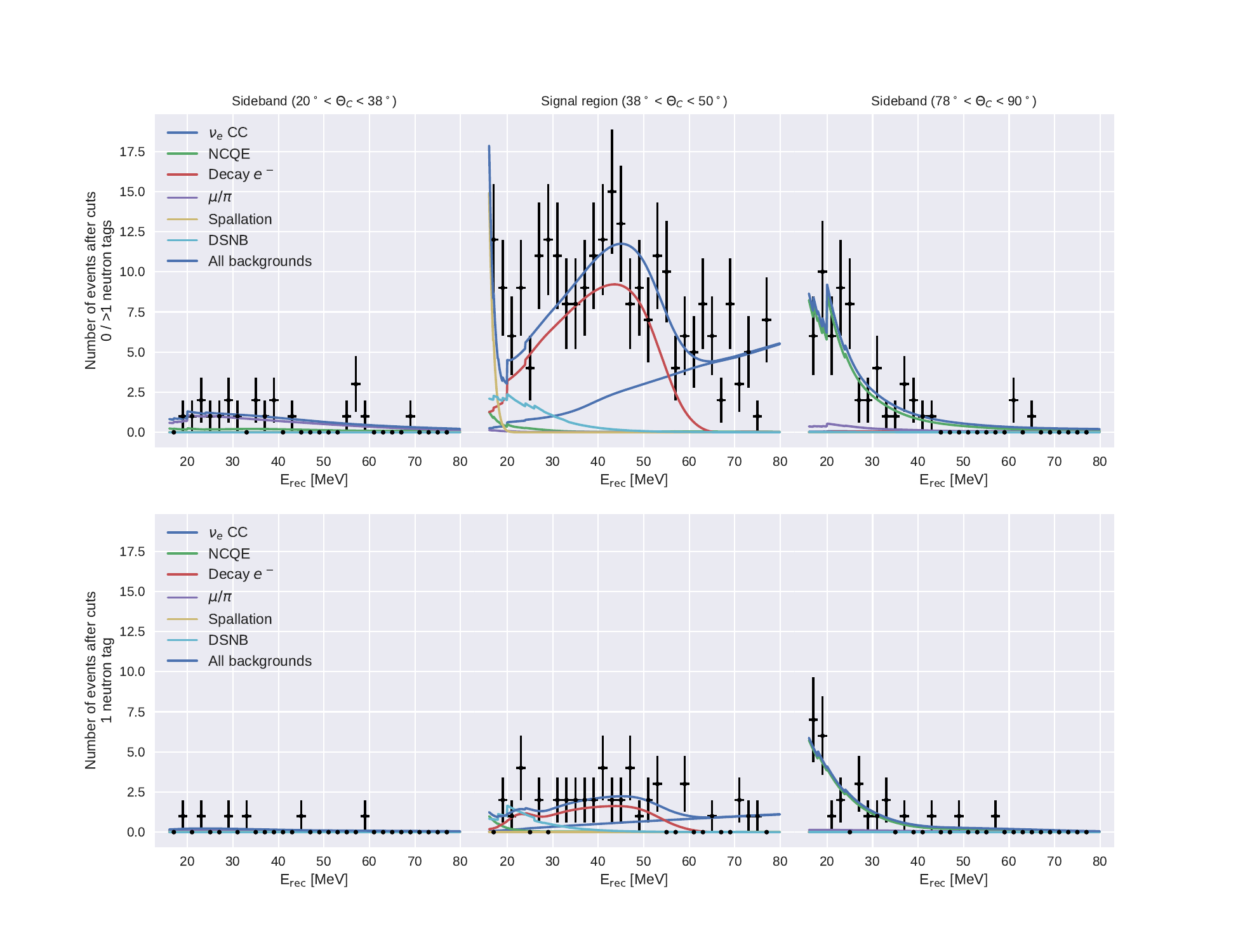}
    \vspace{-15truept}
    \caption{Best-fit signal and background spectra for SK-IV, assuming the DSNB spectrum predicted by the Horiuchi+09 model~\cite{bib:horiuchi09}. The six regions presented here include two signal regions and four sidebands in Table~\ref{tab:sigregions}.}
    \label{fig:horiuchifit4}
\end{figure*}

\begin{figure*}[htbp]
    \centering
    \includegraphics[width=11.cm]{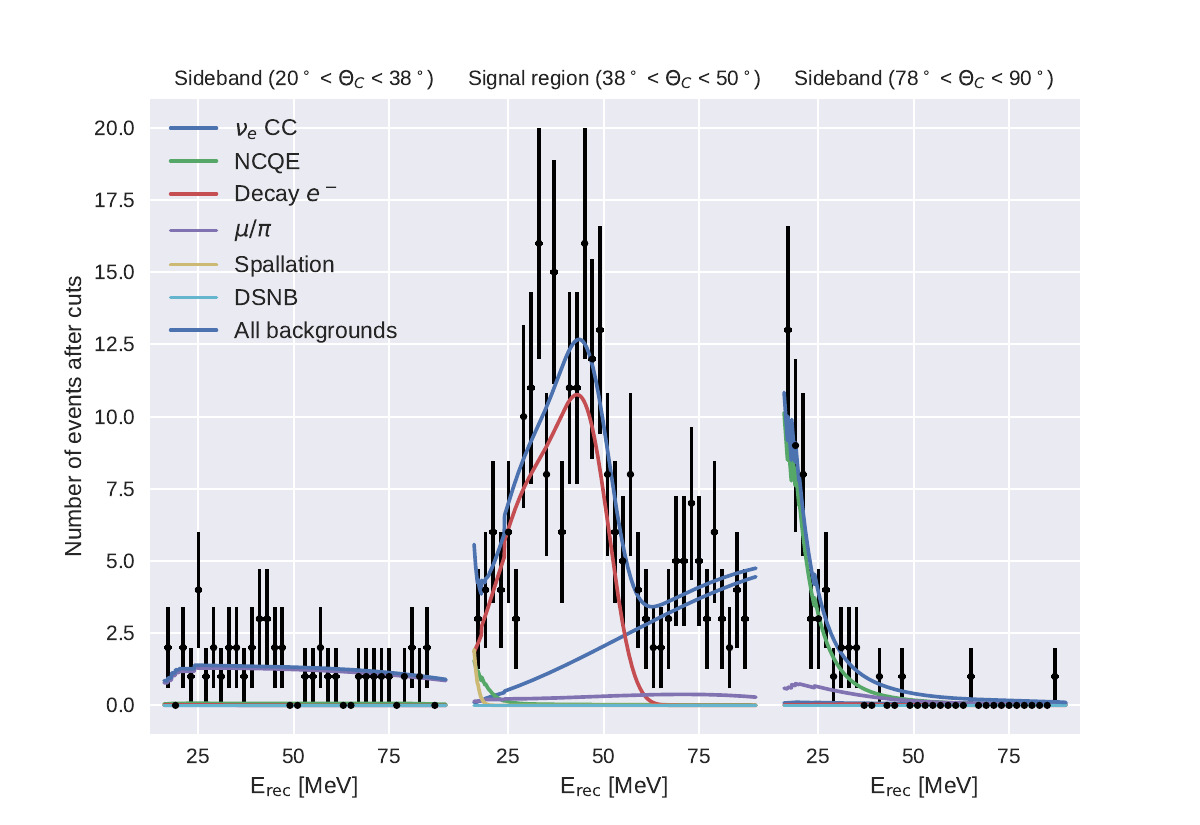}
    \includegraphics[width=11.cm]{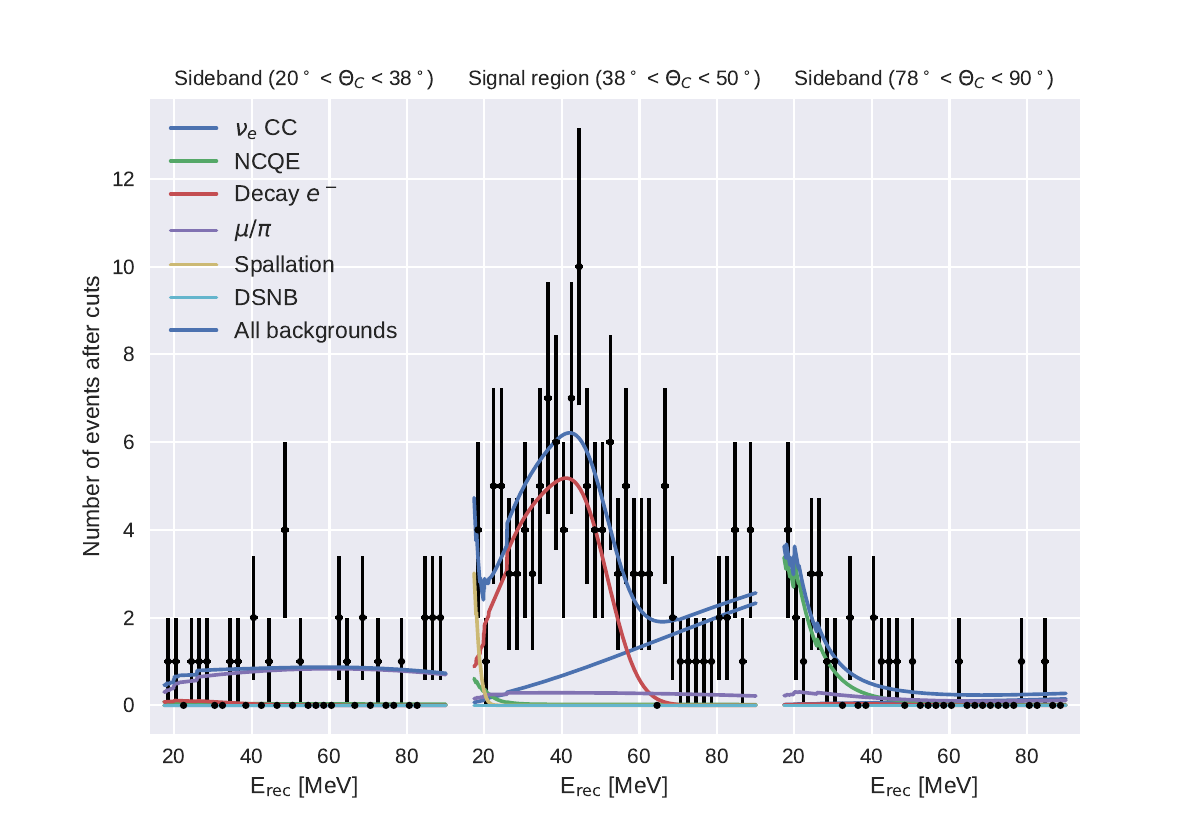}
    \includegraphics[width=11.cm]{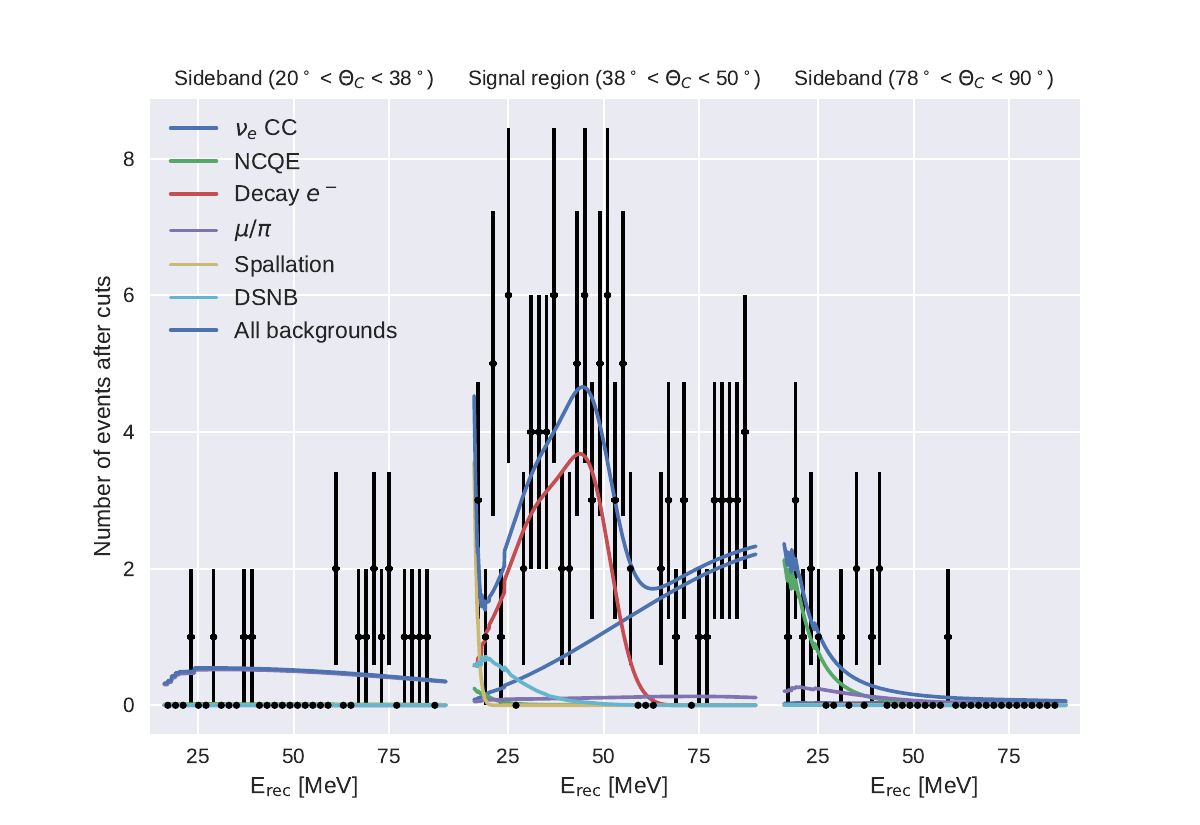}
    \caption{Best-fit signal and background spectra for SK-I (top), II (middle), and III (bottom), assuming the DSNB spectrum predicted by the Horiuchi+09 model~\cite{bib:horiuchi09}. The three regions presented here correspond to different range of the Cherenkov angle, as shown in Table~\ref{tab:sigregions}. Since neutron tagging is not possible in SK-I,II,III, only these three Cherenkov angle regions are considered.}
    \label{fig:horiuchifit123}
\end{figure*}

We apply the procedure described in the previous sections to the observed data collected during phases I to IV of SK for the Horiuchi+09 DSNB model~\cite{bib:horiuchi09}. The associated spectral fits for SK-IV and for SK-I,II,III are shown in Figs.~\ref{fig:horiuchifit4} and \ref{fig:horiuchifit123} for the DSNB signal and for the spallation and atmospheric neutrino background categories introduced in Section~\ref{subsec:shapefit}. Since neutron tagging was not possible in the first three phases of SK, the corresponding fits span only three regions, defined by the low, medium, and high Cherenkov angle ranges defined in Table~\ref{tab:sigregions}. For SK-IV, the low energy bump in the predicted decay electron spectrum for IBD-like events is due to the energy dependence of the neutron tagging cut, that strongly depends on the background composition. As expected, the dominant backgrounds in the low and high Cherenkov angle sidebands are backgrounds with visible muons and pions and NCQE backgrounds, respectively. While most of the signal region is dominated by backgrounds from invisible muons and pions, $\nu_e$ CC contributions become sizable above about 50~MeV. For SK-IV, the background rates are extremely low in the IBD-like region, that will hence be particularly sensitive to IBD signals in spite of the low efficiency of the neutron tagging cut.

In spite of introducing spallation backgrounds in the spectral fits, the results shown in Fig.~\ref{fig:horiuchifit123} for SK-I,II,III are highly similar to the ones shown in Ref.~\cite{bib:sksrn123}. The number of spallation events for each phase never exceed 2, within the upper limit obtained in Ref.~\cite{bib:sksrn123}. Since the current analysis uses a similar procedure with more inclusive muon selection criteria and 20$-$25\% lower signal efficiencies, the spallation cut effectiveness at SK-IV should remain on par with the ones observed at previous SK phases. Yet, an important amount of spallation backgrounds ($\sim$8 events) is predicted below 19.5~MeV at SK-IV. The origin of this important spallation background will be further studied in future SK analyses with an enhanced performance by gadolinium.
Note that this effect was not observed in other SK searches probing a similar energy range. These searches however used specific cuts on e.g. the event timings or opening angles that we do not apply in this analysis.

Even accounting for possible residual spallation, the number of events observed in 17.5$-$19.5~MeV is higher than the best-fit predictions from Fig.~\ref{fig:horiuchifit4}. While this can be explained by statistical fluctuation, it led us to investigating possible non-spallation sources for the excess observed over atmospheric neutrino background predictions in 15.5$-$19.5~MeV. We first ruled out the possibility of a strong DSNB signal, as it would be associated with a steeply falling spectrum and hence to an unrealistically high supernova rate. The hypothesis of an isolated astrophysical event is also strongly disfavored since the observed events are uniformly distributed over the SK-IV period, as discussed in Section~\ref{sec:eventquality}. A radioactive origin for the observed events is also highly unlikely, given their high energies and their uniform spatial distribution.    
The event positions are notably not correlated with the locations of calibration sources. Residual spallation therefore remains the most plausible explanation for the high number of events observed in 15.5$-$19.5~MeV at SK-IV. 

\begin{figure*}[htbp]
    \centering
    \includegraphics[width=12.0cm]{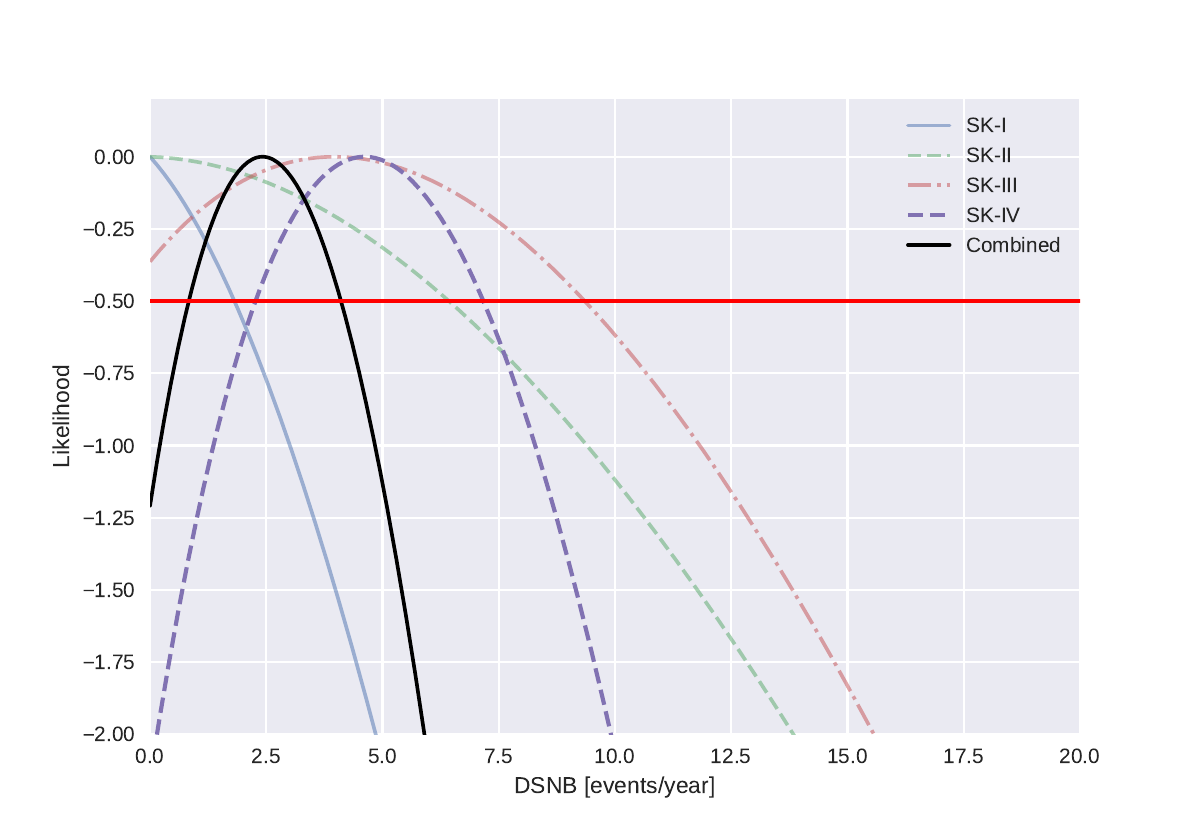}
    \vspace{-5truept}
    \caption{Likelihoods associated with phases I to IV of SK, as well as the combined likelihood. As mentioned in the main text, we can neglect correlations between background systematic uncertainties when combining all phases of SK.}
    \label{fig:likecombination}
\end{figure*}

The likelihoods associated to the SK-I to IV fits for the Horiuchi+09 model are shown in Fig.~\ref{fig:likecombination} as a function of the DSNB rate for the reconstructed equivalent electron kinetic energies $E_{\rm rec} > 15.5$~MeV. This figure also shows the likelihood for the combined analysis, with a total exposure of $22.5\times5823$~kton$\cdot$days. The associated limits on the DSNB flux are shown in Table~\ref{tab:fluxlim}. The expected sensitivities of the SK-I,II,III analysis change by only $\sim$3\% with the inclusion of the spallation backgrounds while the associated observed upper limits on the DSNB flux become tighter. The expected sensitivities of the SK-IV and the combined analyses to the DSNB flux at $90\%$ C.L. are $2.2~\bar{\nu}_e$~cm$^{-2}$$\cdot$$\text{sec}^{-1}$ and $1.5~\bar{\nu}_e$~cm$^{-2}$$\cdot$$\text{sec}^{-1}$, respectively. Hence, the predicted flux for this model, $1.9~\bar{\nu}_e$~cm$^{-2}$$\cdot$$\text{sec}^{-1}$, is well within the reach of the SK-I,II,III,IV analysis. The observed upper limits on the DSNB flux for SK-IV and for the combined analysis are of $4.6~\bar{\nu}_e$~cm$^{-2}$$\cdot$$\text{sec}^{-1}$ and $2.6~\bar{\nu}_e$~cm$^{-2}$$\cdot$$\text{sec}^{-1}$, respectively. The best-fit flux for the combined analysis is now of $1.30^{+0.90}_{-0.85}~\bar{\nu}_e$~cm$^{-2}$$\cdot$$\text{sec}^{-1}$. The 1.5$\sigma$ excess observed over the background prediction is higher than the 0.9$\sigma$ excess from the previous combined analysis~\cite{bib:sksrn123}. 
While not statistically significant, this result is compatible with a wide range of DSNB predictions with a flux comparable to the one of the Horiuchi+09 model~\cite{bib:horiuchi09}.

\section{Discussion}
\label{sec:discussion}
\subsection{Constraints on the DSNB parameter space}
Over the two search strategies presented in this paper, the analysis described in Section~\ref{sec:mianalysis} gives a model-independent differential limit on the $\bar{\nu}_e$ flux while the model-dependent spectral analysis uses data from all phases of SK and yields the tightest constraints on DSNB models. In this section, we present the results of this analysis using both discrete DSNB models ---most of them based on supernova simulations--- and simplified parameterizations. 

\begin{figure*}
    \centering
    \includegraphics[width=8cm]{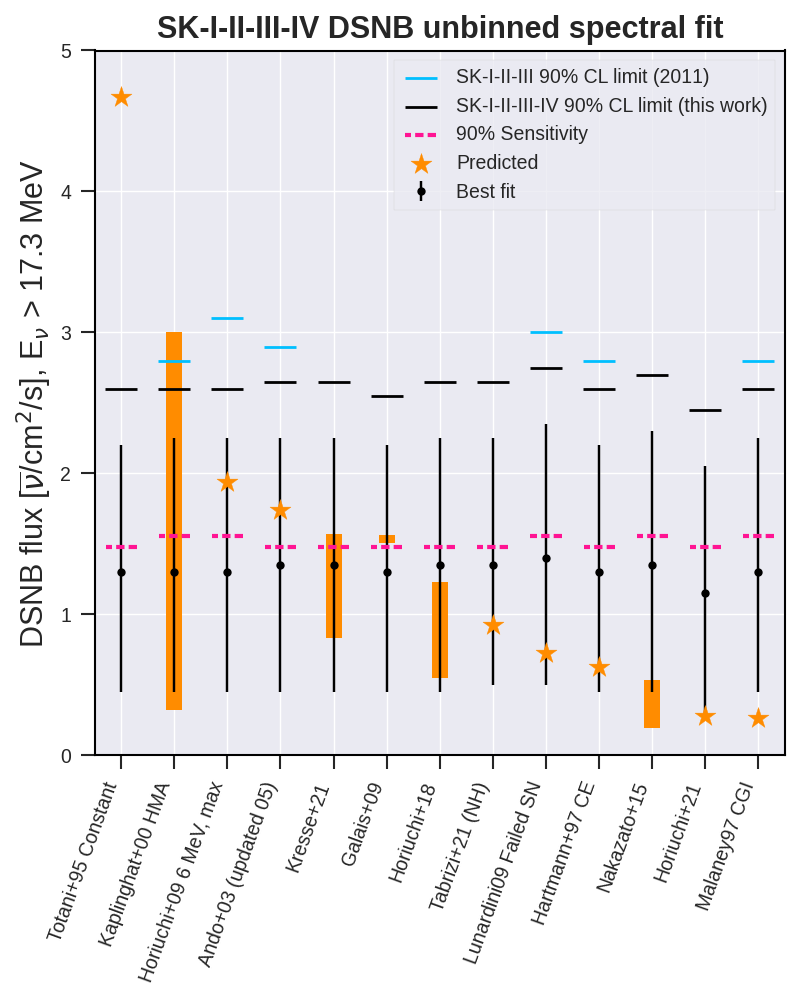}
    \includegraphics[width=8cm]{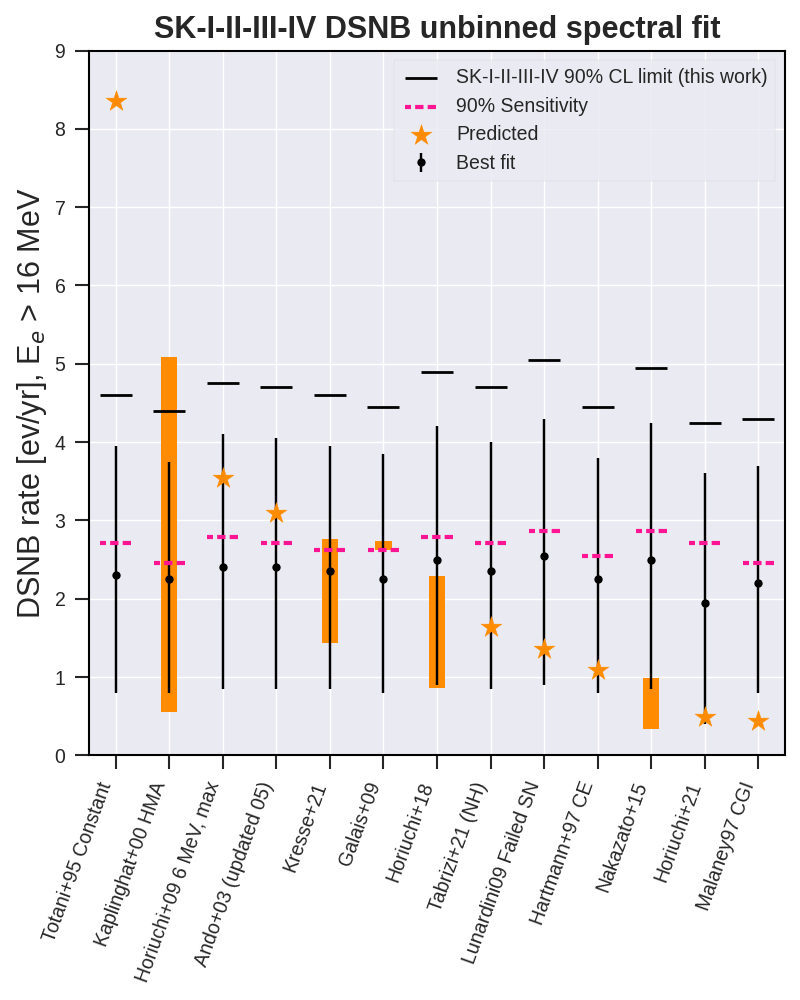}
    \caption{The $90\%$ C.L. upper limits, best-fit values, and expected sensitivities for the DSNB fluxes (left) and rates (right) associated with the models described in Section~\ref{sec:intro}. The best fit rates and fluxes are shown with their associated $1\sigma$ error bars. This figure also shows the predictions for each models, either as a range or as one value. Note the stability of the expected and observed flux limits across all models.}
    \label{fig:speclimitmodels}
\end{figure*}

The $90\%$ C.L. expected and observed upper limits on the DSNB rate and flux for theoretical models are shown in Fig.~\ref{fig:speclimitmodels} for the combined SK-I,II,III,IV analysis, with the corresponding results being tabulated in Table~\ref{tab:fluxlim}. With an exposure of $22.5\times5823$~kton$\cdot$days, the sensitivity of the combined analysis at $90\%$ C.L. is the tightest to date and is on par with predictions from the Ando$+$03, Horiuchi$+$09 ($T_\nu = 6$~MeV), Galais$+$10, and Kresse$+$21 models~\cite{bib:ando03,bib:horiuchi09,bib:galais10,bib:kresse21}. One common feature of these optimistic models is that they use the highest cosmic star formation history allowed by observations. Additionally, the Kresse+21 predictions~\cite{bib:kresse21} are based on neutrino emission models derived from state-of-the-art supernova simulations, and take a wide range of progenitors into account. The ability of this analysis to probe such realistic models makes the prospects for future experiments particularly promising. Currently, an excess of about $1.5\sigma$ has been observed in the data and the resulting $90\%$ C.L. limits on the DSNB flux improve on the constraints obtained in Ref.~\cite{bib:sksrn123} for SK-I,II,III. These results confirm the exclusion of the  Totani$+$95 model~\cite{bib:totani95} and of the most optimistic predictions of the Kaplinghat model~\cite{bib:kaplinghat00}.

\begin{figure}[htbp]
    \centering
    \includegraphics[width=8.5cm]{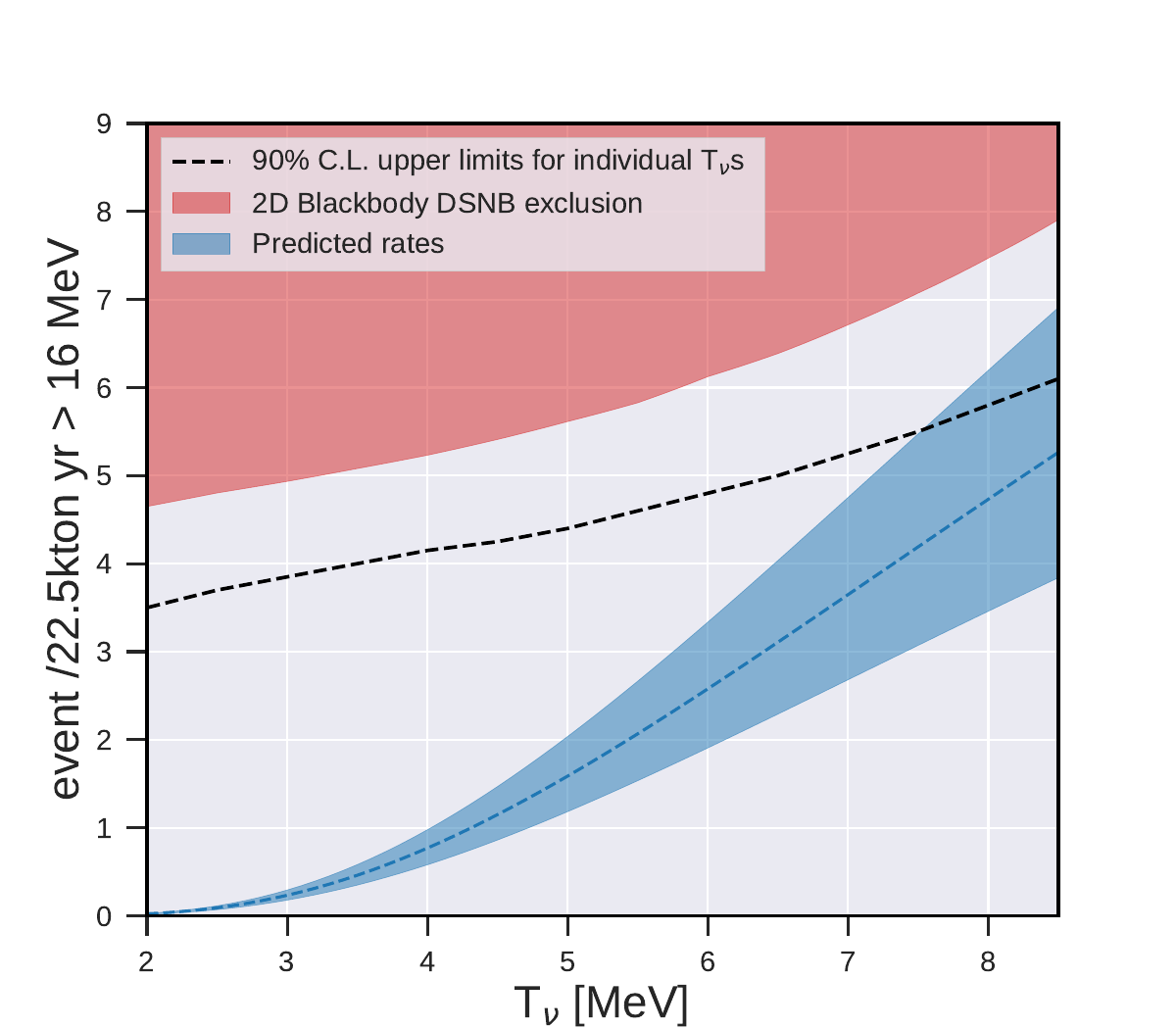}
    \caption{The $90\%$ C.L. excluded region for the DSNB rate for $E_\nu > 17.3$~MeV and the effective neutrino temperature, for the blackbody models described in Section~\ref{sec:signalmodeling} (red surface). To guide the eye, we also show the one-dimensional exclusion limit for individual neutrino temperatures. The blue band shows the predictions for blackbody models, for the low, fiducial, and high star formation rates used in Ref.~\cite{bib:horiuchi09}.}
    \label{fig:blackbody}
\end{figure}

\begin{figure}[htbp]
    \centering
    \includegraphics[width=8.5cm]{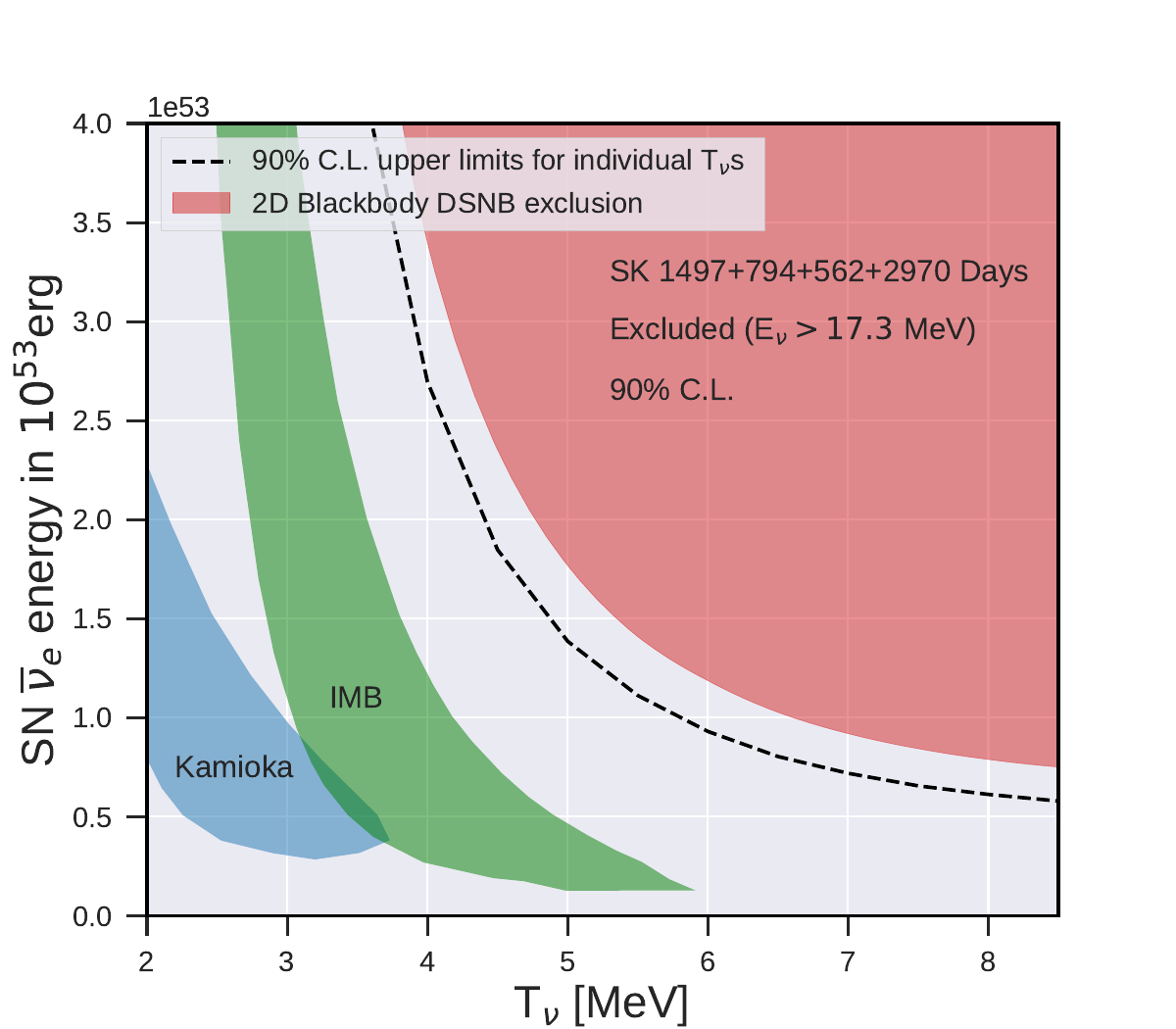}
    \caption{The $90\%$ C.L. excluded region for the total $\bar{\nu}_e$ energy emitted during a supernova and the effective neutrino temperature, for the blackbody models described in Section~\ref{sec:signalmodeling} (red surface). To guide the eye, we also show the one-dimensional exclusion limit for individual neutrino temperatures. We also show the $90\%$ C.L. contours corresponding to the range of parameters allowed for SN1987A.}
    \label{fig:blackbodysn}
\end{figure}

Another striking feature observed in Fig.~\ref{fig:speclimitmodels} is the stability of the observed and expected limits on the DSNB flux over all models. This stability results from the limited sensitivity of the current analysis to the DSNB spectral shape. In order to translate flux limits into constraints on astrophysical parameters, we therefore ignore subtle effects associated with e.g. black-hole formation or multiple classes of progenitors, and use the simple Fermi-Dirac neutrino emission model described in Section~\ref{sec:signalmodeling}, as in Ref.~\cite{bib:sksrn123}. The associated two-dimensional $90\%$ C.L. limit on the neutrino emission temperature and on the DSNB rate for $E_\nu > 17.3$~MeV is shown in Fig.~\ref{fig:blackbody}, for the combined SK-I,II,III,IV analysis. This figure also shows the expected rates for different effective neutrino emission temperatures. Here, the band accounts for the current uncertainty on the cosmic star formation rate, using the low, fiducial, and high estimates from Ref.~\cite{bib:horiuchi09}. Since we used two-dimensional exclusion contours instead of computing limits for individual models, the limits appear weaker than the ones shown in Fig.~\ref{fig:speclimitmodels}. These limits are on par with the SK-I,II,III analysis~\cite{bib:sksrn123}. In addition to the rate limits, whose interpretation is complicated by the high SK analysis threshold, we also show $90\%$ C.L. exclusion contours for the total $\bar{\nu}_e$ energy and the effective neutrino temperature in Fig.~\ref{fig:blackbodysn}. This figure also shows the regions of parameter space allowed by the Kamiokande II and IMB experiments for the supernova SN1987A~\cite{bib:jegerlehner}. While the neutrino temperature for this supernova is particularly low, future experiments should be able to probe at least the associated IMB region.  

\begin{table*}[htbp]
\footnotesize
\centering
\caption{Best-fit values and the $90\%$ C.L. upper limits on the DSNB fluxes (in cm$^{-2}\cdot$sec$^{-1}$) for the theoretical models for phases SK-I to IV as well as for the combined analysis. Here the upper limits are given for $E_\nu > 17.3$~MeV. For the Kresse+21 models, the ``High'', ``Fid'', and ``Low'' predictions correspond to the ``W20-BH2.7-$\alpha$2.0'', ``W18-BH2.7-$\alpha$2.0'', and ``W18-BH2.7-$\alpha$2.0-He33'' models from Ref.~\cite{bib:kresse21} respectively.}
    \vspace{3truept}
\begin{tabular}{lrr|rrrrr|r}
\toprule
& \multicolumn{2}{c}{\textbf{Best fit}}& \multicolumn{5}{c}{\textbf{90\% CL limit}} & \textbf{Pred.}\\
\textbf{Model}  & \textbf{SK4} & \textbf{All} &  \textbf{SK1} & \textbf{SK2} & \textbf{SK3} & \textbf{SK4} & \textbf{All} \\
\hline\scriptsize{Totani+95 Constant} & 2.5$^{+1.4}_{-1.3}$ & 1.3$^{+0.9}_{-0.9}$ & 2.3 & 6.3 & 7.0 & 4.5 & 2.6 & 4.67 \\
\scriptsize{Kaplinghat+00 HMA (max)} & 2.6$^{+1.5}_{-1.3}$ & 1.3$^{+0.9}_{-0.9}$ & 2.3 & 6.7 & 7.1 & 4.7 & 2.6 & 3.00 \\
\scriptsize{Horiuchi+09 \tiny{6 MeV, max}} & 2.6$^{+1.4}_{-1.3}$ & 1.3$^{+0.9}_{-0.9}$ & 2.4 & 6.0 & 7.0 & 4.6 & 2.6 & 1.94 \\
\scriptsize{Ando+03 (updated 05)} & 2.7$^{+1.5}_{-1.4}$ & 1.4$^{+0.9}_{-0.9}$ & 2.3 & 6.6 & 7.2 & 4.7 & 2.7 & 1.74 \\
\scriptsize{Kresse+21 (High, NO)} & 2.7$^{+1.5}_{-1.3}$ & 1.4$^{+0.9}_{-0.9}$ & 2.3 & 6.7 & 7.2 & 4.7 & 2.7 & 1.57 \\
\scriptsize{Galais+09 (NO)} & 2.5$^{+1.4}_{-1.3}$ & 1.3$^{+0.9}_{-0.9}$ & 2.3 & 6.3 & 7.0 & 4.5 & 2.6 & 1.56 \\
\scriptsize{Galais+09 (IO)} & 2.6$^{+1.4}_{-1.3}$ & 1.3$^{+0.9}_{-0.9}$ & 2.3 & 6.4 & 7.0 & 4.5 & 2.6 & 1.50 \\
\scriptsize{Horiuchi+18 \tiny{$\xi_{2.5}=0.1$}} & 2.6$^{+1.4}_{-1.3}$ & 1.4$^{+0.9}_{-0.9}$ & 2.4 & 6.1 & 7.1 & 4.6 & 2.7 & 1.23 \\
\scriptsize{Kresse+21 (High, IO)} & 2.7$^{+1.5}_{-1.3}$ & 1.4$^{+0.9}_{-0.9}$ & 2.3 & 6.7 & 7.1 & 4.7 & 2.7 & 1.21 \\
\scriptsize{Kresse+21 (Fid, NO)} & 2.7$^{+1.5}_{-1.3}$ & 1.4$^{+0.9}_{-0.9}$ & 2.3 & 6.8 & 7.2 & 4.7 & 2.7 & 1.20 \\
\scriptsize{Kresse+21 (Fid, IO)} & 2.7$^{+1.5}_{-1.3}$ & 1.4$^{+0.9}_{-0.9}$ & 2.3 & 6.8 & 7.2 & 4.7 & 2.7 & 1.02 \\
\scriptsize{Kresse+21 (Low, NO)} & 2.7$^{+1.5}_{-1.4}$ & 1.4$^{+0.9}_{-0.9}$ & 2.3 & 6.8 & 7.2 & 4.8 & 2.7 & 0.96 \\
\scriptsize{Tabrizi+21 (NO)} & 2.7$^{+1.5}_{-1.3}$ & 1.4$^{+0.9}_{-0.9}$ & 2.4 & 6.6 & 7.1 & 4.7 & 2.7 & 0.92 \\
\scriptsize{Kresse+21 (Low, IO)} & 2.7$^{+1.5}_{-1.4}$ & 1.4$^{+0.9}_{-0.9}$ & 2.3 & 6.8 & 7.2 & 4.8 & 2.7 & 0.84 \\
\scriptsize{Lunardini09 Failed SN} & 2.8$^{+1.5}_{-1.4}$ & 1.4$^{+0.9}_{-0.9}$ & 2.4 & 6.8 & 7.3 & 4.8 & 2.8 & 0.73 \\
\scriptsize{Hartmann+97 CE} & 2.6$^{+1.4}_{-1.3}$ & 1.3$^{+0.9}_{-0.9}$ & 2.3 & 6.5 & 7.1 & 4.6 & 2.6 & 0.63 \\
\scriptsize{Nakazato+15 (max, IO)} & 2.7$^{+1.5}_{-1.4}$ & 1.4$^{+1.0}_{-0.9}$ & 2.4 & 6.5 & 7.2 & 4.8 & 2.7 & 0.53 \\
\scriptsize{Horiuchi+18 \tiny{$\xi_{2.5}=0.5$}} & 2.7$^{+1.5}_{-1.4}$ & 1.3$^{+0.9}_{-0.9}$ & 2.2 & 7.1 & 7.1 & 4.8 & 2.6 & 0.55 \\
\scriptsize{Horiuchi+21} & 2.1$^{+1.3}_{-1.2}$ & 1.2$^{+0.9}_{-0.9}$ & 3.4 & 4.3 & 5.9 & 3.9 & 2.5 & 0.28 \\
\scriptsize{Malaney97 CGI} & 2.7$^{+1.5}_{-1.3}$ & 1.3$^{+0.9}_{-0.9}$ & 2.3 & 6.8 & 7.1 & 4.7 & 2.6 & 0.26 \\
\scriptsize{Nakazato+15 (min, NO)} & 2.8$^{+1.5}_{-1.4}$ & 1.4$^{+1.0}_{-0.9}$ & 2.3 & 6.8 & 7.2 & 4.8 & 2.7 & 0.19 \\
\botrule
\end{tabular}
\label{tab:fluxlim}
\end{table*}

\begin{table*}[htbp]
\footnotesize
\centering
\caption{Best-fit values and the $90\%$ C.L. upper limits on the DSNB rates (in events$\cdot$year$^{-1}$) for the theoretical models for phases SK-I to IV as well as for the combined analysis. Here the upper limits are given for $E_\nu> 17.3$~MeV. For the Kresse+21 models, the ``High'', ``Fid'', and ``Low'' predictions correspond to the ``W20-BH2.7-$\alpha$2.0'', ``W18-BH2.7-$\alpha$2.0'', and ``W18-BH2.7-$\alpha$2.0-He33'' models from Ref.~\cite{bib:kresse21} respectively.}
    \vspace{3truept}
\begin{tabular}{lrr|rrrrr|r}
\toprule
& \multicolumn{2}{c}{\textbf{Best fit}}& \multicolumn{5}{c}{\textbf{90\% CL limit}} & \textbf{Pred.}\\
\textbf{Model}  & \textbf{SK4} & \textbf{All} &  \textbf{SK1} & \textbf{SK2} & \textbf{SK3} & \textbf{SK4} & \textbf{All} \\
\hline\scriptsize{Totani+95 Constant} & 4.5$^{+2.5}_{-2.2}$ & 2.3$^{+1.6}_{-1.5}$ & 4.2 & 11.2 & 12.2 & 7.9 & 4.6 & 8.35 \\
\scriptsize{Kaplinghat+00 HMA (max)} & 4.4$^{+2.4}_{-2.2}$ & 2.2$^{+1.5}_{-1.4}$ & 3.9 & 11.2 & 11.7 & 7.7 & 4.4 & 5.09 \\
\scriptsize{Horiuchi+09 \tiny{6 MeV, max}} & 4.6$^{+2.5}_{-2.3}$ & 2.4$^{+1.7}_{-1.6}$ & 4.4 & 11.7 & 12.5 & 8.2 & 4.8 & 3.54 \\
\scriptsize{Ando+03 (updated 05)} & 4.7$^{+2.5}_{-2.4}$ & 2.4$^{+1.6}_{-1.6}$ & 4.2 & 11.8 & 12.4 & 8.2 & 4.7 & 3.09 \\
\scriptsize{Kresse+21 (High, NO)} & 4.5$^{+2.5}_{-2.3}$ & 2.4$^{+1.6}_{-1.5}$ & 4.1 & 11.7 & 12.2 & 8.1 & 4.6 & 2.76 \\
\scriptsize{Galais+09 (NO)} & 4.4$^{+2.4}_{-2.2}$ & 2.2$^{+1.6}_{-1.4}$ & 4.0 & 11.0 & 11.9 & 7.8 & 4.5 & 2.74 \\
\scriptsize{Galais+09 (IO)} & 4.4$^{+2.4}_{-2.2}$ & 2.2$^{+1.6}_{-1.4}$ & 4.0 & 11.0 & 11.9 & 7.7 & 4.5 & 2.62 \\
\scriptsize{Horiuchi+18 \tiny{$\xi_{2.5}=0.1$}} & 4.8$^{+2.6}_{-2.4}$ & 2.5$^{+1.7}_{-1.6}$ & 4.5 & 12.1 & 12.8 & 8.4 & 4.9 & 2.29 \\
\scriptsize{Kresse+21 (High, IO)} & 4.6$^{+2.5}_{-2.3}$ & 2.4$^{+1.6}_{-1.5}$ & 4.1 & 11.7 & 12.2 & 8.1 & 4.6 & 2.14 \\
\scriptsize{Kresse+21 (Fid, NO)} & 4.5$^{+2.5}_{-2.2}$ & 2.3$^{+1.5}_{-1.5}$ & 4.0 & 11.7 & 11.9 & 7.9 & 4.5 & 2.06 \\
\scriptsize{Kresse+21 (Fid, IO)} & 4.5$^{+2.5}_{-2.2}$ & 2.3$^{+1.5}_{-1.5}$ & 4.0 & 11.6 & 11.9 & 7.9 & 4.5 & 1.75 \\
\scriptsize{Kresse+21 (Low, NO)} & 4.5$^{+2.5}_{-2.2}$ & 2.2$^{+1.6}_{-1.4}$ & 3.9 & 11.6 & 11.8 & 7.9 & 4.5 & 1.65 \\
\scriptsize{Tabrizi+21 (NO)} & 4.6$^{+2.5}_{-2.3}$ & 2.4$^{+1.6}_{-1.5}$ & 4.2 & 11.8 & 12.3 & 8.2 & 4.7 & 1.64 \\
\scriptsize{Kresse+21 (Low, IO)} & 4.5$^{+2.5}_{-2.2}$ & 2.2$^{+1.6}_{-1.4}$ & 4.0 & 11.6 & 11.9 & 7.9 & 4.5 & 1.43 \\
\scriptsize{Lunardini09 Failed SN} & 5.0$^{+2.7}_{-2.5}$ & 2.6$^{+1.7}_{-1.7}$ & 4.5 & 12.7 & 13.1 & 8.8 & 5.1 & 1.36 \\
\scriptsize{Hartmann+97 CE} & 4.4$^{+2.4}_{-2.2}$ & 2.2$^{+1.6}_{-1.4}$ & 4.0 & 11.2 & 11.9 & 7.7 & 4.5 & 1.09 \\
\scriptsize{Nakazato+15 (max, IO)} & 5.0$^{+2.7}_{-2.5}$ & 2.5$^{+1.8}_{-1.6}$ & 4.5 & 12.1 & 13.0 & 8.7 & 5.0 & 0.99 \\
\scriptsize{Horiuchi+18 \tiny{$\xi_{2.5}=0.5$}} & 4.2$^{+2.3}_{-2.1}$ & 2.1$^{+1.5}_{-1.4}$ & 3.5 & 11.4 & 10.9 & 7.4 & 4.1 & 0.87 \\
\scriptsize{Horiuchi+21} & 3.7$^{+2.2}_{-2.1}$ & 2.0$^{+1.6}_{-1.6}$ & 6.0 & 7.5 & 10.2 & 6.8 & 4.2 & 0.49 \\
\scriptsize{Malaney97 CGI} & 4.3$^{+2.4}_{-2.1}$ & 2.2$^{+1.5}_{-1.4}$ & 3.8 & 11.2 & 11.3 & 7.6 & 4.3 & 0.44 \\
\scriptsize{Nakazato+15 (min, NO)} & 4.7$^{+2.5}_{-2.4}$ & 2.4$^{+1.6}_{-1.6}$ & 4.0 & 11.8 & 12.1 & 8.2 & 4.6 & 0.34 \\
\botrule
\end{tabular}
\label{tab:ratelim}
\end{table*}

\subsection{Prospects at future experiments}
Recent doping of SK with gadolinium (SK-Gd) as well as the advent of Hyper-Kamiokande (HK) in the more distant future open new promising perspectives for the DSNB search. Using gadolinium will notably allow to considerably improve neutron tagging and hence eliminate most spallation backgrounds and lower the threshold of the analysis. As shown in Fig.~\ref{fig:finalspectrumwide}, however, multiple backgrounds involving neutrons can still dominate over the DSNB up to about $15.5$~MeV. Reactor antineutrinos, in particular, provide an irreducible background up to $9.5$~MeV. Up to about $11.5$~MeV, $^9$Li decays might then dominate, and spallation reduction will remain a key component of future analyses. Locating muon-induced showers using neutrons will become especially powerful at SK-Gd. After eliminating spallation, atmospheric NCQE interactions will dominate over a wide energy range and will call for new dedicated reduction techniques. At SK-Gd, notably, the high visibility of the neutron capture signal will allow to locate neutron capture without using the positron vertex, which could allow to use the neutron travel distance as a discriminating observable. Finally, while HK may not be doped with Gd, this considerably larger detector coupled with the accelerator neutrino beams will allow to improve the characterization of the NCQE rate, spectral shape, and neutron multiplicity, that are currently associated with large systematic uncertainties~\cite{bib:t2kncqe1to9}. 
In the future, further studies using the upcoming Intermediate Water Cherenkov Detector (IWCD) will allow to considerably reduce neutron multiplicity uncertainties by studying monochromatic beams, and will hence be a key piece of the DSNB search program at HK.
The DSNB study will also be extended by searches at lower energies, achieved at future large-scale liquid scintillation detectors, such as JUNO~\cite{bib:junosrn}, giving deeper insights into spectrum shape over the wide energy range in combination with HK~\cite{bib:moller}.

\section{Conclusion}
\label{sec:conclusion}

In this paper, we presented two analyses taking advantage of the longest data taking phase of SK, the improved data reduction, and better background estimates. We developed improved spallation reduction and neutron tagging algorithms with respect to the previous analyses of SK-IV data, and constrained the DSNB flux in the 7.5$-$29.5~MeV and 15.5$-$79.5~MeV energy ranges. With the model-independent analysis, we achieved the world's tightest upper limit on the extraterrestrial electron antineutrino flux. For the latter region, we notably extended the spectral analysis developed for SK-I,II,III~\cite{bib:sksrn123}, and combined the results of the first four SK phases to reach an exposure of $22.5\times5823$~kton$\cdot$days. This exposure yields SK a world-leading sensitivity to the DSNB flux at $90\%$ C.L., comparable to the fluxes of several realistic models. A 1.5$\sigma$ excess has been observed across all the models probed and the final $90\%$ C.L. limits on the DSNB flux are on par with the ones obtained in previous SK analyses~\cite{bib:hzhang,bib:sksrn123}. The last 20 years of data taking at SK have hence brought us to the doorstep of the DSNB parameter space. The enhanced neutron tagging capabilities at SK-Gd and the larger detector volume achieved at HK will allow to explore this parameter space, setting meaningful constraints on astrophysical observables, with the realistic prospect of a groundbreaking discovery.

\section*{Acknowledgment}
We gratefully acknowledge the cooperation of the Kamioka Mining and Smelting Company. The Super-Kamiokande experiment has been built and operated from funding by the Japanese Ministry of Education, Culture, Sports, Science and Technology, the U.S. Department of Energy, and the U.S. National Science Foundation. Some of us have been supported by funds from the National Research Foundation of Korea NRF-2009-0083526 (KNRC) funded by the Ministry of Science, ICT, and Future Planning and the Ministry of Education (2018R1D1A3B07050696, 2018R1D1A1B07049158), the Japan Society for the Promotion of Science, the National Natural Science Foundation of China under Grants No. 11620101004, the Spanish Ministry of Science, Universities and Innovation (grant PGC2018-099388-B-I00), the Natural Sciences and Engineering Research Council (NSERC) of Canada, the Scinet and Westgrid consortia of Compute Canada, the National Science Centre, Poland (2015/18/E/ST2/00758), the Science and Technology Facilities Council (STFC) and GridPPP, UK, the European Union’s Horizon 2020 Research and Innovation Programmeunder the Marie Sklodowska-Curie grant  agreement  no. 754496, H2020-MSCA-RISE-2018 JENNIFER2 grant agreement no.822070, and  H2020-MSCA-RISE-2019 SK2HK grant agreement no. 872549.

\appendix

\section{Spallation cut}
\label{sec:appendixspall}

\subsection{Neutron cloud cut}

For each cloud, we define a specific coordinate system whose origin is the projection of the cloud barycenter on the muon track. For a given DSNB candidate reconstructed vertex, $\ell_l$ is defined as the distance to this origin along the muon track while $\ell_t$ is the transverse distance to the track. For a given cloud multiplicity, a DSNB candidate observed less than $dt$ after the associated muon, and within the ellipse defined by $\ell_l$ and $\ell_t$, is classified as spallation background.
Conditions for the neutron cloud cut are summarized in Table~\ref{tab:cloudtab} for different multiplicities of the cloud. 

\begin{table}[htbp]
    \centering
    \caption{Conditions for the neutron cloud cut as a function of the cloud multiplicity. A number followed by the `$+$' symbol represents a lower bound on this multiplicity.}
    \vspace{+3truept}
    \begin{tabular}{ c | c | c | c | c | c | c | c } \hline \hline
    Multiplicity  & 2$+$ & 2$+$ & 2 & 3 & 4$-$5 & 6$-$9 & 10$+$ \\ \hline
    $dt$ [sec]  & 0.1 & 1 & 30 & 60 & 60 & 60 & 60 \\ \hline
    $\ell_l$ [cm] & 1200 & 800 & 383 & 548 & 603 & 712 & 766  \\ \hline
    $\ell_t$ [cm] & 1200 & 800 & 219 & 268 & 379 & 490 & 548 \\ \hline \hline
    \end{tabular}
    \label{tab:cloudtab}
\end{table}

\subsection{Rectangular cuts}

As mentioned in Section~\ref{subsec:spallreduc}, we applied a series of rectangular cuts based on spallation variables. 
The underlying idea is to remove isotopes with their half-lives dependent on the energy region and muon type. In particular, for events with a number of tagged neutrons different from one in the spectral analysis, we apply specific rectangular cuts to entirely $\mathcal{O}(0.01)$~sec half-lives. The criteria are summarized in Tables~\ref{tab:spallreccut} and \ref{tab:spallreccut2} for events with and without exactly one tagged neutron respectively. 

    \begin{table*}
    \centering
    \caption{Summary on the rectangular spallation cuts for events with exactly one tagged neutron. When a given cut has requirements on both $dt$ and $\ell_t$, at least one should be fulfilled.}
    \vspace{-5truept}
    \begin{tabular}{c c c c} \\ \hline\hline
     Muon type & Energy range & $dt$ requirement & $\ell_t$ requirement \\ \hline
     \multirow{2}{*}{All} & $E_{\rm rec} < 23.5$~MeV & $dt > 0.001$~sec & - \\
     & $E_{\rm rec} < 23.5$~MeV & $dt > 0.1$~sec & $\ell_t > 400$~cm \\ \hline
     Misfit & $15.5 < E_{\rm rec} < 19.5$~MeV & $dt > 1.5$~sec & - \\ \hline
     Well-fitted single through-going & $15.5 < E_{\rm rec} < 17.5$~MeV & $dt > 7$~sec & $\ell_t > 150$~cm \\ \hline
     Stopping & $15.5 < E_{\rm rec} < 19.5$~MeV & $dt > 0.05$~sec & - \\ \hline
     Poorly-fitted stopping & $15.5 < E_{\rm rec} < 17.5$~MeV & $dt > 6$~sec & - \\ \hline
     All multiple & $15.5 < E_{\rm rec} < 19.5$~MeV & $dt > 0.05$~sec & - \\ \hline\hline
    \end{tabular}
    \label{tab:spallreccut}
    \end{table*}

    \begin{table*}
    \centering
    \caption{Summary on the rectangular spallation cuts for events with tagged neutrons different from one. When a given cut has requirements on both $dt$ and $\ell_t$, at least one should be fulfilled. Here tighter rectangular cuts are applied to the muon categories generating most of the spallation ---single through-going and multiple muons--- to remove contributions from short-lived isotopes.}
    \vspace{-5truept}
    \begin{tabular}{c c c c} \\ \hline\hline
     Muon type & Energy range & $dt$ requirement & $\ell_t$ requirement \\ \hline
     \multirow{2}{*}{All} & $E_{\rm rec} < 23.5$~MeV & $dt > 0.001$~sec & - \\
      & $E_{\rm rec} < 23.5$~MeV & $dt > 0.1$~sec & $\ell_t > 400$~cm \\
     \hline
     Misfit & $15.5 < E_{\rm rec} < 19.5$~MeV & $dt > 1.5$~sec & - \\ 
     \hline
     \multirow{2}{*}{Well-fitted single through-going} & $15.5 < E_{\rm rec} < 17.5$~MeV & $dt > 7$~sec & $\ell_t > 150$~cm \\
     & $17.5 < E_{\rm rec} < 19.5$~MeV & $dt > 7$~sec & $\ell_t > 100$~cm \\
     \hline
     Stopping & $15.5 < E_{\rm rec} < 19.5$~MeV & $dt > 0.05$~sec & - \\
     \hline
     All multiple & $15.5 < E_{\rm rec} < 19.5$~MeV & $dt > 0.05$~sec & - \\
     \hline
     \multirow{2}{*}{Well-fitted multiple} & $15.5 < E_{\rm rec} < 19.5$~MeV & $dt > 0.1$~sec & - \\
     & $15.5 < E_{\rm rec} < 19.5$~MeV & $dt > 1$~sec & $\ell_t > 400$~cm \\
     \hline\hline
    \end{tabular}
    \label{tab:spallreccut2}
    \end{table*}

\subsection{Spallation PDFs and log-likelihood ratio}

\begin{figure*}[htbp]
  \begin{minipage}{0.45\hsize}
   \begin{center}
    \includegraphics[clip,width=8.cm]{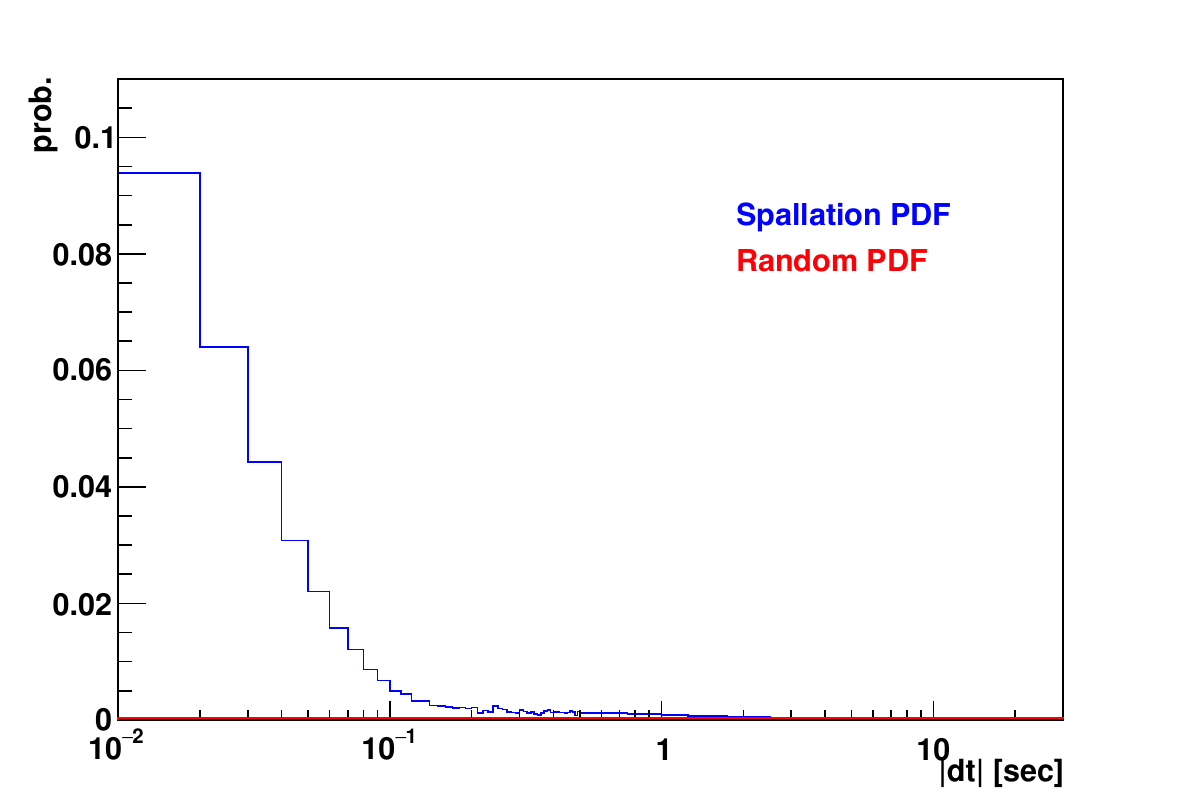}
   \end{center}
  \end{minipage}
  \begin{minipage}{0.45\hsize}
   \begin{center}
    \includegraphics[clip,width=8.cm]{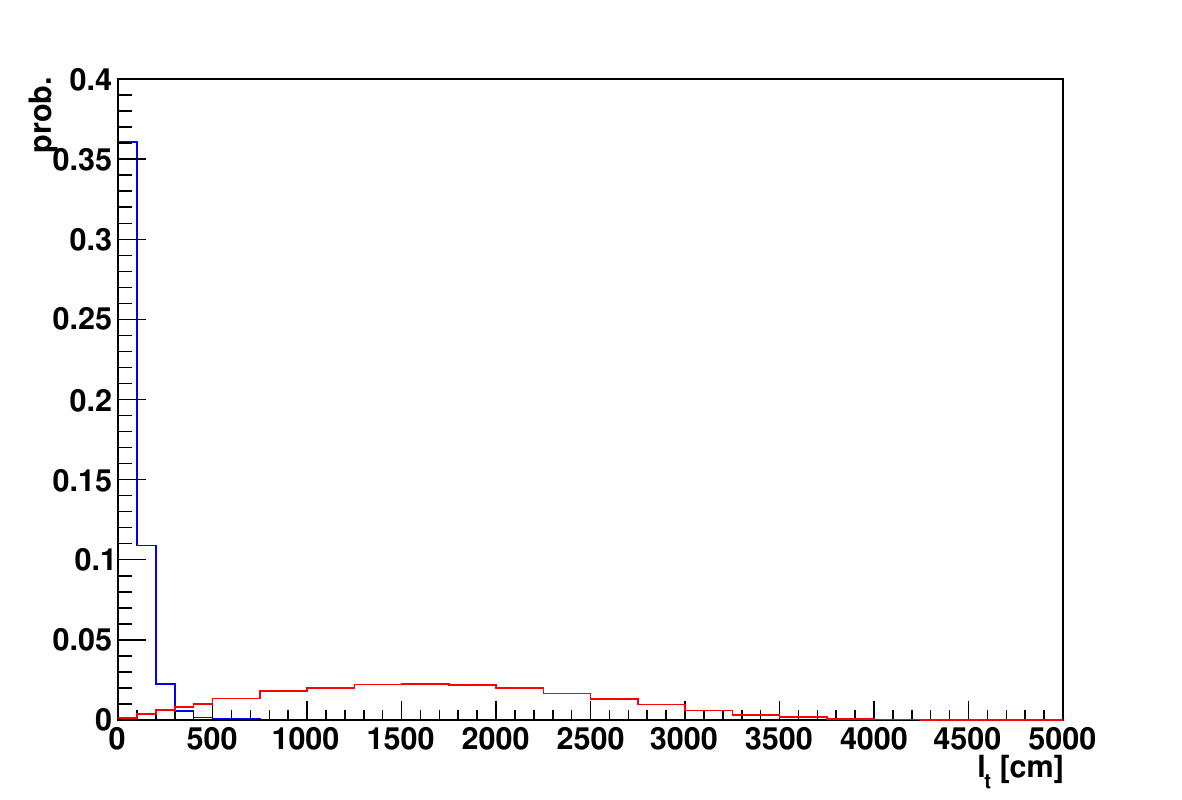}
   \end{center}
  \end{minipage}
  \\
  \noindent
  \begin{minipage}{0.45\hsize}
   \begin{center}
    \includegraphics[clip,width=8.cm]{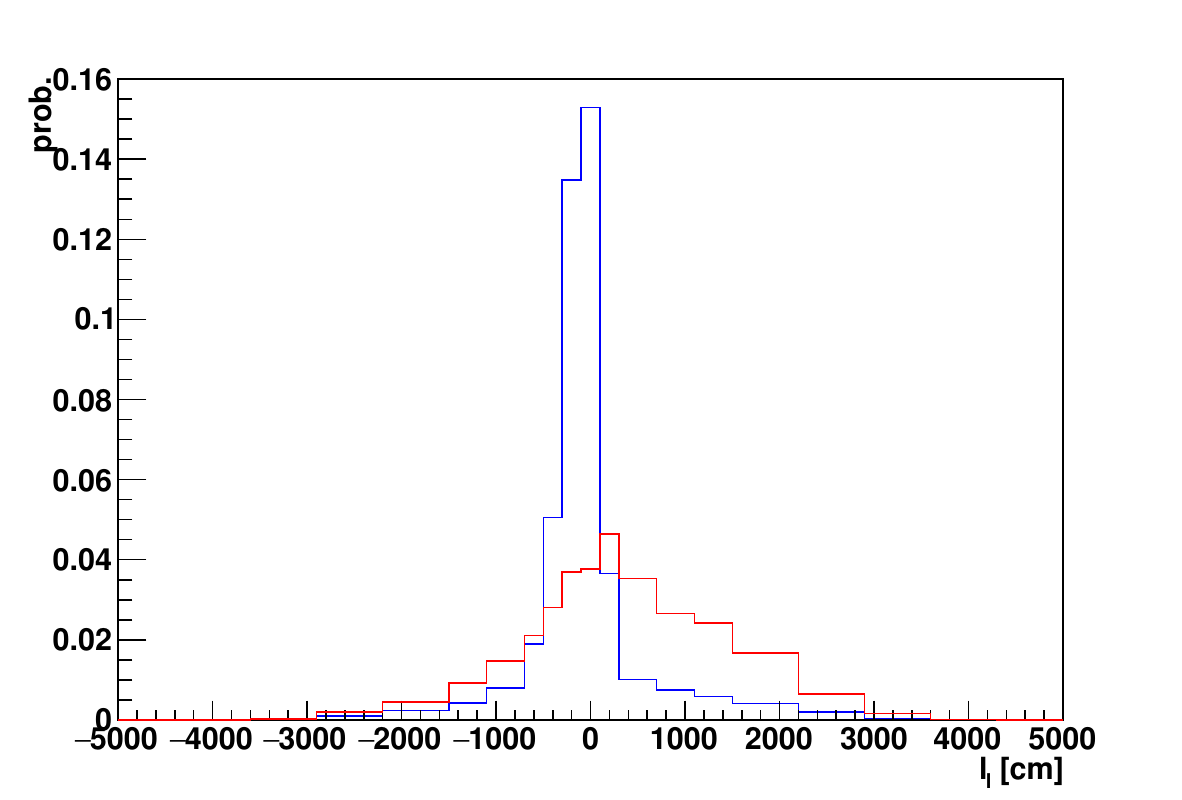}
   \end{center}
  \end{minipage}
  \begin{minipage}{0.45\hsize}
   \begin{center}
    \includegraphics[clip,width=8.cm]{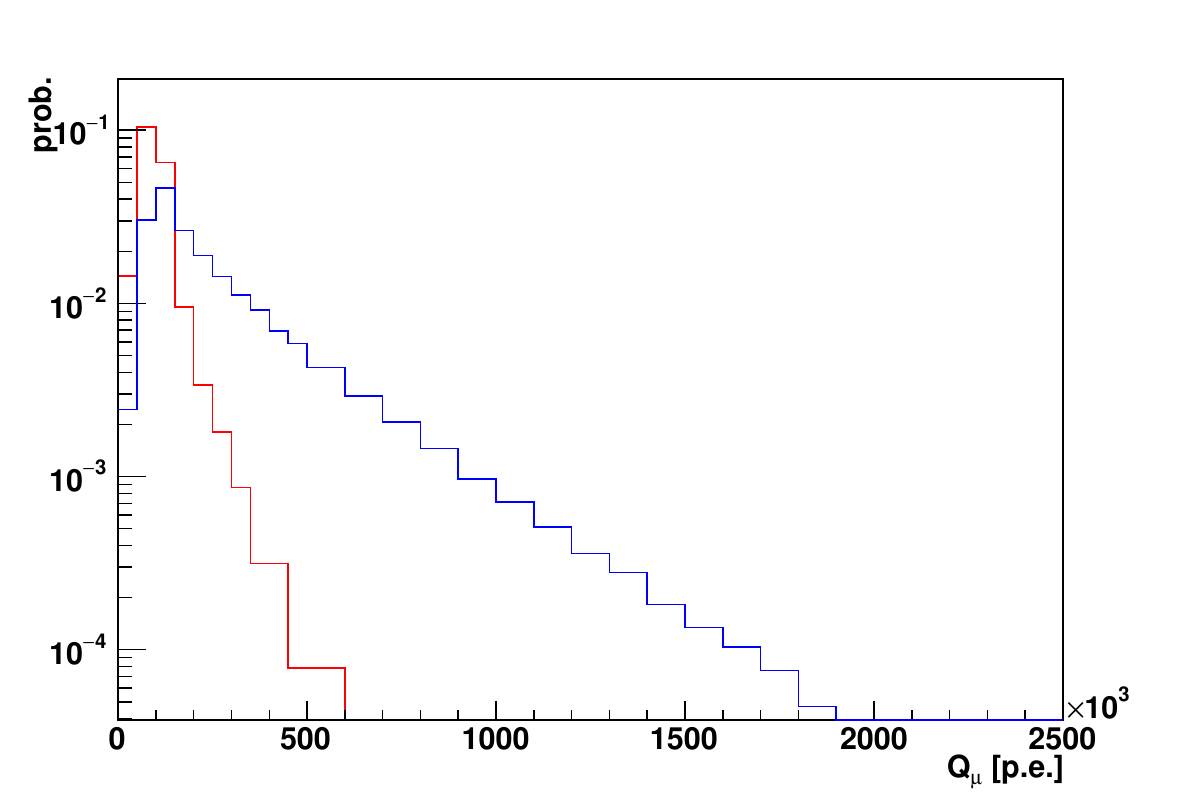}
   \end{center}
  \end{minipage}
  \vspace{-5truept}
  \begin{center}
   \includegraphics[clip,width=8.cm]{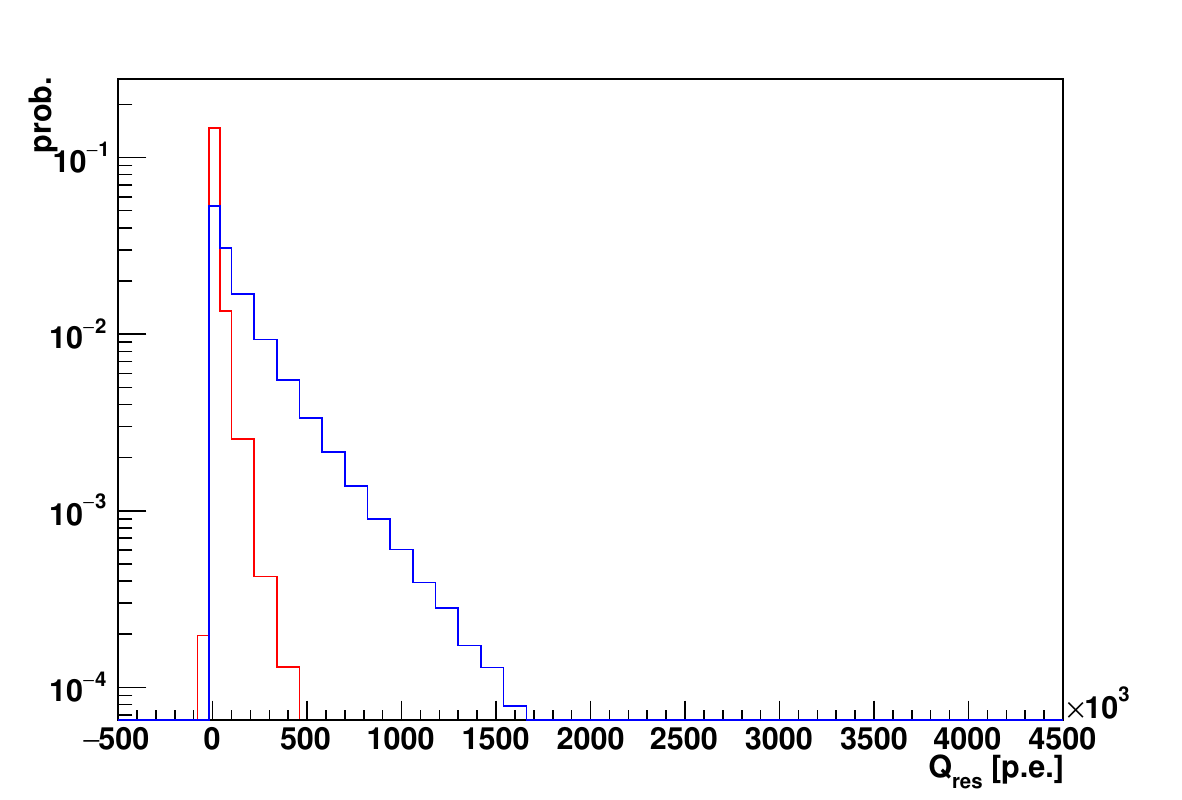}
  \end{center}
  \vspace{-5truept}
  \caption{Spallation and random PDFs for $dt$ (top left), $\ell_t$ (top right), $\ell_l$ (middle left), $Q_\mu$ (middle right), and $Q_{\rm res}$ (bottom), for the pre (blue) and post (red) samples. Results from the single through-going muons for the $dt$ region of 0$-$0.05~sec and the $\ell_t$ region of 0$-$300~cm are shown here.} 
  \label{fig:pdf_all_single_short_0}
  \end{figure*}

\begin{figure}[htbp]
  \begin{center}
    \includegraphics[clip,width=9.5cm]{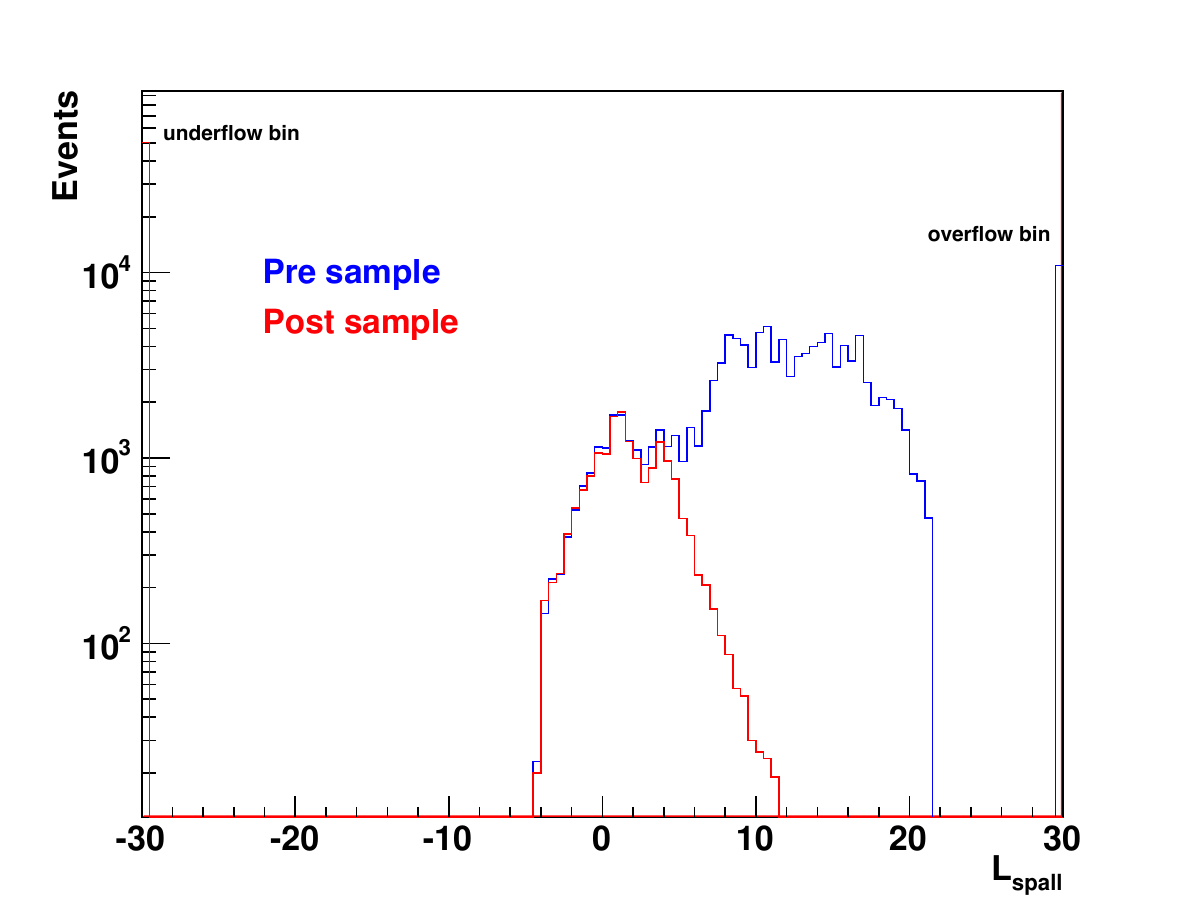}
  \end{center}
  \vspace{-10truept}
  \caption{Spallation log-likelihood ratio of the single through-going muon for $7.5 < E_{\rm rec} < 9.5$~MeV. The values smaller or larger than $-$30 or 30 are filled in the underflow or overflow bins, respectively.}
  \label{fig:spaloglike}
  \end{figure}

When producing the spallation and random PDFs, we separate the $dt$ and $\ell_t$ bins to account for the possible correlations, 0$-$0.05, 0.05$-$0.5, and 0.5$-$30~sec for $dt$ (determined by the isotope half-lives in Fig.~\ref{fig:spall_products}), and 0$-$300, 300$-$1000, 1000$-$5000~cm for $\ell_t$. Fig.~\ref{fig:pdf_all_single_short_0} shows PDFs for each spallation variable from single through-going muons. Here the results from the most spallation-rich region ($0 < dt < 0.05$~sec and $0 < \ell_t < 300$~cm) are shown. 
We calculate the spallation log-likelihood ratio with the procedure described in Section~\ref{subsec:spallreduc}. An example distribution is shown for single through-going muons in Fig.~\ref{fig:spaloglike}. 
The cut criteria are tuned depending on the reconstructed energy as well as the $dt$ and $\ell_t$ bins. We use the common binning for $dt$, 0$-$0.05, 0.05$-$0.5, and 0.5$-$30~sec, independent of the muon type. In contrast, for $\ell_t$, we separate bins only for most impactful muon types, that is, single through-going and multiple muons: 0$-$200, 200$-$300, 300$-$500, 500$-$1000, and $>$1000~cm for single through-going, and 0$-$100, 100$-$200, 200$-$300, 300$-$500, 500$-$700, 700$-$1000, 1000$-$2000, 2000$-$3000, $>$3000~cm for multiple muons.

\subsection{Spallation cut performance estimation}
\label{sec:spacutperf}
In this section we present our methodology to estimate the impact of spallation cuts on non-spallation ``signal'' events (DSNB, solar, reactor, and atmospheric neutrino interactions), as well as on spallation backgrounds. In what follows, we treat $^9$Li decays separately as they provide the dominant $\beta$$+$$n$ signature associated with spallation.

As explained in Section~\ref{subsec:spallreduc}, random event efficiencies for multiple spallation and neutron cloud cuts 
are estimated to be $98$\% and $99.8$\%, respectively.
We estimate the signal efficiencies for the rectangular and likelihood cuts described above by evaluating their impact on the post-sample, as described in the previous section. The associated efficiency ($\epsilon_{\rm random}$) can thus be expressed as: 

\begin{align}
  \label{eq:eff_random}
   \epsilon_{\rm random} = N_{\rm post,after}/N_{\rm before}, 
\end{align}

\noindent 
where $N_{\rm before}$ and $N_{\rm post,after}$ are the numbers of events in the post-sample before and after spallation cuts. We validate this efficiency by evaluating the impact of spallation cuts on events from solar neutrino elastic interactions, using the opening angle $\theta_{\rm sun}$ defined in Section~\ref{sec:background} as a discriminator. 
%

To estimate the background rejection power of the spallation cuts, we use a sample of DSNB candidate events verifying $\cos\theta_{\rm sun} < 0$. In order to suppress remaining radioactivity and atmospheric neutrino events, we also apply cuts on the effective distance to the wall and the Cherenkov angle, that will be  defined in Section~\ref{subsec:posreduc}, and require $E_{\rm rec} <15.5$~MeV.
The spallation remaining rate ($\epsilon_{\rm spall}$) is then estimated as the fraction of events remaining after spallation cuts, after the expected number of remaining atmospheric neutrino events from Monte-Carlo simulation has been subtracted. 

%
%
The procedure outlined above allows to accurately estimate the spallation remaining rate for $E_{\rm rec} < $ 15.5~MeV ---where spallation largely dominates even after cuts, but not to assess the impact of spallation cuts on individual isotopes. Using the spallation-dominated sample described above, we therefore refine our approach as follows. To compute the remaining fraction of a specific isotope, we attribute a random $dt$, drawn from the isotope decay time distribution, to each pair in the sample, leaving all other observables untouched. This is equivalent to renormalizing the spallation likelihood from Equation~\ref{eq:spalike} by the isotope's exponential decay time distribution. Then, we apply the likelihood cuts described above to the pairs with the modified $dt$, and compute the fraction of remaining events, $\epsilon_{\mathrm{pre,iso}}$. For each event, this fraction can be expressed as:

\begin{align}
    \epsilon_{\mathrm{pre,iso}} = \epsilon_{\mathrm{iso,likeli}}^{\rm (pair)} \times \epsilon_{\mathrm{random,iso}}
\end{align}

\noindent 
where $\epsilon_{\mathrm{iso,likeli}}^{\rm (pair)}$ is the probability for a pair formed by a spallation event and its parent muon to pass the likelihood cuts. Conversely,  $\epsilon_{\mathrm{random,iso}}$ is the probability for all other (uncorrelated) pairs between the event and preceding muons to pass these cuts. This probability can be computed using a post-sample analogous to the one described in Section~\ref{sec:spaloglike} and, since we reassigned $dt$ for all pairs in each event, will be artificially lower than the signal efficiency $\epsilon_\mathrm{random}$. The probability for a given isotope decay to pass the likelihood cuts ($\epsilon_{\mathrm{iso,likeli}}$) is thus given by:

\begin{align}
    \epsilon_{\mathrm{iso,likeli}} = \epsilon_{\mathrm{pre,iso}}/\epsilon_{\mathrm{random,iso}}\times \epsilon_\mathrm{random}.
\end{align}

\noindent 
The fraction of remaining events after both preselection and likelihood cuts ($\epsilon_{\mathrm{iso,spall}}$) can then be estimated as:

\begin{align}
    &\epsilon_{\mathrm{iso,spall}} = \epsilon_{\mathrm{pre,iso}}/\epsilon_{\mathrm{random,iso}}\times \epsilon_\mathrm{random} \times \epsilon_\mathrm{presel},\\
    & \ \ \ \ \ \ \ \ \ \ \mathrm{where}\quad \epsilon_\mathrm{presel} = \frac{\epsilon_\mathrm{spall}}{\epsilon_\mathrm{likeli}}.
\end{align}

\noindent 
Here, $\epsilon_\mathrm{spall}$ is the spallation remaining rate for all isotopes after spallation cuts and $\epsilon_\mathrm{likeli}$ is the rate after likelihood cuts only. This ratio yields $\epsilon_\mathrm{presel} \sim 70$\% while the preselection cut alone removes about half of the background, highlighting the significant overlap between preselection and likelihood cuts.

This procedure is used throughout this study to estimate the remaining fractions of various isotopes. For $E_{\rm rec} < $15.5~MeV, we will notably use it to estimate the impact of spallation cuts on the $^9$Li background. Above 15.5~MeV, we will estimate the remaining fractions of $^8$Li and $^8$B, two isotopes that are particularly difficult to remove due to their long half-lives. This estimate will allow us to model the spallation background spectrum for the spectral analysis. 
%

In order to estimate the remaining amounts of individual isotopes without relying on a simulation we made two major assumptions. First, that the time difference between spallation events and their parent muon is not correlated with other spallation observables. Second, that the impact of spallation cuts above 15.5~MeV can be estimated using a sample of lower energy events. Both assumptions would be verified if the $\ell_t$, $\ell_l$, $Q_\mu$, and $Q_{res}$ spallation observables described in Section~\ref{sec:spaloglike} were similarly distributed for all isotopes. This is however not the case as their values will depend on the isotope production mode. To estimate the effect of this isotope dependence, we split the sample described above into four 2-MeV bins from 7.5 to 15.5~MeV, that will have different isotope compositions. For a given spallation cut, we then compute the spallation remaining rates for the four subsamples and take the largest deviation from the average value as our systematic uncertainty.
This uncertainty depends on the isotope considered but typically lies around 50\%.

\section{$^{\bf 241}$Am/Be calibration study for neutron tagging}
\label{sec:appendixntag}

As described in Section~\ref{sec:ntag_ambe}, in order to validate the efficiency of the neutron tagging procedure on real data, we applied neutron tagging to data collected in the presence an $^{241}$Am/Be source, which, surrounded by a BGO scintillator, produces a prompt signal and a delayed neutron capture signal as part of its radioactive decay chain. However, the data collected with the $^{241}$Am/Be source is not fully modeled by the simulation software; notably, no modeling of the BGO scintillator is available. The Monte-Carlo simulation that we use to model $^{241}$Am/Be events is hence similar to the IBD simulation. Because of this limitation, a direct comparison between the neutron tagging efficiency in simulation and calibration data would be overly pessimistic, as scintillation in the BGO can produce prompt events that are not accompanied by the production of a neutron, thus artificially reducing the observed preselection efficiency of the neutron candidates (see Section~\ref{sec:ntagpresel}). To account for this, we treated uncertainty associated with the preselection separately from that of the BDT selection, following the procedure described in Ref.~\cite{bib:hayatontag}. 

\begin{figure}[htbp]
    \centering
    \includegraphics[width=\linewidth]{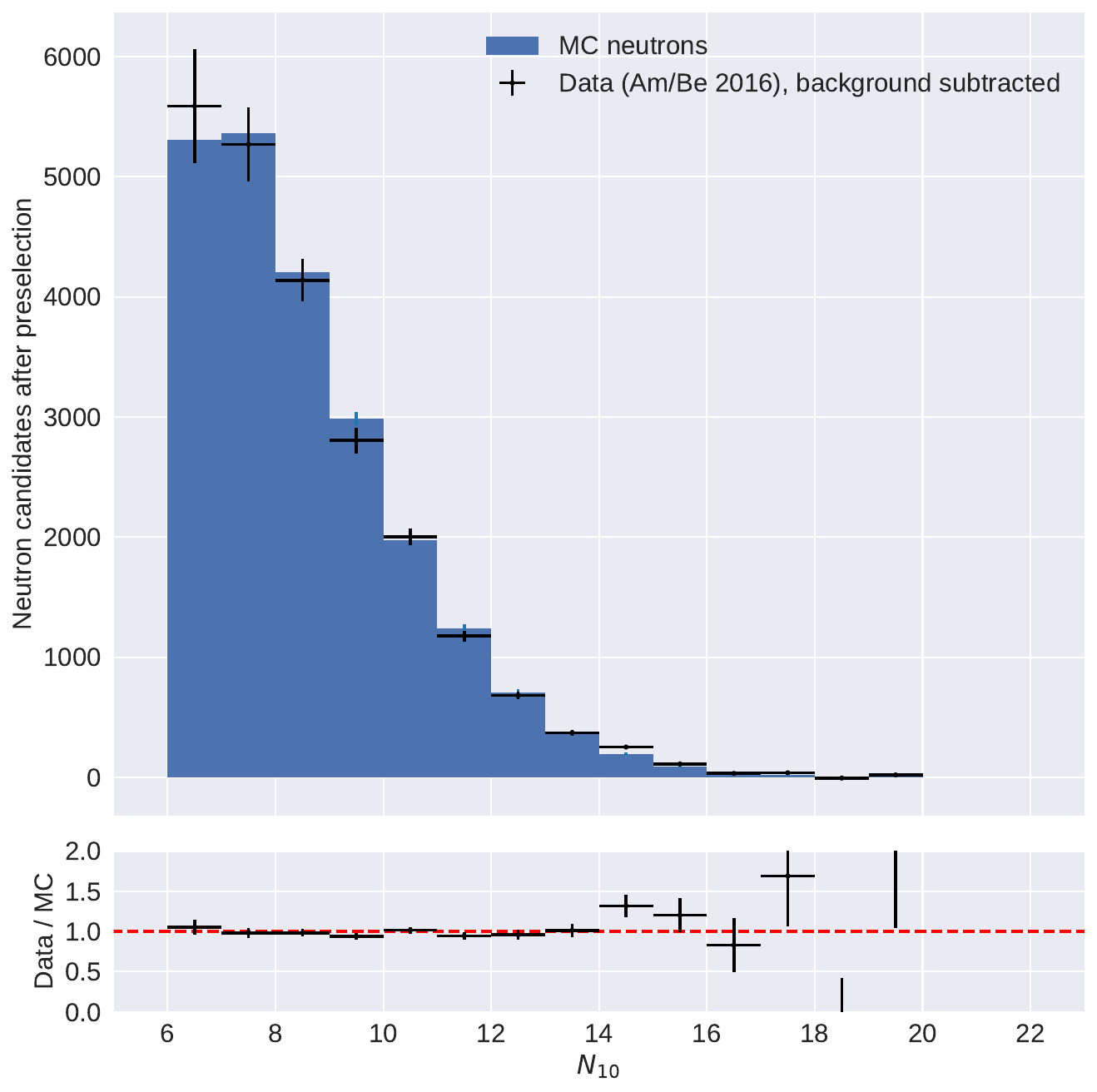}
    \caption{$N_{10}$ observed and expected distributions for neutrons. The expected distribution corresponds to the PMT photon detection efficiency for which the $N_{10}$ fits the observations best. The observed distribution has been obtained by subtracting the $N_{10}$ background distribution from the total observed distribution. This background distribution has been obtained from data, using random triggers. The lower predicted number of events for $N_{10} = 6$ is due to harsher continuous dark noise reduction cuts, as mentioned in Section~\ref{sec:ntagpresel}.}
    \label{fig:ambe_presel}
\end{figure}

The preselection efficiency is based on the number of hits in a given time window and is hence particularly sensitive to the PMT photon detection efficiency (quantum efficiency, QE). Hence, we determine the associated systematic uncertainty by determining the value of QE for which the predicted $N_{10}$ distribution fits the observations best. The $N_{10}$ predicted and observed distributions are shown in Fig.~\ref{fig:ambe_presel} for the best-fit QE, that is found to be 2.1\% larger than the QE used in the IBD simulation from Section~\ref{subsec:signalsimu}. The preselection efficiency associated to this best-fit QE is 3.0\% larger than the efficiency predicted by our simulation. Moreover, the 1$\sigma$ uncertainty on this best-fit value, found using a $\chi^2$ fit, leads to a 2.1\% uncertainty on the value of the preselection efficiency. The combined preselection uncertainty is hence of 3.7\%.

After the uncertainty on the preselection efficiency $\epsilon_{\rm IS}$, we evaluate the uncertainty associated with the BDT selection cuts by computing the relative efficiencies $\epsilon_{\rm rel} = \epsilon_{\rm BDT + IS}/\epsilon_{IS}$ for the data and the Monte-Carlo simulation for different values of the BDT discriminant. We determine these efficiencies by fitting the timing distributions of neutron candidates by an exponential plus a constant. Results for different BDT cuts are shown in Table~\ref{tab:ambe_ratios} for the different $^{241}$Am/Be datasets considered here. The maximum discrepancy between data and Monte-Carlo lies around 12\% irrespective of the BDT cut.

Combining the 12\% BDT reduction uncertainty to the 3.7\% preselection uncertainty leads to an overall uncertainty of 12.5\% on the neutron tagging efficiency. This uncertainty also include possible discrepancies in neutron kinetic energy distributions, like the ones between $^9$Li decays and IBD events.


\begin{table}[htbp]
    \centering
    \caption{Observed (`Data') and predicted (`MC') relative efficiencies (in \%) for different BDT cuts and for the $^{241}$Am/Be calibration runs considered in this study. Data was taken in 2009 and 2016, at three different locations in the tank: in the center of the detector, near the barrel wall (Y12) and the top wall (Z15).}
    \vspace{+3truept}
    \begin{tabular}{c|c|c|ccc}
        \toprule
        BDT cut & Position & Year & Data  & MC & (Data $-$ MC)/Data\\
        \hline
        \multirow{6}{*}{0.40} & \multirow{2}{*}{Center} & 2009 & 69.4  & 76.2 & $-$9.8\\
        && 2016 & 80.0 & 72.3 &9.5 \\\cline{2-6}
        &\multirow{2}{*}{Y12} & 2009 & 74.6 & 77.4 & $-$3.7\\
        && 2016 & 83.8 & 77.6 &7.2 \\\cline{2-6}
        &\multirow{2}{*}{Z15} & 2009 & 87.6 & 77.1 & 12.0\\
        && 2016 & 80.2 & 74.1 & $-$7.7\\\hline\hline
        \multirow{6}{*}{0.90} & \multirow{2}{*}{Center} & 2009 & 50.0  & 55.3 & $-$10.6\\
        && 2016 & 56.1 & 52.8 & 5.8 \\\cline{2-6}
        &\multirow{2}{*}{Y12} & 2009 & 56.4 & 59.6 & $-$5.7\\
        && 2016 & 60.8 & 59.2 &2.6 \\\cline{2-6}
        &\multirow{2}{*}{Z15} & 2009 & 65.8 & 58.3 & 11.4\\
        && 2016 & 57.0 & 55.4 & 2.8\\\hline\hline
         \multirow{6}{*}{0.99} & \multirow{2}{*}{Center} & 2009 & 33.6  & 35.5 & $-$5.8\\
        && 2016 & 34.9 & 33.2 & 4.8 \\\cline{2-6}
        &\multirow{2}{*}{Y12} & 2009 & 36.9 & 37.8 & $-$2.4\\
        && 2016 & 38.2 & 37.1 &2.8 \\\cline{2-6}
        &\multirow{2}{*}{Z15} & 2009 & 38.8 & 34.5 & 10.5\\
        && 2016 & 31.9 & 32.2 & $-$0.1\\
        \botrule
    \end{tabular}
    \label{tab:ambe_ratios}
\end{table}


\section{Neutrons produced from atmospheric neutrino interactions}
\begin{figure}[htbp]
    \centering
    \includegraphics[width=9.5cm]{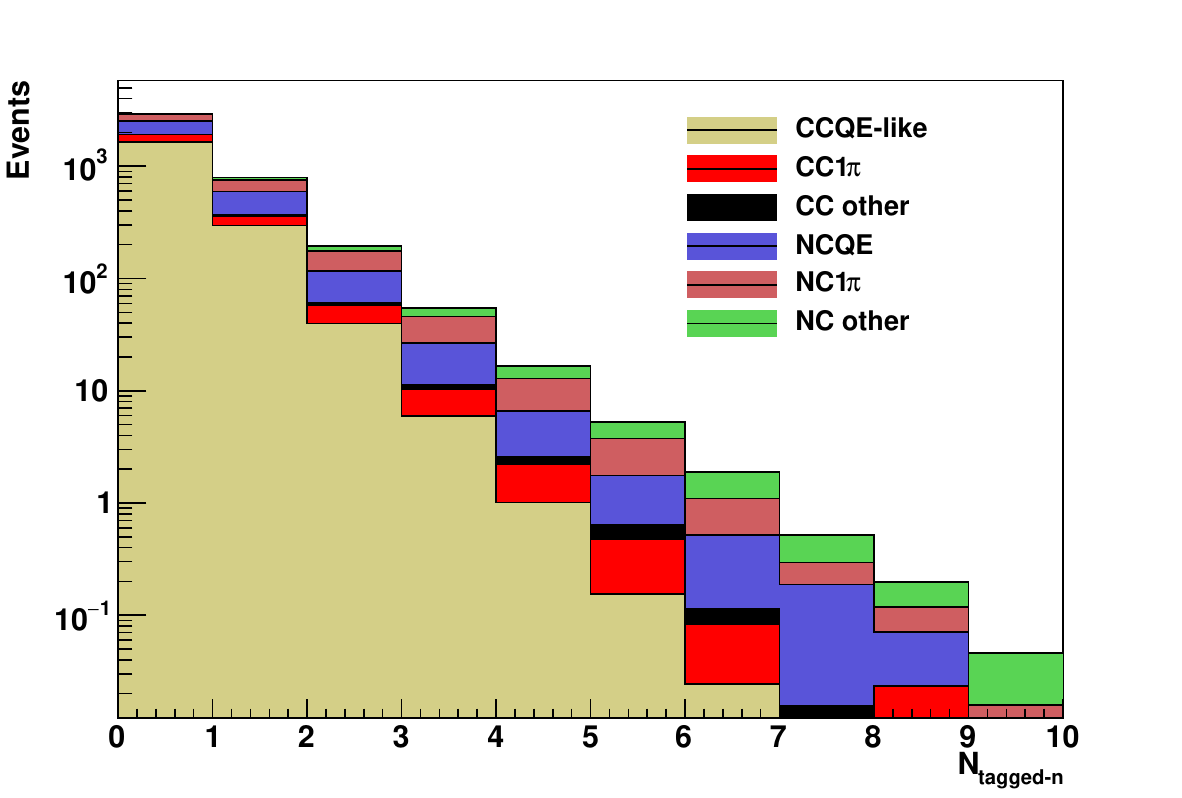}
    \caption{Predicted neutron multiplicity for atmospheric neutrino backgrounds. Here we use the atmospheric neutrino Monte-Carlo simulation in the 7.5$-$79.5~MeV range, after the noise reduction cuts.}
    \label{fig:atmneutrons}
\end{figure}

The neutron multiplicities predicted by the Monte-Carlo simulation for atmospheric neutrinos are shown in Fig.~\ref{fig:atmneutrons}. Note that a significant number of atmospheric neutrino interactions produce one or more neutrons. However, these neutrons are more energetic than for the IBDs of DSNB neutrinos and sometimes travel further away from their production vertex. Since the current neutron tagging algorithm requires neutrons to be close to their associated positron vertex, it can also be used to reduce neutron-producing atmospheric backgrounds. A detailed study of neutrons associated with atmospheric neutrino backgrounds has been performed in Ref.~\cite{bib:hayatontag}.

\section{Spectral fitting}
\label{sec:appendixspecfit}
\subsection{Modeling of atmospheric neutrino backgrounds}
\begin{figure*}[htbp]
  \centering
\begin{minipage}{0.45\textwidth}
 \begin{tabular}{@{}cccc@{}}
& $20^\circ<\theta_{\rm C}<38^\circ$ & $38^\circ<\theta_{\rm C}<50^\circ$ & $78^\circ<\theta_{\rm C}<90^\circ$\\
\rotatebox{90}{\hspace{10pt} 0 or $>$1 ntags}& \includegraphics[width=.32\linewidth]{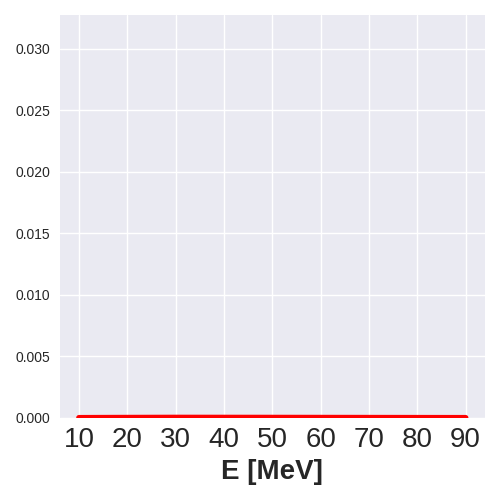}&  \includegraphics[width=.32\linewidth]{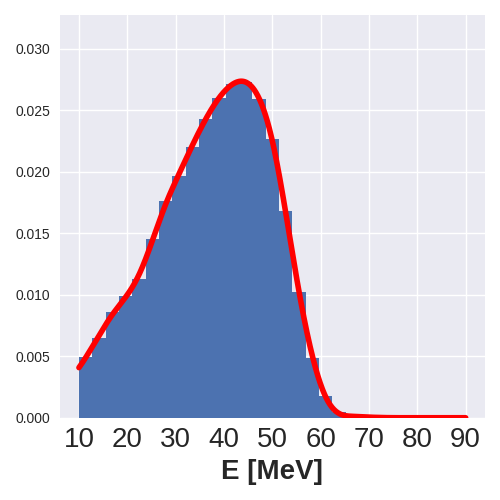}&
\includegraphics[width=.32\linewidth]{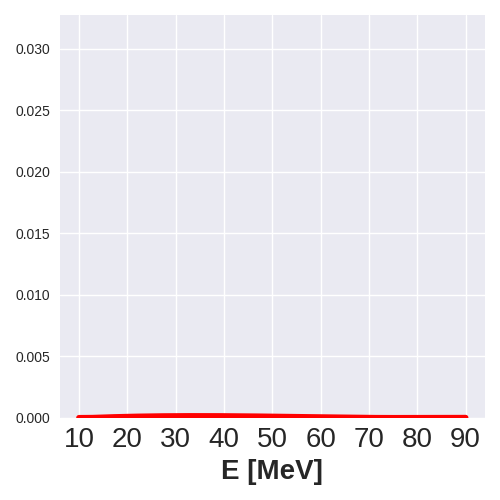} \\
\rotatebox{90}{\hspace{20pt}1 ntag}& \includegraphics[width=.32\linewidth]{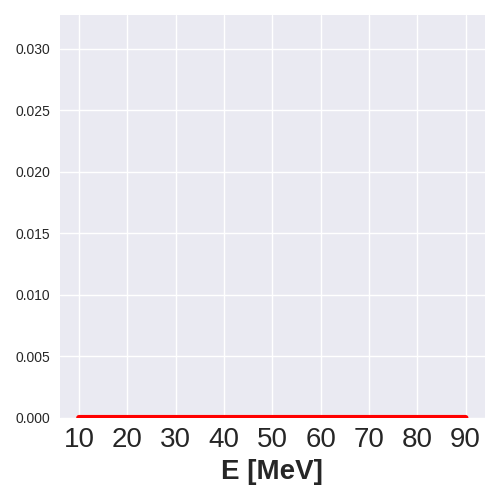}&  \includegraphics[width=.32\linewidth]{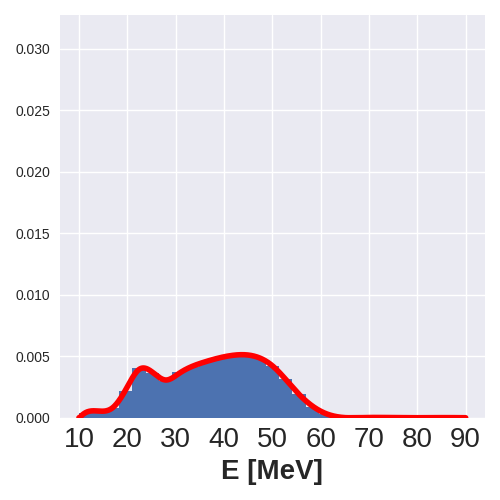}&
\includegraphics[width=.32\linewidth]{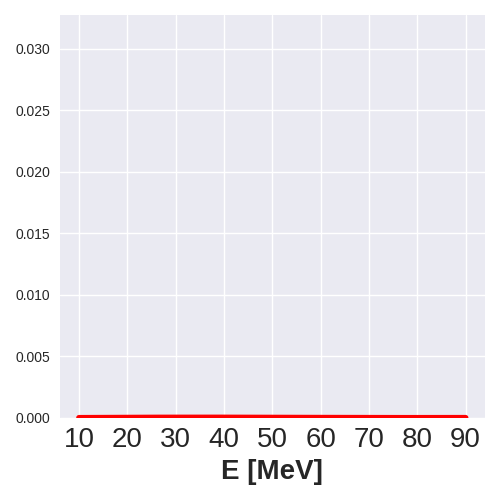} \\
\end{tabular}
\end{minipage}
\hfill
\begin{minipage}{0.45\textwidth}
 \begin{tabular}{@{}cccc@{}}
& $20^\circ<\theta_{\rm C}<38^\circ$ & $38^\circ<\theta_{\rm C}<50^\circ$ & $78^\circ<\theta_{\rm C}<90^\circ$\\
\rotatebox{90}{\hspace{10pt} 0 or $>$1 ntags}&
\includegraphics[width=.32\linewidth]{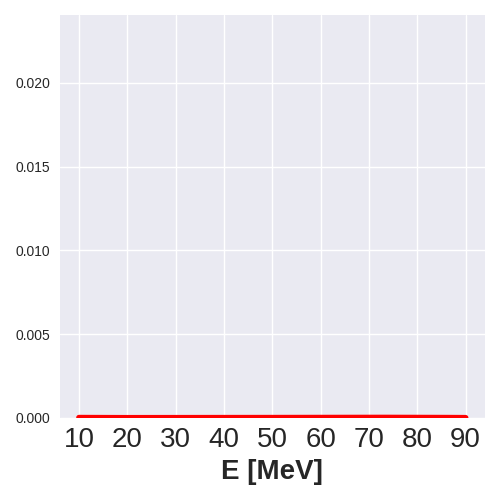}&  \includegraphics[width=.32\linewidth]{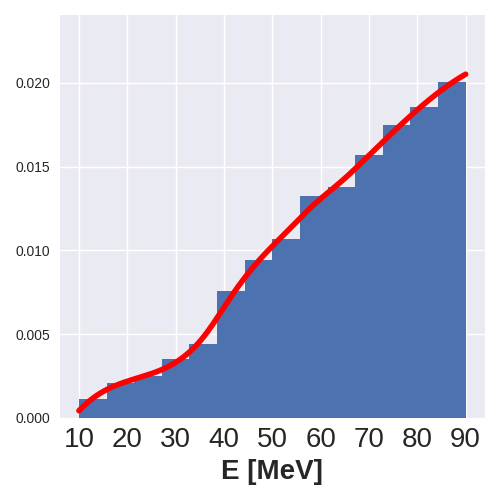}&
\includegraphics[width=.32\linewidth]{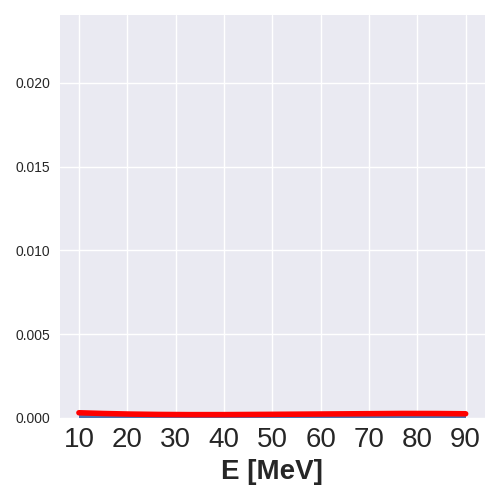} \\
\rotatebox{90}{\hspace{20pt}1 ntag}&
\includegraphics[width=.32\linewidth]{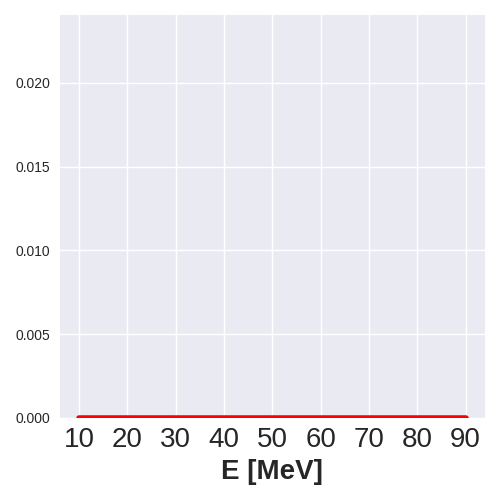}&  \includegraphics[width=.32\linewidth]{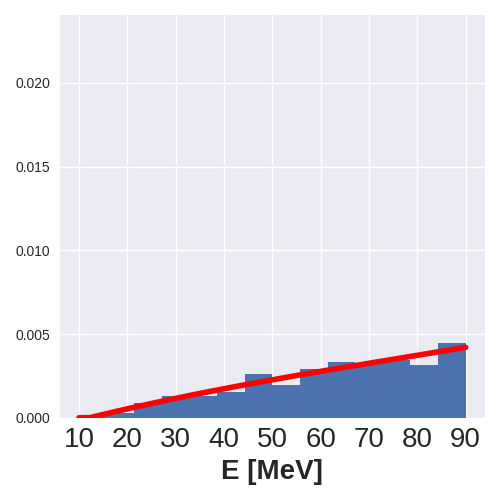}&
\includegraphics[width=.32\linewidth]{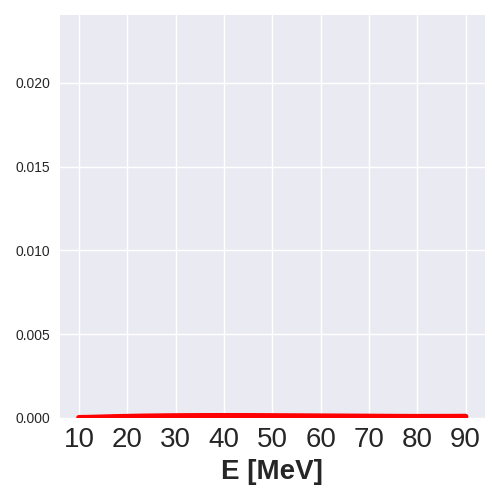} \\
\end{tabular}
\end{minipage}
\begin{minipage}{0.45\textwidth}
 \begin{tabular}{@{}cccc@{}}
& $20^\circ<\theta_{\rm C}<38^\circ$ & $38^\circ<\theta_{\rm C}<50^\circ$ & $78^\circ<\theta_{\rm C}<90^\circ$\\
\rotatebox{90}{\hspace{10pt} 0 or $>$1 ntags}& \includegraphics[width=.32\linewidth]{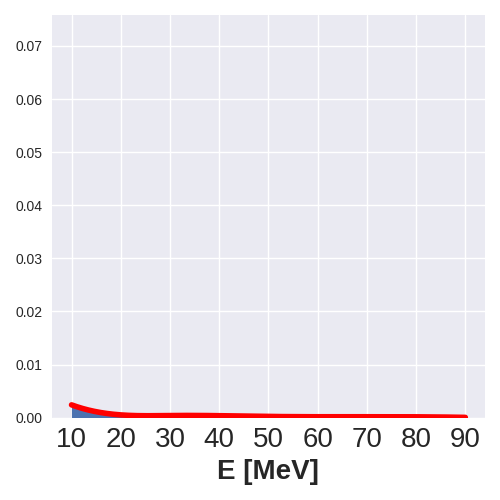}&  \includegraphics[width=.32\linewidth]{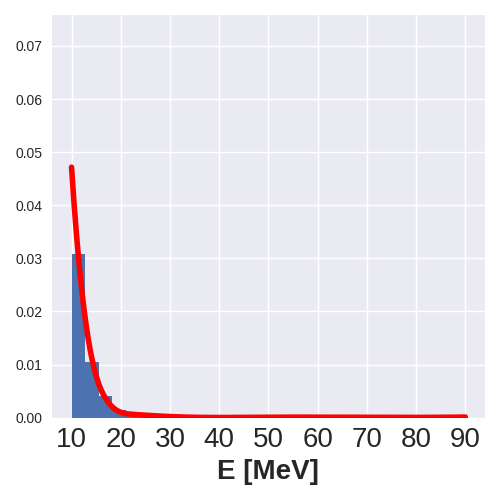}&
\includegraphics[width=.32\linewidth]{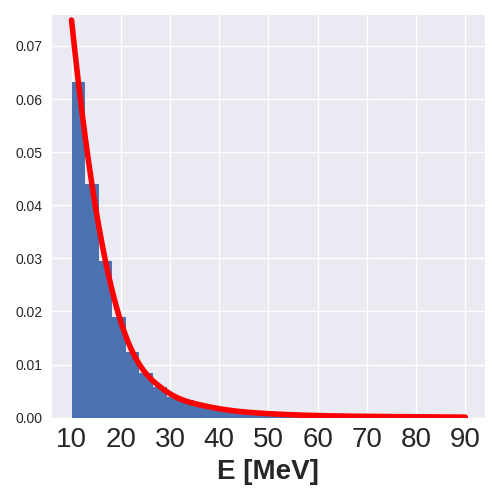} \\
\rotatebox{90}{\hspace{20pt}1 ntag}& \includegraphics[width=.32\linewidth]{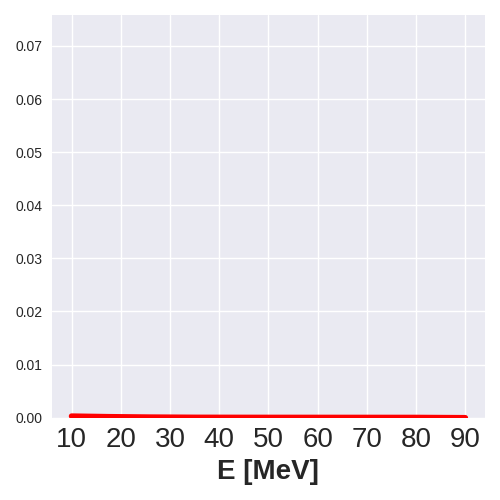}&  \includegraphics[width=.32\linewidth]{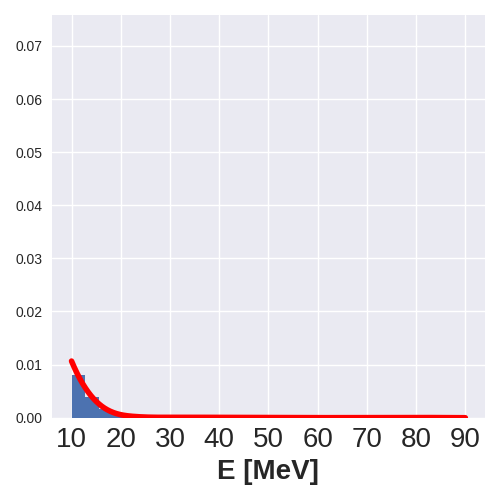}&
\includegraphics[width=.32\linewidth]{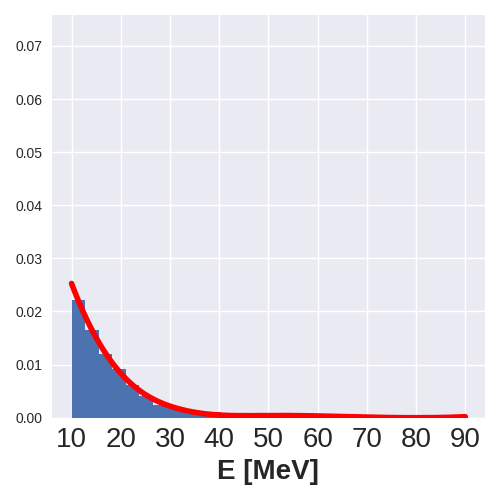} \\
\end{tabular}
\end{minipage}
\hfill
\begin{minipage}{0.45\textwidth}
 \begin{tabular}{@{}cccc@{}}
& $20^\circ<\theta_{\rm C}<38^\circ$ & $38^\circ<\theta_{\rm C}<50^\circ$ & $78^\circ<\theta_{\rm C}<90^\circ$\\
\rotatebox{90}{\hspace{10pt} 0 or $>$1 ntags}
& \includegraphics[width=.32\linewidth]{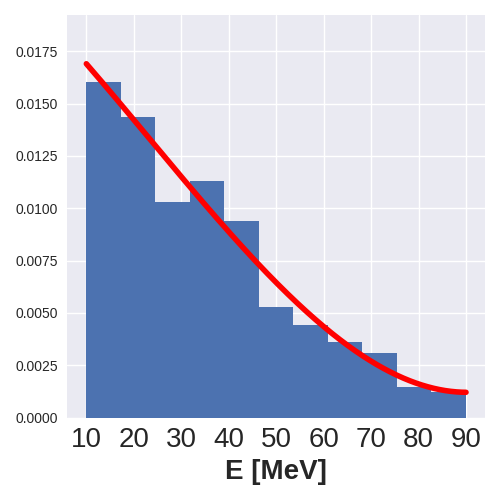}&  \includegraphics[width=.32\linewidth]{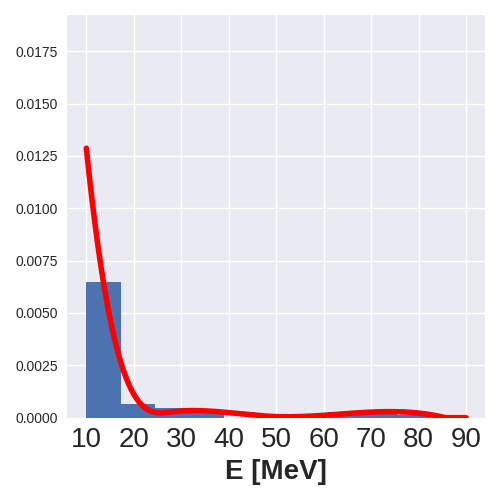}&
\includegraphics[width=.32\linewidth]{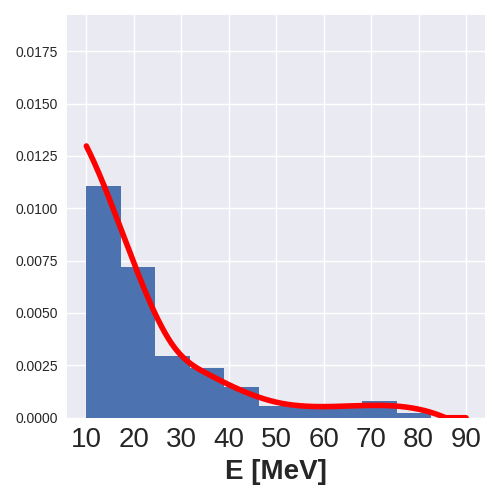} \\
\rotatebox{90}{\hspace{20pt}1 ntag}&
\includegraphics[width=.32\linewidth]{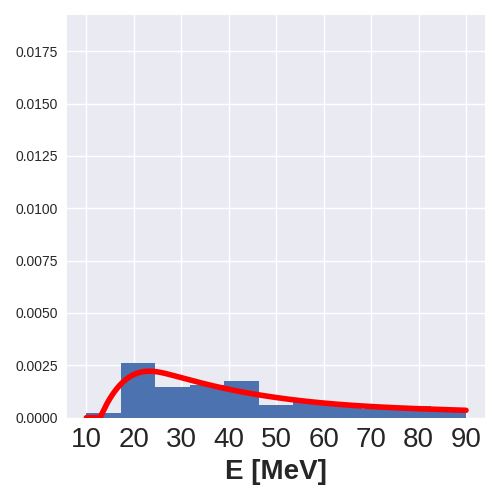}&  \includegraphics[width=.32\linewidth]{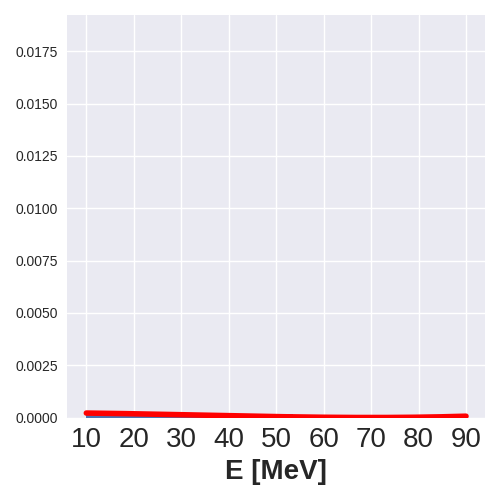}&
\includegraphics[width=.32\linewidth]{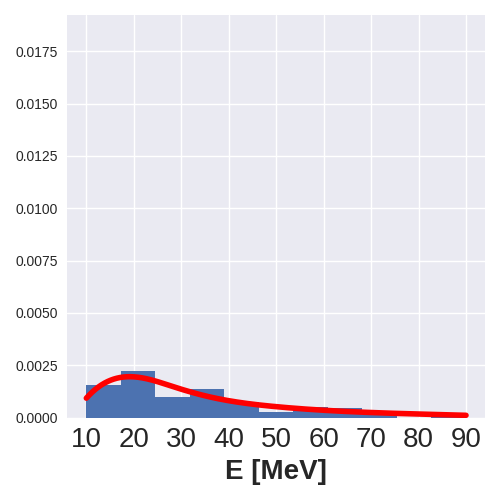} \\
\end{tabular}
\end{minipage}
  \caption{Probability distribution functions used for the fitting of atmospheric backgrounds in the unbinned spectral fit of the data. As detailed in Section \ref{sec:spectral}, the backgrounds are separated into four categories, and for each category a PDF is defined over three Cherenkov angle regions (two sidebands and one signal region) and two neutron tagging regions (corresponding to $=$1 or $\neq$1 tagged neutrons). Each PDF is normalized to 1 over all regions. The blue histograms show the background spectra obtained from Monte-Carlo simulation or directly from data for the decay electron, while the red lines are the fitted PDFs, obtained by fitting the binned spectra with a piecewise polynomial. Top left: electrons from $\mu$ or $\pi$ decays. Top right: $\nu_e$ CC interactions. Bottom left: NCQE interactions. Bottom right: $\mu/\pi$-producing backgrounds.}
  \label{fig:specpdfs}
\end{figure*}

The probability distribution functions of the atmospheric neutrino background spectra used in the spectral analysis are displayed in Fig. \ref{fig:specpdfs}. As mentioned in Section~\ref{sec:spectral}, we divide atmospheric neutrino backgrounds into four categories: $\mu/\pi$, $\nu_e$ CC, decay electrons, and NCQE interactions. The decay electron spectrum is obtained from data while the other spectra are obtained from the atmospheric neutrino Monte-Carlo simulation.

In the extended maximum likelihood fit described in Section~\ref{sec:spectral}, atmospheric background fluxes are treated as nuisance parameters. In Table~\ref{tab:atmpred}, we compare the predicted numbers of events in the four atmospheric background categories to the best-fit values obtained when considering the Horiuchi+09 model~\cite{bib:horiuchi09}. Since, as discussed in Section~\ref{sec:spectral}, atmospheric events from e.g.~pion-producing NC interactions could be categorized as either NCQE or $\mu/\pi$, we quote ranges rather than numbers for these two categories. For decay electrons and $\nu_e$ CC interactions, the predicted and observed numbers of events are within at most 23\% of each other.

\begin{table}[htbp]
    \centering
    \caption{Predicted and best-fit numbers of atmospheric events using the Horiuchi+09 model~\cite{bib:horiuchi09}. The predicted numbers of events in the NCQE and $\mu/\pi$ categories are quoted as ranges, since certain types of NC events could belong to either category after reconstruction.}
    \vspace{+3truept}
    \begin{tabular}{c|cccc}
    \toprule
         & Decay-e & $\nu_e$ CC & NCQE & $\mu/\pi$ \\
         \hline
         Predicted & 218.5 & 121.4 & 81.9$-$138.4 & 10.6$-$67.2\\
         Observed & 169.5 & 109.9 & 88.2 & 27.9\\
         \botrule
    \end{tabular}
    \label{tab:atmpred}
\end{table}

\subsection{Modeling of spallation backgrounds}
\begin{figure}[htbp]
    \centering
    \includegraphics[width=\linewidth]{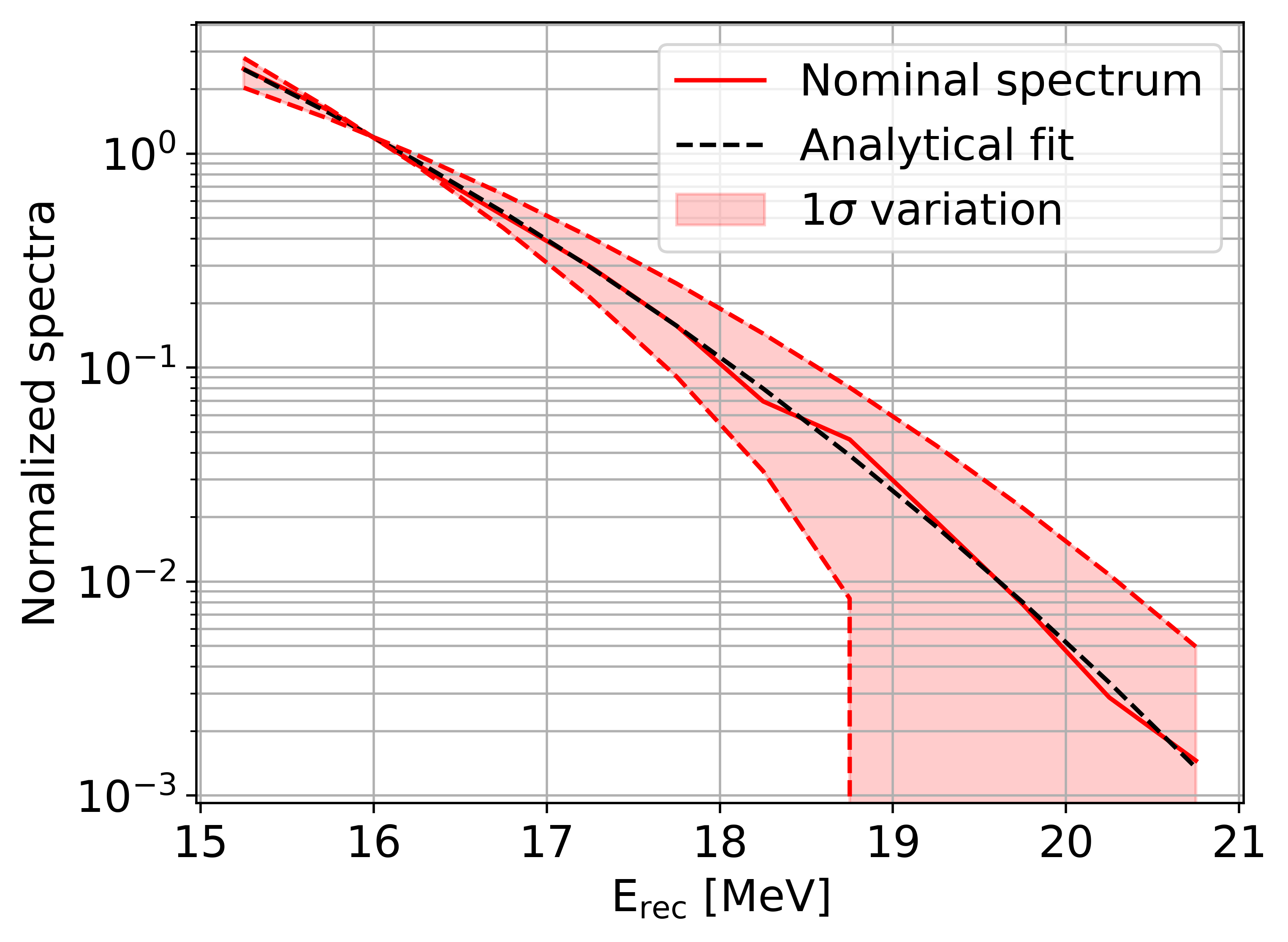}
    \caption{Nominal spallation spectrum (solid red) and its associated analytical fit (dashed black) for SK-IV. The 1$\sigma$ systematic uncertainty is shown as a red band.}
    \label{fig:spallspectra}
\end{figure}

To model the spallation background PDF we vary the isotope composition of the background as described in Equation~\ref{eq:spalike} and take the nominal spectrum as the average between the two extreme slopes. We parameterize this average by the following function:

\begin{align}
    S_{\rm spall}(E_{\rm rec}) &= N \exp\left(-\frac{(E_{\rm rec} + 0.511~\text{MeV})^\alpha}{\beta}\right),
    \label{eq:spallexp}
\end{align}

\noindent 
where $N$ is a normalization factor and $\alpha,\beta$ are determined by a fit and depend on the SK phase considered. Then, as described in Section~\ref{sec:spallsys}, we both vary the isotope composition of the spallation spectrum and distort it to account for energy scale and resolution uncertainties. We then fit the resulting extreme spectra by the function from Equation~\ref{eq:spallexp} times a third order polynomial:

\begin{align}
    S'_{\rm spall}(E_{\rm rec}) =& N' \exp\left(-\frac{(E_{\rm rec} + 0.511~\text{MeV})^\alpha}{\beta}\right) \nonumber \\
    &\times \left[1 + \mathcal{P}_3(E_{\rm rec})\right],
\end{align}

\noindent 
and thus parameterize the 1$\sigma$ systematic error distorsions. The nominal spectrum and its associated analytical fit are shown in Fig.~\ref{fig:spallspectra} for SK-IV with its 1$\sigma$ uncertainty range. Note that, in order to account for the steepness of the spallation spectrum while avoiding unphysical results, we impose $S'_{\rm spall}(E_{\rm rec}) = 0$ for energies larger than the minimal $E_{\rm rec}$ for which the distorted PDF is zero.

\bibliography{reference}

\end{document}